\documentclass[aps,prb,twocolumn]{revtex4} %,preprint

\usepackage{graphicx}% Include figure files
\usepackage{dcolumn}% Align table columns on decimal point
\usepackage{bm}% bold math
\usepackage{amsmath}

\newcommand{\comment}[1]{}
\begin{document}
%\preprint{}   % Preprint number in upper right corner
\renewcommand{\theequation}{\arabic{section}.\arabic{equation}}

\title{Quantum Statistical Mechanics as an Exact Classical Expansion
with Results for  Lennard-Jones Helium}

%\author{}
%\thanks{}

\author{Phil Attard}
%\affiliation{\protect\texttt{phil.attard1@gmail.com}}

\date{10 Oct., 2016. v2. \ phil.attard1@gmail.com}
%\\ ..  Begun: 21.07.2016. Projects/QSM-LJ16/QSMLJ.tex}

\begin{abstract}
The quantum states representing classical phase space are given,
and these are used to formulate quantum statistical mechanics
as a formally exact double perturbation expansion
about classical statistical mechanics.
%in terms of classical phase space averages.
One series of quantum contributions arises
from the non-commutativity of the position and momentum operators.
Although the formulation of the  quantum states differs,
the present results for separate averages of position operators
and of momentum operators
agree with  Wigner (1932) and Kirkwood (1933).
The second series arises from wave function symmetrization,
and is given in terms of $l$-particle permutation loops
in an infinite order re-summation.
%The even $l$ terms are of opposite sign for bosons and fermions.
The series gives analytically the known exact result %for the grand potential
for the quantum ideal gas to all orders.
The leading correction corrects a correction given by Kirkwood.

The first four quantum corrections to the grand potential
%due to symmetrization %, $\Omega_{2,0}$ and $\Omega_{3,0}$,
%and to non-commutativity %, $\Omega_{1,2'}$,
are calculated  for a Lennard-Jones fluid
using the hypernetted chain closure.
For helium on liquid branch isotherms,
the corrections range from several times to 1\%
of the total classical pressure,
with the effects of non-commutativity being significantly larger
in magnitude than those of wave function symmetrization.
All corrections are found to be negligible for argon
at the densities and temperatures studied.
The calculations are computationally trivial
as the method avoids having
to compute eigenfunctions, eigenvalues,
and numerical symmetrization.
\end{abstract}

%\keywords{}

\maketitle

%%%%%%%%%%%%%%%%%%%%%%%%%%%%%%%%%%%%%%%%%%%%%%%%%%%%%%%%%%%%%%%%%%%%%%%%%%
%
\section{Introduction}
\setcounter{equation}{0} \setcounter{subsubsection}{0}
%\renewcommand{\theequation}{\Alph{section}.\arabic{equation}}
%
%%%%%%%%%%%%%%%%%%%%%%%%%%%%%%%%%%%%%%%%%%%%%%%%%%%%%%%%%%%%%%%%%%%%%%%%%%

Quantum mechanics is fine in theory.
But for applications to real-world condensed matter systems,
quantum mechanics has proven computationally intractable.
The two main impediments
are the need to obtain the eigenvalues and eigenfunctions of the system,
and the need to symmetrize the wave function.
The scaling of both problems with the size of the system is prohibitive.

Faced with these challenges,
workers have had two choices.
They could choose to focus on a small system
and hope to develop techniques to solve it exactly up to a certain level,
or else they could look to use approximate approaches on larger systems.
As an example of the former,
Hernando and Vin\`i\v cek\cite{Hernando11}
found the first 50 eigenfunctions for five Lennard-Jones particles
using a non-uniform discretization of space and time.
The algorithm is said to be considerably more efficient
than the ${\cal O}(e^N)$ scaling of standard finite difference methods.
Nevertheless the method requires diagonalization of an  $M \times M$
sparse matrix,
with $M \approx 10^5$ for five particles in one dimension,
and apparently scaling as ${\cal O}(N^3)$.
This is \emph{prior} to wave function symmetrization.
As an example of an approximate approach,
Georgescu  and Mandelshtam\cite{Georgescu11}
approximated the wave functions with unsymmetrised Gaussian wave packets
in a variational method and
with them studied the ground state of
Lennard-Jones clusters of up to 6,500 atoms.
%Neither approach included interactions with a reservoir.
This work estimated that the scaling with system size was reduced from
${\cal O}(N^3)$ to ${\cal O}(N^2)$.
Again this is without wave function symmetrization.
The reader is referred to either article
for a comprehensive review of
the state of the art in computational methods.\cite{Hernando11,Georgescu11}

In contrast, the exact numerical simulation of classical systems
is almost trivial,
with $10^3$--$10^4$ particles being routine,
and with the computational burden scaling linearly with system size
(if a potential cut-off and neighbor tables are used).\cite{Allen87}
Monte Carlo simulations in particular account exactly for interactions
with thermodynamic reservoirs,
as do molecular dynamic approaches that include a stochastic thermostat.
\cite{NETDSM}
There have been a number of semi-classical approaches to quantum systems
that invoke various forms for the wave functions or density matrix
(see, for example, Ch.~10 of Ref.~\onlinecite{Allen87}),
with mixed results.

These observations suggest that it would probably be most efficient
to avoid the wave function or density matrix altogether,
if this were at all possible,
and to instead treat quantum systems
as some sort of perturbation expansion of classical systems.

To this end there have been attempts
to formulate quantum mechanics in terms of
a classical phase space representation of quantum states,
with most attention paid to the quasi-probability distribution
first introduced by  Wigner,
and shortly later modified by Kirkwood.
\cite{Wigner32,Kirkwood33,Groenewold46,Moyal49,Zachos05}
The Wigner phase space quasi-probability distribution has
found applications in quantum optics
\cite{Barnett03,Gerry05,Najarbashi16,Park16}
and quantum information theory.\cite{Braunstein05,Weedbrook12}
%It has also been used to discuss
%the quantum to classical transition.\cite{Zurek03}
The Wigner and Kirkwood quasi-probability functions
are similar but not identical,
which can be said of all the phase space quasi-probability functions
that have been proposed.\cite{Praxmeyer02,Dishlieva08}
%It is puzzling that there should be such ambiguity
%in such a fundamental physical concept
%as the quantum representation of classical phase space.
% In particular, Wigner's expression arises from his axiomatic assertion
%that the quasi-probability distribution should evolve via Liouville's equation,
%which Kirkwood's distribution does not do.\cite{Praxmeyer02}
%It has been argued by the present author that Liouville's equation
%for an open system is an approximation, not a theorem,
%that neglects the stochastic interactions with the environment or reservoir
%(\S\S 3.7.5, 7.5.3 of Ref.~\onlinecite{NETDSM}).

The phase space quasi-probability distribution
given by Wigner\cite{Wigner32} neglected wave function symmetrization,
whereas that given by Kirkwood\cite{Kirkwood33}
accounted for the leading order correction
(ie.\ the single transposition of a pair of particles).
The sometimes dramatic effects that are observed to arise from
particle statistics in some systems suggests
that the approximation of Wigner, Kirkwood, %\cite{Wigner32,Kirkwood33}
and followers may not always capture
the most important quantum corrections to the classical.

The present paper also
%like the work of Wigner\cite{Wigner32} and Kirkwood,\cite{Kirkwood33}
seeks the quantum states that correspond to classical phase space,
with the aim of developing a perturbation expansion
for quantum statistical mechanics that has classical statistical mechanics
as the leading order term.
The twin motivations are
the relative computational simplicity of classical systems,
and also the observation that quantum effects are but a small
perturbation for most terrestrial condensed matter systems.

The reasons for focussing on quantum statistical mechanics
rather than on quantum mechanics
are both practical
(real world problems are thermodynamic in nature,
and at the molecular level
large systems are most efficiently treated statistically),
and, more importantly, conceptual.
Quantum statistical mechanics
applies to open quantum systems,
and in these the wave function has collapsed into mixed quantum states
with no interference.
\cite{Zurek03,Davies76,Breuer02,Weiss08,Zeh01,QSM1,QSM}
This marks the distinction between the quantum and the classical.
With  quantum statistical mechanics formulated
as the  sum over states of the Maxwell-Boltzmann operator,\cite{Neumann27}
one sees the beginning of the connection with classical statistical
mechanics,
which is essentially the integral over classical phase space
of Maxwell-Boltzmann probability.

So the end-point point is clear
---a theory of condensed matter with classical statistical mechanics
as the leading order term.
As also is the starting point ---quantum statistical mechanics.
This paper maps the route between the two.

In \S \ref{Sec:Sym-chi-Xi}
the grand partition function is formally written
as the sum over unique states
by invoking an overlap factor for double counting
due to wave function symmetrization.
In \S \ref{Sec:Basis} the basis wave functions
that lead to classical phase space are given,
and in \S \ref{Sec:Xi-qp}
the grand partition function is formulated  in terms of them.
In \S \ref{Sec:MBtilde} the Maxwell-Boltzmann operator
is recast as a function suitable for expansion in powers of Planck's constant,
which accounts for the non-commutativity of the position
and momentum operators,
and in \S \ref{Sec:OverlapFac}
this is incorporated into the  grand partition function
as a weight for phase space.
In \S \ref{Sec:Xi-Omega}
the permutations that comprise the overlap factor
for wave function symmetrization
are written as a sum over distinct permutation loops,
which allows the grand partition function to be resummed
and the grand potential to be written as a sum
of classical equilibrium averages over phase space of the loops.
In \S \ref{Sec:Stat-Av} the statistical average of a quantum operator
is expressed in the same formalism.
In \S \ref{Sec:compare} the formalism
is compared with known results,
and in \S \ref{Sec:numerical} numerical results
for the quantum corrections in liquid helium are given.

\comment{ %%%%%%%%%%%%%%%%%%%%%%%%%%%%%%%%%%%%%%%%%%%%%%
%The present paper gives an account of the journey,
%and it describes several prominent landmarks visible
%upon the completion of the trip.
The first challenge is to identify precisely the quantum states
that give rise to classical phase space.
Next one must show that the expectation value
of the Maxwell-Boltzmann operator
in those states is equal in value to the classical  Maxwell-Boltzmann
probability function in phase space.
In doing this one has to account for the non-commutativity
of position and momentum operators.
One also has to account for
the symmetry of the wave function
with respect to particle interchange.
It turns out that properly formulating symmetrization
is the key to developing the perturbation expansion.
Having completed these tasks,
it will be shown that one has
a formulation of quantum statistical mechanics
with classical statistical mechanics as the leading term,
and with successive quantum corrections that remain tractable to compute.
} % end comment %%%%%%%%%%%%%%%%%%%%%%%%%%%%%%%%%%%%%%%%%%%%%

%%%%%%%%%%%%%%%%%%%%%%%%%%%%%%%%%%%%%%%%%%%%%%%%%%%%%%%%%%%%%%%%%%%%%%%%%%
%
\section{Formal Analysis} \label{Sec:Symmetry}
\setcounter{equation}{0} \setcounter{subsubsection}{0}
%
%%%%%%%%%%%%%%%%%%%%%%%%%%%%%%%%%%%%%%%%%%%%%%%%%%%%%%%%%%%%%%%%%%%%%%%%%%

\subsection{Symmetrization, Overlap, and the Grand Partition Function}
\label{Sec:Sym-chi-Xi}

%\subsubsection{Entropy Eigenfunctions}

The grand partition function for a quantum system is
\cite{Neumann27,Messiah61,Merzbacher70,QSM1,QSM,Attard16}
\begin{eqnarray} \label{Eq:Xi=sum'-n}
\Xi & = & \mbox{TR } z^{ N} e^{-\beta \hat {\cal H}}
\nonumber \\ & = &
\sum_{N=0}^\infty z^N \sum_{\bf m} \!' e^{-\beta {\cal H}_{\bf m}}
\nonumber \\ & = &
\sum_{N=0}^\infty \frac{ z^N }{N!}
\sum_{\bf m} \chi_{\bf m}  e^{-\beta {\cal H}_{\bf m}}.
\end{eqnarray}
Here $N$ is the number of particles,
the fugacity is $z \equiv e^{\beta \mu}$,
where $\mu$ is the chemical potential,
$\beta = 1/k_\mathrm{B}T$  is sometimes called the inverse temperature,
with $k_\mathrm{B}$ being Boltzmann's constant and $T$ the temperature,
and $\hat{\cal H}$ is the Hamiltonian or energy operator.
%The states ${\bf m}$ are entropy states.
The first equality is the conventional expression,
\cite{Neumann27,Messiah61,Merzbacher70,QSM1,QSM}
and perhaps also the second;\cite{Messiah61,Attard16}
the final equality, not so much.\cite{Attard16}

In the second equality the prime on the summation
indicates that the sum is over distinct entropy states.\cite{Messiah61,Attard16}
Some authors, specifically Kirkwood,\cite{Kirkwood33}
%as is discussed in the detailed comparison below,
and some text book writers, specifically Pathria,\cite{Pathria72}
neglect this point.
Since the entropy operator is proportional to the energy operator,
$\hat S_\mathrm{r} = -\hat {\cal H}/T$,
entropy states are the same as energy states,
entropy eigenvalues are proportional to energy eigenvalues,
$S_{\mathrm{r},{\bf m}} = -{\cal H}_{\bf m}/T$,
and the entropy eigenfunctions $\phi_{\bf m}$
are also energy eigenfunctions.
(The grand partition function is not restricted to entropy states,
but it is these that originally give rise to the form,
and it is these that are shortly used to introduce the expectation value.)

In the third equality the sum over all entropy states has been invoked,
with the factor  $\chi_{\bf m}/N!$,
which is explained next,
properly accounting for the fact that some states are counted multiple times.
\cite{Attard16}

In quantum mechanics,
under particle interchange
the wave function is fully symmetric for identical bosons
and fully anti-symmetric for identical fermions.
Hence the basis wave functions must be symmetrized
from the more general basis wave functions
by writing
\cite{Messiah61,Merzbacher70,Attard16}
\begin{equation} \label{Eq:zeta^SA}
\phi_{{\bf m}}^\mathrm{S/A}({\bf r})
\equiv
\frac{1}{\sqrt{ N!\chi^\pm(\phi_{\bf m}) } }
\sum_{\hat{\mathrm P}} (\pm 1)^p
 \, \phi_{{\bf m}}(\hat{\mathrm P}{\bf r}) .
\end{equation}
Here ${\bf r}$ are the particle coordinates,
which is to say that the position representation is invoked here and below.
Below $\phi_{{\bf m}}$ will be taken to be an entropy (equivalently, energy)
eigenfunction, although this symmetrization equation in fact holds
for any wave function.
The superscript  S signifies symmetric,
and it applies for bosons using $(+ 1)^p=1$ on the right hand side.
The superscript A signifies anti-symmetric,
and it applies for fermions using $(- 1)^p$ on the right hand side.
Here $\hat{\mathrm P}$ is the permutation operator,
and $p$ is its parity
(ie.\ the number of pair transpositions
that comprise the permutation).
One can equivalently permute instead the state label ${\bf m}$.

The prefactor including $\chi_{\bf m}$ ensures the correct normalization,
\begin{equation}
\langle \phi^\mathrm{S/A}_{\bf m'}\, | \, \phi^\mathrm{S/A}_{\bf m} \rangle
= \delta_{{\bf m}',{\bf m}} ,
\end{equation}
where the Kronecker-$\delta$ function appears.
%This holds even if ${\bf m}'$ is a permutation of ${\bf m}$
%that is the same as  ${\bf m}$:
%${\bf m}' \equiv \hat{\mathrm P} {\bf m} = {\bf m}$,
%$\hat{\mathrm P}\ne \hat{\mathrm I}$.
%This can occur if different particles are in the same one-particle state.
Inserting in this the expression for the symmetrized wave function
and rearranging gives the overlap factor as
\begin{eqnarray} \label{Eq:chi_m}
\chi^\pm(\phi_{\bf m})
& = &
\frac{1}{N!}
\sum_{\hat{\mathrm P},\hat{\mathrm P}'} (\pm 1)^{p+p'}
\langle \phi_{{\bf m}}(\hat {\mathrm P}{\bf r})
| \phi_{{\bf m}}(\hat {\mathrm P}'{\bf r}) \rangle
\nonumber \\  & = &
\sum_{\hat {\mathrm P}}
(\pm 1)^{p}
\langle \phi_{{\bf m}}(\hat {\mathrm P}{\bf r})
| \phi_{{\bf m}}({\bf r}) \rangle
\nonumber \\ & = &
\sum_{\hat{\mathrm P}} (\pm 1)^{p}
\int \mathrm{d}{\bf r}\;
\phi_{{\bf m}}(\hat{\mathrm P}{\bf r})^*
\phi_{{\bf m}}({\bf r}) .
\end{eqnarray}
This shows that in each case the overlap factor depends upon
the chosen basis wave functions.
This quantity is called the overlap factor
because it tells how much symmetrization
counts the same microstate multiple times.
The formula for the overlap factor, Eq.~(\ref{Eq:chi_m}),
holds for multi-particle states,
as well as for when the entropy microstate consists of
one-particle states.
For a detailed discussion see Ref.~\onlinecite{Attard16}.

In general for identical particles the Hamiltonian operator
is unchanged by a permutation of the particles,
\begin{equation}
\hat {\cal H}({\bf r})
=
\hat {\cal H}(\hat{\mathrm P}{\bf r}).
\end{equation}
In this work the Hamiltonian operator is taken to be
\begin{equation}
\hat {\cal H}({\bf r})
=
U({\bf r}) - \frac{\hbar^2}{2m} \nabla^2 ,
\end{equation}
where $U({\bf r})$ is the potential energy,
$\hbar$ is Planck's constant divided by $2\pi$,
and $m$ is the particle mass.

Let the basis wave functions $\phi_{\bf m}$ be entropy eigenfunctions.
Hence they are also energy eigenfunctions,
\begin{equation}
\hat {\cal H}({\bf r}) \phi_{\bf m}({\bf r})
= {\cal H}_{\bf m} \phi_{\bf m}({\bf r}) .
\end{equation}
Obviously  the symmetrized basis wave functions are also
eigenfunctions of the entropy operator
with unchanged eigenvalues.

The grand partition function given above can be recast
explicitly in terms of the entropy eigenfunctions,
\begin{eqnarray} \label{Eq:Xi(phi)}
\Xi^\pm & = &
\sum_{N} \frac{z^N}{N!}
\sum_{\bf m} \chi^\pm(\phi_{\bf m}) e^{-\beta {\cal H}_{\bf m}}
\nonumber \\  & = &
\sum_{N} \frac{z^N}{N!} \sum_{\bf m}
\sum_{\hat {\mathrm P}}
(\pm 1)^{p}
\langle \phi_{\hat {\mathrm P}{\bf m}} | \phi_{{\bf m}} \rangle
 e^{-\beta {\cal H}_{\bf m}}
\nonumber \\ & = &
\sum_{N} \frac{z^N}{N!}
\sum_{\bf m} \sum_{\hat{\mathrm P}} (\pm 1)^p
\langle \phi_{\hat {\mathrm P}{\bf m}} |
 e^{-\beta \hat {\cal H}} |\phi_{\bf m} \rangle .
\end{eqnarray}
This formulation of the grand partition function
as a (permuted) expectation value is a small
but significant modification of the original.\cite{Attard16}
Although derived for entropy basis functions,
the expectation value holds for any set of basis functions,
which has the  advantage that one does not have
to find the entropy eigenfunctions.
This formulation as an  expectation value
provides the starting point for the present paper.

%%%%%%%%%%%%%%%%%%%%%%%%%%%%%%%%%%%%%%%%%%%%%%%%%%%%%%%%%%%%%%%%%%%%%%%
\subsection{Basis Functions} \label{Sec:Basis}

It is an important conceptual point,
in this section and in the rest of the paper,
that there is a distinction between the coordinate positions, ${\bf r}$,
and the configurations positions ${\bf q}$.
The coordinate positions appear as the argument of the wave function
and operators in the position representation.
The coordinate positions appear in classical phase space.
Essentially, the conceptual distinction between the two
is that the coordinate positions ${\bf r}$ are a representation,
and the configuration positions ${\bf q}$ are a quantum state.

The predecessors of the present author,
specifically Wigner\cite{Wigner32} and Kirkwood,\cite{Kirkwood33}
treat these as the same thing.\cite{fn1}
The present paper keeps them distinct.
To be sure, as is detailed below,
a zero-width limit is invoked in which certain wave functions
become Dirac-$\delta$ functions, $\delta({\bf r}-{\bf q})$,
and in this limit the two quantities become equal to each other.
But the point to emphasize is that this limit is invoked \emph{after}
certain mathematical manipulations.
It is essential to keep in mind the passage to this limit
in order to achieve the final results.

%%%%%%%%%%%%%%%%%%%%%%%%%%%%%
\subsubsection{Wave Packets}

In Ref.~\onlinecite{Attard16}
the problem of quantum statistical mechanics
was approached by using minimum uncertainty wave packets
as the basis for the basis functions.
These are
\begin{equation}
\zeta_{{\bf q}{\bf p}}({\bf r})
= \frac{1}{(2\pi\xi^2)^{3N/4}}
e^{-{\varepsilon}_{{\bf q}}({\bf r})^2 /4\xi^2}
e^{ -{\bf p}\cdot {\bm \varepsilon}_{{\bf q}}({\bf r})/i\hbar} ,
\end{equation}
where ${\bm \varepsilon}_{{\bf q}}({\bf r}) \equiv {\bf r}-{\bf q}$.
For $N$ particles in three dimensions,
the vectors are  3N-dimensional,
with ${\bf r}$ being the coordinate positions,
${\bf q}$ being the configuration positions,
and ${\bf p}$ being the configuration momenta.

The motivation for using these as the starting point
of the analysis was that they are approximately
entropy (energy) eigenfunctions,
\begin{eqnarray} \label{Eq:hatH-zeta}
\lefteqn{
\hat{\cal H}({\bf r}) \, \zeta_{{\bf q}{\bf p}}({\bf r})
} \nonumber  \\
& = &
\left\{ U({\bf r})
+ \frac{1}{2m} {\bf p} \cdot {\bf p}
+\frac{3N\hbar^2}{4m\xi^2}
\right. \nonumber \\ &&  \left. \mbox{ }
 - \frac{\hbar^2}{8m\xi^4}
{\bm \varepsilon}_{{\bf q}}({\bf r}) \cdot  {\bm \varepsilon}_{{\bf q}}({\bf r})
 - \frac{\hbar^2}{2mi\hbar\xi^2}
{\bm \varepsilon}_{{\bf q}}({\bf r}) \cdot  {\bf p}
\right\} \zeta_{{\bf q}{\bf p}}({\bf r})
\nonumber \\ & \approx &
\left\{ {\cal H}({\bf \Gamma}) +\frac{3N\hbar^2}{4m\xi^2}
\right\} \zeta_{{\bf q}{\bf p}}({\bf r}) ,
\end{eqnarray}
where a point in classical phase space is
${\bf \Gamma} \equiv \{{\bf q},{\bf p}\}$.
The second equality assumes that the wave packet is sharply peaked so that
terms involving powers of
${\bm \varepsilon}_{{\bf q}}({\bf r})$ can be neglected.
This includes the terms in the expansion of
$U({\bf r}) = U({\bf q})
+ {\bm \varepsilon}_{{\bf q}}({\bf r}) \cdot {\bf U}'({\bf q})
+ \ldots $.
The classical Hamiltonian is
\begin{equation}
{\cal H}({\bf \Gamma})
= \frac{1}{2m} {\bf p} \cdot {\bf p} + U({\bf q}) .
\end{equation}
Since one expects that any expansion of quantum statistical mechanics
will have classical statistical mechanics as the leading term,
this provided two further motivations for using wave packets,
namely that the classical Hamiltonian is the approximate  eigenvalue,
and points in classical phase space are the index of the wave packets.

Despite these advantages wave packets have several disadvantages.
The first is that they are approximate, not exact, entropy eigenfunctions.
The algebra required to systematically modify them
so as to reduce the error in the approximation rapidly proliferates,
as does the computational complexity.\cite{Attard16}
The second is that they do not form an exact basis set
because they  overlap for finite wave packet widths $\xi$
and they are therefore non-orthogonal.
With the wisdom of hindsight,
the latter problem appears prohibitive.

Nevertheless, the original motivation
---that minimum uncertainty wave packets
are close to entropy eigenfunctions,
and that their indeces are points in classical phase space---
remain compelling.
It turns out to be exceedingly fruitful to instead consider
two closely related basis sets.

%%%%%%%%%%%%%%%%%%%%%%%%%%%
\subsubsection{Plane Waves}

One half of the wave packets given above are plane waves.
Plane waves localize the momenta of the particles.
They are in fact kinetic energy eigenfunctions.
These motivate considering plane waves as a basis set,
\begin{eqnarray}
\zeta_{\bf p}({\bf r})
& =&
\frac{1}{V^{N/2}}
e^{-{\bf p}\cdot{\bf r}/i\hbar}
\nonumber \\ & = &
 \prod_{j=1}^N \frac{ e^{-{\bf p}_j \cdot {\bf r}_j/i\hbar} }{ V^{1/2} } ,
\end{eqnarray}
with the configuration momentum being
${\bf p} = 2 \pi \hbar {\bf n} /V^{1/3}$,
$n = 0, \pm 1, \pm 2, \ldots$,
and $V$ the volume.
The quantization arises from imposing periodic boundary conditions.
From this the width of a momentum state is evidently
\begin{equation}
\Delta_p = 2 \pi \hbar /V^{1/3}.
\end{equation}
This is used below in taking the continuum limit,
after which quantization of the momentum index becomes irrelevant.

Plane waves are orthonormal,
%\begin{equation}
%\langle \zeta_{{\bf p}'} | \zeta_{\bf p} \rangle =\delta_{{\bf p}',{\bf p}} ,
%\end{equation}
\begin{eqnarray}
\langle \zeta_{{\bf p}'} | \zeta_{\bf p} \rangle
& = &
\frac{1}{V^{N}}
\int \mathrm{d}{\bf r} \;
e^{-{\bf r} \cdot ( {\bf p}-{\bf p}')/i\hbar}
\nonumber \\ & = &
\delta_{{\bf p}',{\bf p}} ,
\end{eqnarray}
as follows from the boundary conditions.
They also form a complete set,
\begin{eqnarray}
\sum_{\bf p}  \zeta_{\bf p}({\bf r}')^* \zeta_{\bf p}({\bf r})
& = &
\frac{V^{-N}}{(2\pi \hbar V^{-1/3})^{3N}}
\int \mathrm{d}{\bf p} \;
e^{-{\bf p} \cdot ( {\bf r}-{\bf r}')/i\hbar}
\nonumber \\ & = &
 \delta( {\bf r}-{\bf r}') ,
\end{eqnarray}
since $\int \mathrm{d}x \, e^{ikx} = 2\pi \delta(k)$.
In bra-ket notation completeness is written
$ \sum_{\bf p} | \zeta_{\bf p} \rangle \, \langle \zeta_{\bf p} |
= \hat{\mathrm I}$.

%%%%%%%%%%%%%%%%%%%%%%%%%%%
\subsubsection{Gaussians}

The other half of wave packets  are Gaussians,
which localize the positions of the particles.
Here they are taken to be
\begin{eqnarray}
\zeta_{\bf q}({\bf r})
& = &
\frac{e^{-({\bf r} -{\bf q})^{2}/4\xi^2}}{(2\pi\xi^2)^{3N/4}}
\nonumber \\ & = &
 \prod_{j=1}^N
\frac{ e^{-({\bf r}_j - {\bf q}_j)^2/4\xi^2} }{ (2\pi\xi^2)^{3/4} }.
\end{eqnarray}
(Using $4\xi^2$ rather than say $2\xi^2$ in the exponent
is an evolutionary accident;
its parent  is the minimal uncertainty wave packet.
There is nothing significant in it
and it in no way affects the following results.)

The overlap or orthonormality of such Gaussians is
\begin{eqnarray}
\langle \zeta_{\bf q} | \zeta_{{\bf q}'}\rangle
& = &
\frac{1}{(2\pi\xi^2)^{3N/2}}
\int \mathrm{d}{\bf r} \;
e^{-({\bf r} -{\bf q})^{2}/4\xi^2}
e^{-({\bf r} -{\bf q}')^{2}/4\xi^2}
\nonumber \\ & = &
e^{ -({\bf q}-{\bf q}')^2/8\xi^2}
\nonumber \\ & \equiv &
\delta_\xi({\bf q},{\bf q}').
\end{eqnarray}
This is a soft Kronecker-$\delta$
that becomes an exact Kronecker-$\delta$
in the limit $\xi \rightarrow 0$.

The completeness of the set of Gaussians is manifest as
\begin{eqnarray}
\lefteqn{
\sum_{\bf q} | \zeta_{\bf q}({\bf r}') \rangle \,
\langle \zeta_{\bf q}({\bf r})  |
} \nonumber \\
& = &
\frac{1}{\Delta_q^{3N}} \int \mathrm{d}{\bf q}\;
\zeta_{{\bf q}}({\bf r})^* \, \zeta_{{\bf q}}({\bf r}')
\nonumber \\ & = &
\frac{(2\pi\xi^2)^{-3N/2}}{\Delta_q^{3N}}
\int \mathrm{d}{\bf q}\;
e^{-({\bf r}-{\bf q})^2/4\xi^2}
e^{-({\bf r}'-{\bf q})^2/4\xi^2}
\nonumber \\ & = &
\frac{1}{\Delta_q^{3N}}
 e^{-({\bf r}'-{\bf r})^2/8\xi^2}  .
\end{eqnarray}
Normalization (ie.\ demanding that this integrate to unity)
fixes the spacing of the configuration position states as
\begin{equation}
\Delta_q = \sqrt{8\pi\xi^2}.
\end{equation}
The completeness now is
\begin{eqnarray} \label{Eq:zetaq-complete}
\sum_{\bf q} | \zeta_{\bf q}({\bf r}') \rangle \,
\langle \zeta_{\bf q}({\bf r})  |
& = &
\frac{e^{-({\bf r}'-{\bf r})^2/8\xi^2}}{(8\pi\xi^2)^{3N/2}}
\nonumber \\ & \equiv &
 \delta_\xi({\bf r}'-{\bf r}) .
\end{eqnarray}
This is a  soft Dirac-$\delta$ function that is normalized
and that becomes an exact Dirac-$\delta$ function
in the limit $\xi \rightarrow 0$.
Hence
\begin{equation}
\lim_{\xi \rightarrow 0}
\sum_{\bf q} | \zeta_{\bf q} \rangle \, \langle \zeta_{\bf q} |
= \hat{\mathrm I} .
\end{equation}

One concludes that Gaussians form a basis set
that is orthonormal and complete in the limit that the width of the Gaussian
goes to zero.
The fact that this Gaussian becomes
a generalized function in the zero width limit is no cause for concern.
The real test of the theory is whether or not its final
formulation is finite and well-behaved in the zero width limit.

A Gaussian can be expanded in terms of plane waves (and \emph{vice versa}).
One has
\begin{equation} \label{Eq:zq=gqp*zp}
\zeta_{\bf q}({\bf r})
=
\sum_{\bf p} g_{\bf p}({\bf q})
\zeta_{\bf p}({\bf r}) ,
\end{equation}
with the coefficients being
\begin{eqnarray}  \label{Eq:zq=gqp}
g_{\bf p}({\bf q})
& = &
\langle \zeta_{\bf p} | \zeta_{\bf q} \rangle
\nonumber \\ & = &
\frac{V^{-N/2}}{(2\pi\xi^2)^{3N/4}}
\int \mathrm{d}{\bf r} \,
e^{-({\bf r} -{\bf q})^{2}/4\xi^2}
e^{{\bf p}\cdot{\bf r}/i\hbar}
\nonumber \\ & = &
\frac{V^{-N/2}}{(2\pi\xi^2)^{3N/4}}
\int \mathrm{d}{\bf r} \,
e^{-({\bf r} - {\bf q} - 2\xi^2{\bf p}/i\hbar )^{2}/4\xi^2}
\nonumber \\ & & \times \mbox{ }
e^{{\bf q} \cdot {\bf p}/i\hbar }
e^{- \xi^2 p^2/\hbar^2 }
%\nonumber \\ & = &
%\frac{V^{-N/2}}{(2\pi\xi^2)^{3N/4}} (4\pi\xi^2)^{3N/2}
%e^{{\bf q} \cdot {\bf p}/i\hbar } e^{- \xi^2 p^2/\hbar^2 }
\nonumber \\ & = &
\frac{(8\pi\xi^2)^{3N/4}}{V^{N/2}}
e^{- \xi^2 p^2/\hbar^2} e^{ {\bf q} \cdot {\bf p}/i\hbar } .
%okpa 28Aug16
\end{eqnarray}
It will sometimes prove useful
to write this as the product of individual single particle factors
\begin{eqnarray}
g_{\bf p}({\bf q})
& = &
\prod_{j=1}^N
\frac{ (8\pi\xi^2)^{3/4}}{V^{1/2}}
e^{- \xi^2 p_j^2/\hbar^2} e^{ {\bf q}_j \cdot {\bf p}_j/i\hbar }
\nonumber \\ & \equiv &
\prod_{j=1}^N
g_{{\bf p}_j}({\bf q}_j) .
\end{eqnarray}
The limit ${\xi \rightarrow 0}$ will be invoked
during the analysis,  as expedient.

Finally, it seems straightforward to include
spin in the theory
by multiplying the various basis functions
by the product of single particle spin functions.
This has not been done here
in order to keep the theory as simple as possible
and to keep the focus on the main aim of a phase space formulation
of quantum statistical mechanics.

%\newpage
%%%%%%%%%%%%%%%%%%%%%%%%%%%%%%%%%%%%%%%%%%%%%%%%%%%%
\subsection{Phase Space Representation of Quantum States} \label{Sec:Xi-qp}

The fact that plane waves form an orthonormal, complete basis set
means that the grand partition function, Eq.~(\ref{Eq:Xi(phi)}),
may be written as a sum over them.
This could simply be written down directly,
since the trace operation is universal and it can be expressed
as the sum over the set of any basis states.
However it is worthwhile to illustrate the use of the completeness property,
and  in full the derivation is
\begin{eqnarray}
\Xi^\pm & = &
\sum_{N} \frac{z^N}{N!}
\sum_{\bf m} \sum_{\hat{\mathrm P}} (\pm 1)^p
\langle \phi_{\hat{\mathrm P}{\bf m}} |
e^{-\beta \hat {\cal H}} |\phi_{\bf m} \rangle
\nonumber \\ & = &
\sum_{N} \frac{z^N}{N!}
\sum_{\bf m} \sum_{{\bf p},{\bf p}'}  \sum_{\hat{\mathrm P}} (\pm 1)^p
\nonumber \\ && \mbox{ } \times
\langle \phi_{\hat{\mathrm P}{\bf m}}
| \zeta_{\bf p} \rangle\, \langle \zeta_{\bf p}|
e^{-\beta \hat {\cal H}}
| \zeta_{{\bf p}'} \rangle\, \langle \zeta_{{\bf p}'} | \phi_{\bf m} \rangle
\nonumber \\ & = &
\sum_{N} \frac{z^N}{N!}
\sum_{{\bf p},{\bf p}'}  \sum_{\hat{\mathrm P}} (\pm 1)^p
%\nonumber \\ && \mbox{ } \times
\langle \zeta_{\bf p}| e^{-\beta \hat {\cal H}}
| \zeta_{{\bf p}'} \rangle\,
\langle \zeta_{{\bf p}'} | \zeta_{\hat{\mathrm P}{\bf p}} \rangle
\nonumber \\ & = &
\sum_{N} \frac{z^N}{N!}
\sum_{{\bf p}}  \sum_{\hat {\mathrm P}} (\pm 1)^p
%\nonumber \\ && \mbox{ } \times
\langle \zeta_{\bf p}| e^{-\beta \hat {\cal H}}
| \zeta_{\hat {\mathrm P}{\bf p}} \rangle .
\end{eqnarray}

In the same way the completeness of the Gaussian basis functions,
Eq.~(\ref{Eq:zetaq-complete}),
and their expansion in terms of plane waves, Eq.~(\ref{Eq:zq=gqp*zp}),
enable this to be re-written
\begin{eqnarray} \label{Eq:Xipm-1}
\Xi^\pm & = &
\sum_{N} \frac{z^N}{N!}
\sum_{\bf p} \sum_{\hat{\mathrm P}} (\pm 1)^p
\langle \zeta_{{\bf p}} |e^{-\beta \hat {\cal H}}
|\zeta_{\hat {\mathrm P}{\bf p}} \rangle
\nonumber \\ & = &
\sum_{N}\frac{z^N  }{N!}
 \sum_{{\bf q}}\sum_{\bf p} \sum_{\hat {\mathrm P}} (\pm 1)^p
%\nonumber \\ && \mbox{ } \times
\langle \zeta_{\bf p}
| \zeta_{{\bf q}} \rangle \, \langle \zeta_{{\bf q}} |
e^{-\beta \hat {\cal H}}
| \zeta_{\hat {\mathrm P}{\bf p}} \rangle
\nonumber \\ & = &
\sum_{N}\frac{z^N  }{N!}
\sum_{{\bf q},{\bf p}} g_{\bf p}({\bf q})
\sum_{\hat {\mathrm P}} (\pm 1)^p
\langle  \zeta_{{\bf q}}|
e^{-\beta \hat {\cal H}}
| \zeta_{\hat {\mathrm P}{\bf p}}\rangle
\nonumber \\ & \equiv &
\sum_{N}\frac{z^N  }{N!}
\sum_{{\bf q},{\bf p}} g_{\bf p}({\bf q})
\chi^\pm_{\cal H}(\zeta_{\bf q},\zeta_{\bf p}) .
\end{eqnarray}
It is emphasized that this formulation is formally exact in the limit
$\xi \rightarrow 0$.

The immediate and obvious merit of formulating the problem
in terms of this asymmetric expectation value
is that the sum over states has become
a sum over points in classical phase space.
The continuum limit is
\begin{equation}
\sum_{{\bf q},{\bf p}} \Rightarrow
\frac{1}{(\Delta_p\Delta_q)^{3N}} \int \mathrm{d}{\bf \Gamma} .
\end{equation}
The volume elements are those derived above
for plane waves,  $\Delta_p = 2 \pi \hbar /V^{1/3}$,
and for Gaussians, $\Delta_q = \sqrt{8\pi\xi^2}$.

One sees from this that
the combination of plane waves and Gaussian basis sets
in an asymmetric expectation value
provides a way of representing classical phase space.
The plane waves localize the momenta,
and the Gaussians localize the positions.
This way of representing phase space
is distinctly different to the way advocated by Wigner,\cite{Wigner32}
Kirkwood,\cite{Kirkwood33} and followers.
\cite{Groenewold46,Moyal49,Barnett03,Gerry05,Zachos05,Praxmeyer02,Dishlieva08}

The final equality for the grand partition function
defines the weighted overlap factor,
\begin{eqnarray} \label{Eq:chi-zeta}
\chi^\pm_{\cal H}(\zeta_{\bf q},\zeta_{\bf p})
& \equiv &
\sum_{\hat{\mathrm P}} (\pm 1)^{p}
\langle \zeta_{\hat{\mathrm P}{\bf q}} |e^{-\beta \hat {\cal H}}
| \zeta_{{\bf p}} \rangle
 \\ & = &
\sum_{\hat{\mathrm P}} (\pm 1)^{p}
\int \mathrm{d}{\bf r}\;
\zeta_{{\bf q}}(\hat{\mathrm P}{\bf r})^*
e^{-\beta \hat {\cal H}({\bf r}) }\zeta_{{\bf p}}({\bf r}) . \nonumber
\end{eqnarray}
Because one is summing over all permutations and states,
it makes no difference to which wave function index or coordinate argument
the permutator is applied.

It is worth mentioning that here the Maxwell-Boltzmann operator
is explicitly shown as acting on the plane wave basis function.
One could have formulated the problem with the opposite
asymmetry so that it acted on the Gaussian basis function
(or else just use its properties as a Hermitian operator).
In view of the results obtained in the following sub-section,
the present formulation is the simplest.

%The present formulation is similar to that given
%in Ref.~\onlinecite{Attard16} with the replacement
%$\chi_n^\pm e^{-\beta {\cal H}_{\bf n}}
%\Rightarrow \chi^\pm_{\cal H}(\zeta_{\bf q},\zeta_{\bf p})$.
%The present, more general and asymmetric formulation
%reduces to the original form when both basis functions
%are the same exact or approximate entropy eigenfunctions,
%as they were in Ref.~\onlinecite{Attard16}.

%%%%%%%%%%%%%%%%%%%%%%%%%%%%%%%%%%%%%%%%%%%%%%%%%%%%%%%%%%%%%%%%%%%%
\subsection{Expansion of the Maxwell-Boltzmann Operator} \label{Sec:MBtilde}

Wigner\cite{Wigner32},
in developing his pseudo-probability distribution for phase space
as a type of Fourier transform of a mixture of wave functions,
gave a useful result for the transformation of
the Maxwell-Boltzmann position coordinate matrix.
That result may be exploited here for the action
of the Maxwell-Boltzmann operator on the plane wave basis functions.

In general one has for the operator equation
\begin{equation}
e^{{\bf p}\cdot{\bf r}/i\hbar}
f(\hat{ O})
e^{-{\bf p}\cdot{\bf r}/i\hbar}
=
f(\hat{ \tilde O})
, \;\;
\hat{ \tilde O}
\equiv
e^{{\bf p}\cdot{\bf r}/i\hbar}
\hat O
e^{-{\bf p}\cdot{\bf r}/i\hbar} ,
\end{equation}
as can be confirmed by writing $f(x)$ as a power series.
Hence the action  of the Maxwell-Boltzmann
operator on a plane wave basis function
can be written as
\begin{eqnarray} \label{Eq:eBHz=zeBH}
e^{-\beta \hat{{\cal H}} } \zeta_{\bf p}({\bf r})
&=&
\zeta_{\bf p}({\bf r}) e^{-\beta \hat{\tilde{\cal H}} } 1
\nonumber \\ & \equiv &
\zeta_{\bf p}({\bf r})
e^{-\beta {\cal H}({\bf r},{\bf p}) } W({\bf r},{\bf p}).
\end{eqnarray}
The 1 written explicitly here is a reminder
that the operator acts on the constant unit wave function.
This arises because the left hand side is
a wave function, and one cannot have on the right hand side
an operator hanging with nothing to act on.
The modified energy operator that is induced by this is
\begin{eqnarray}
%\lefteqn{
\hat{\tilde{\cal H}}
%} \nonumber \\
& \equiv &
e^{{\bf p} \cdot {\bf r}/i\hbar}
\hat{\cal H}
e^{-{\bf p} \cdot {\bf r}/i\hbar}
\nonumber \\ & = &
e^{{\bf p} \cdot {\bf r}/i\hbar}
\left[ \frac{-\hbar^2}{2m} \nabla^2 + U({\bf r}) \right]
e^{-{\bf p} \cdot {\bf r}/i\hbar}
\nonumber \\ & = &
\frac{p^2}{2m} + U({\bf r})
- \frac{i\hbar}{m} {\bf p} \cdot \nabla - \frac{\hbar^2}{2m} \nabla^2 .
%\nonumber \\ & \equiv &
%\frac{p^2}{2m}  + U({\bf r}) + \hat{\cal D}.
\end{eqnarray}
%Obviously $ \hat{\cal D} 1 = 0$,
%which fact will be used in what follows without comment.

In the above a phase function $W({\bf r},{\bf p})$
was defined that satisfies
\begin{equation}  \label{Eq:W-defn}
e^{-\beta{\hat{\tilde{\cal H}}}} 1
=
e^{-\beta {\cal H}({\bf r},{\bf p}) } W({\bf r},{\bf p}).
\end{equation}
By inspection, $W(\beta=0) = 1$.
This will provide the basis for an expansion in powers of $\hbar$,
\begin{equation}
W = \sum_{n=0}^\infty W_n \hbar^n .
\end{equation}
The term $W_0=1$ is the classical term,
and the terms $W_n$, $n \ge 1$ may be called
the quantum corrections due to non-commutativity.
They arise from the action of the kinetic energy operator
on the potential energy
and they involve gradients of the potential energy.
An alternative formulation of this particular quantum correction
is given in appendices~\ref{Sec:W=exp(w)} and \ref{Sec:W=exp(w)2}.

The temperature derivative of the left hand side is
\begin{eqnarray}
\frac{\partial }{\partial \beta} e^{-\beta{\hat{\tilde{\cal H}}}} 1
& = &
- {\hat{\tilde{\cal H}}} e^{-\beta{\hat{\tilde{\cal H}}}} 1
\nonumber \\ & = &
- {\hat{\tilde{\cal H}}}
\left\{  e^{-\beta {\cal H} } W \right\},
\end{eqnarray}
and that of the right hand side is
\begin{eqnarray}
\frac{\partial }{\partial \beta}
\left\{
e^{-\beta {\cal H} } W
\right\}
& = &
\left\{\frac{\partial W }{\partial \beta} - {\cal H} \right\}
e^{-\beta {\cal H} } W .
\end{eqnarray}
Equating these and rearranging gives
\begin{eqnarray} \label{Eq:dW/dB}
%\lefteqn{
\frac{\partial W}{\partial \beta }
%} \nonumber \\
& = &
{\cal H}  W
-
e^{\beta {\cal H} }
{\hat{\tilde{\cal H}}} \left\{e^{-\beta {\cal H} } W\right\}
%\nonumber \\ & = &
%- e^{\beta {\cal H} } \hat {\cal D} \left\{e^{-\beta {\cal H} } W\right\}
\nonumber \\ & = &
\frac{i\hbar}{m} e^{\beta U}  {\bf p} \cdot \nabla
 \left\{e^{-\beta U } W\right\}
\nonumber \\  && \mbox{ }
+ \frac{\hbar^2}{2m} e^{\beta U }
\nabla^2  \left\{e^{-\beta U } W\right\} .
\end{eqnarray}
Integrating, with $W(\beta =0 ) = 1$,
\begin{eqnarray}
W
& = &
1
+ \frac{i\hbar}{m} \int_0^\beta \mathrm{d} \beta' \;
e^{\beta' U }  {\bf p} \cdot \nabla
\left\{e^{-\beta' U } W(\beta') \right\}
\nonumber \\  && \mbox{ }
+ \frac{\hbar^2}{2m} \int_0^\beta \mathrm{d} \beta' \;
e^{\beta' U }
 \nabla^2  \left\{e^{-\beta' U } W(\beta')  \right\}.
\nonumber \\
\end{eqnarray}
This is essentially identical to an expression given
by Kirkwood
for the case of the identity permutation.\cite{Kirkwood33}
(In the present paper
wave function symmetrization is treated differently to Kirkwood.)

Successive substitution gives
$W_0 = 1$,
the first order correction as
\begin{eqnarray} \label{Eq:W1}
W_1
& = &
\frac{i}{m} \int_0^\beta \mathrm{d} \beta' \;
e^{\beta' U }  {\bf p} \cdot \nabla
\left\{e^{-\beta' U } W_0(\beta') \right\}
\nonumber \\ & = &
\frac{i}{m} \int_0^\beta \mathrm{d} \beta' \;
(-\beta') {\bf p} \cdot \nabla U
\nonumber \\ & = &
\frac{-i\beta^2}{2m}  {\bf p} \cdot \nabla U ,
\end{eqnarray}
and the second order correction as
\begin{eqnarray} \label{Eq:W2}
W_2
& = &
\frac{i}{m} \int_0^\beta \mathrm{d} \beta' \;
e^{\beta' U }  {\bf p} \cdot \nabla
\left\{e^{-\beta' U } W_1(\beta') \right\}
\nonumber \\  && \mbox{ }
+ \frac{1}{2m} \int_0^\beta \mathrm{d} \beta' \;
e^{\beta' U }
 \nabla^2  \left\{e^{-\beta' U } W_0(\beta')  \right\}
\nonumber \\ & = &
\frac{i}{m} \int_0^\beta \mathrm{d} \beta' \;
\left[  (-\beta') ({\bf p} \cdot \nabla U) \times
\frac{-i\beta'^2}{2m}  {\bf p} \cdot \nabla U
\right. \nonumber \\ && \left. \mbox{ }
+ \frac{-i\beta'^2}{2m}  ( {\bf p} \cdot \nabla )^2 U
\right]
\nonumber \\  && \mbox{ }
+ \frac{1}{2m} \int_0^\beta \mathrm{d} \beta' \;
\left\{ -\beta'  \nabla^2  U + \beta'^2 \nabla U \cdot \nabla U  \right\}
\nonumber \\ & = &
\frac{-\beta^4}{8m^2} ({\bf p} \cdot \nabla U)^2
% \nonumber \\ && \mbox{ }
+ \frac{\beta^3}{6m^2}  ( {\bf p} \cdot \nabla )^2 U
\nonumber \\ && \mbox{ }
- \frac{\beta^2}{4m} \nabla^2  U
+ \frac{\beta^3}{6m} \nabla U \cdot \nabla U .
\end{eqnarray}
These are essentially the same as Kirkwood's Eq.~(16).\cite{Kirkwood33}
In general
\begin{eqnarray}
W_n
& = &
\frac{i}{m} \int_0^\beta \mathrm{d} \beta' \;
e^{\beta' U }  {\bf p} \cdot \nabla
\left\{e^{-\beta' U } W_{n-1}(\beta') \right\}
 \\  && \mbox{ }
+ \frac{1}{2m} \int_0^\beta \mathrm{d} \beta' \;
e^{\beta' U }
 \nabla^2  \left\{e^{-\beta' U } W_{n-2}(\beta')  \right\} .\nonumber
\end{eqnarray}
This is essentially the same as Kirkwood's Eq.~(17).\cite{Kirkwood33}

It is shown below that $W_1$ plus  $W_2$
(and neglecting symmetrization corrections)
give a configuration position probability density, \S \ref{Sec:negl-sym},
and an average kinetic energy, \S \ref{Sec:<KE>},
that are the same as that given by Wigner\cite{Wigner32}
and by Kirkwood.\cite{Kirkwood33}
It is shown in \S \ref{Sec:Omega-21}
that  including the first correction for symmetrization and $W_1$
essentially agrees with a result given by Kirkwood\cite{Kirkwood33}
(apart from a factor of 2 that arises because Kirkwood
does not correct for double counting).

%%%%%%%%%%%%%%%%%%%%%%%%%%%%%%%%%%%%%%%
\subsection{Overlap Factor} \label{Sec:OverlapFac}

With the above expression
that writes the action of the Maxwell-Boltzmann operator
on a plane wave basis function
as a sum of quantum corrections,
Eq.~(\ref{Eq:eBHz=zeBH}),
the weighted overlap factor, Eq.~(\ref{Eq:chi-zeta}), becomes
\begin{eqnarray}
\lefteqn{
\chi^\pm_{\cal H}(\zeta_{\bf q},\zeta_{\bf p})
} \nonumber \\
& \equiv &
\sum_{\hat{\mathrm P}} (\pm 1)^{p}
\langle \zeta_{\hat{\mathrm P}{\bf q}} |e^{-\beta \hat {\cal H}}
| \zeta_{{\bf p}} \rangle
\nonumber \\ & = &
\sum_{\hat{\mathrm P}} (\pm 1)^{p}
\int \mathrm{d}{\bf r}\;
\zeta_{\hat{\mathrm P}{\bf q}}({\bf r})^*
e^{-\beta \hat {\cal H}({\bf r}) }\zeta_{{\bf p}}({\bf r})
\nonumber \\ & = &
\sum_{\hat{\mathrm P}} (\pm 1)^{p}
\int \mathrm{d}{\bf r}\;
\zeta_{\hat{\mathrm P}{\bf q}}({\bf r})^*
\zeta_{{\bf p}}({\bf r})
%\nonumber \\ & &   \mbox{ }\times
e^{-\beta {\cal H}({\bf r},{\bf p})}  W({\bf r},{\bf p})
\nonumber \\ & = &
\sum_{\hat{\mathrm P}} (\pm 1)^{p}
e^{-\beta {\cal H}(\hat{\mathrm P}{\bf q},{\bf p})}
W(\hat{\mathrm P}{\bf q},{\bf p})
%\nonumber \\ & &   \mbox{ }\times
\int \mathrm{d}{\bf r}\;
\zeta_{\hat{\mathrm P}{\bf q}}({\bf r})^*
\zeta_{{\bf p}}({\bf r})
\nonumber \\ & = &
e^{-\beta {\cal H}({\bf q},{\bf p})} W({\bf q},{\bf p})
%\nonumber \\ & &   \mbox{ }\times
\sum_{\hat{\mathrm P}} (\pm 1)^{p}
\int \mathrm{d}{\bf r}\;
\zeta_{\hat{\mathrm P}{\bf q}}({\bf r})^*
\zeta_{{\bf p}}({\bf r})
\nonumber \\ & \equiv &
e^{-\beta {\cal H}({\bf q},{\bf p})}  W({\bf q},{\bf p})
\chi^\pm(\zeta_{\bf q},\zeta_{\bf p}).
\end{eqnarray}
In the fourth equality the $\delta$-function  Gaussian
in the limit $\xi\rightarrow 0$ allows
the Maxwell-Boltzmann factor
and the quantum correction $W({\bf r},{\bf p})$
to be evaluated at ${\bf r} = {\bf q}$.
In the fifth equality,
the symmetry of potential energy with respect to permutations
of the position configuration arguments has been used.
The Hamiltonian is now a function of the configuration
positions and momenta, which is just a point in classical phase space,
${\cal H}({\bf \Gamma}) \equiv {\cal H}({\bf q},{\bf p}) $.
Similarly for the quantum correction factor,
$W({\bf \Gamma}) \equiv W({\bf q},{\bf p}) $.
The final equality defines the unweighted overlap factor
\begin{eqnarray}
\chi^\pm(\zeta_{\bf q},\zeta_{\bf p})
& \equiv &
\sum_{\hat{\mathrm P}} (\pm 1)^{p}
\langle \zeta_{\hat{\mathrm P}{\bf q}} |  \zeta_{{\bf p}} \rangle
\nonumber \\ & = &
\sum_{\hat{\mathrm P}} (\pm 1)^{p}
\int \mathrm{d}{\bf r}\;
\zeta_{{\bf q}}(\hat{\mathrm P}{\bf r})^*
\zeta_{{\bf p}}({\bf r}) .
\end{eqnarray}
As above, it makes no difference
 which index or position coordinate argument
the permutator is applied to.

In view of this result the grand partition function may now be written
\begin{eqnarray}
\Xi^\pm & = &
\sum_{N}\frac{z^N  }{N!}
\sum_{\bf \Gamma} g_{\bf p}({\bf q})
\sum_{\hat{\mathrm P}} (\pm 1)^p
\langle  \zeta_{{\bf q}}|
e^{-\beta \hat {\cal H}}
| \zeta_{\hat{\mathrm P}{\bf p}}\rangle
\nonumber \\ & = &
\sum_{N}\frac{z^N  }{N!}
\sum_{{\bf \Gamma}} g_{\bf p}({\bf q})
 e^{ -\beta {\cal H}({\bf \Gamma})}  W({\bf \Gamma})
\sum_{\hat{\mathrm P}} (\pm 1)^p
\langle \zeta_{\hat{\mathrm P}{\bf q}} | \zeta_{\bf p}\rangle
\nonumber \\ & = &
\sum_{N}\frac{z^N  }{N!}
\sum_{{\bf \Gamma}}
 e^{ -\beta {\cal H}({\bf \Gamma})}  W({\bf \Gamma})
g_{\bf p}({\bf q})
\chi^\pm(\zeta_{\bf q},\zeta_{\bf p}) .
\end{eqnarray}
In the continuum limit this becomes
\begin{eqnarray}  \label{Eq:Xi-pm-cont}
\Xi^\pm & = &
\sum_{N}\frac{z^N  }{N!( \Delta_q \Delta_p )^{3N}}
\int \mathrm{d}{\bf \Gamma} \;
 e^{ -\beta {\cal H}({\bf \Gamma})}  W({\bf \Gamma})
\nonumber \\ && \mbox{ } \times
g_{\bf p}({\bf q})
\chi^\pm(\zeta_{\bf q},\zeta_{\bf p}) ,
\end{eqnarray}
where the volume element of phase space is
$\Delta_q\Delta_p =(8\pi\xi^2)^{1/2}h/V^{1/3}$.
At this stage, both the overlap factor and the Gaussian expansion
coefficients still depend on the width $\xi$.

%%%%%%%%%%%%%%%%%%%%%%%%%%%%%%%%%%%%%%%
\subsection{Grand Partition Function and Grand Potential} \label{Sec:Xi-Omega}

With the conversion to the unweighted overlap factor,
the present formulation of the problem
is rather close to that given originally.\cite{Attard16}
However, some of the steps in the original derivation
that were regarded as approximate at the time
can actually be shown to be exact in the present approach.
For this reason it is worthwhile deriving
the expansion for the grand potential in full in the present formulation,
abbreviating the discussion somewhat compared to the original.

%%%%%%%%%%%%%%%%%%%%%%%%%%%%%%%%%%%%%%%%%%%%%%%%%%%%%%%%%%%
\subsubsection{Permutation Expansion of the Overlap Factor}

Any permutation can be cast as the product of disconnected loops.
A loop is the cyclic permutation of a set of particles,
which is just a sequence of connected pair transpositions.
Because the partition function is the sum over all microstates,
the nodes of the loops can be re-labeled as convenient.
A monomer is a one-particle loop
(the identity permutation).
A dimer is a two-particle loop (a single transposition),
for example, $ 1 \rightarrow 2 \rightarrow 1$,
which is equivalent to the  single transposition $\hat{\mathrm P}_{21}$.
The trimer, or three-particle loop,
$ 1 \rightarrow 2 \rightarrow 3 \rightarrow 1$,
is equivalent to  the double transposition
$\hat{\mathrm P}_{32} \hat{\mathrm P}_{21} $.
In general an $l$-mer is an $l$-particle loop.
An $l$-mer of particles $1, 2, \ldots, l$ in order
can be written as the application of $l-1$ successive transpositions,
$\hat{\mathrm P}^{(l)} \equiv \hat{\mathrm P}_{l,l-1}
\ldots \hat{\mathrm P}_{32} \hat{\mathrm P}_{21}$.

The parity of a loop is the parity of the number of nodes minus one.
That is, an $l$-mer has parity $l-1$,
and its symmetrization factor is $(\pm 1)^{l-1}$,
with the upper sign for bosons and the lower sign for fermions.

The permutation operator breaks up into loops
\begin{eqnarray}
\sum_{\hat{\mathrm P} } (\pm1)^p\; \hat{\mathrm P}
& = &
\hat {\mathrm I}
\pm \sum_{i,j} \!' \; \hat{\mathrm P}_{ij}
+ \sum_{i,j,k} \!' \; \hat{\mathrm P}_{ij} \hat{\mathrm P}_{jk}
\nonumber \\ & & \mbox{ }
+ \sum_{i,j,k,l} \!\!' \; \hat{\mathrm P}_{ij} \hat{\mathrm P}_{kl}
\pm \ldots
\end{eqnarray}
The prime on the sums restrict them to unique loops,
with each index being different.
The first term is just the identity.
The second term is a dimer loop,
the third term is a trimer loop,
and the fourth term is the product of two different dimers.

The overlap factor,
$\chi^\pm(\zeta_{\bf q},\zeta_{\bf p})$ $= \sum_{\hat{\mathrm P} }(\pm 1)^p$
$ \langle \zeta_{\hat{\mathrm P}{\bf q}} | \zeta_{\bf p} \rangle$,
is the sum of the expectation values of these loops.

The monomer overlap factor is just the complex
conjugate of the expansion coefficient, Eq.~(\ref{Eq:zq=gqp}),
\begin{equation}
\chi^{(1)}({\bf \Gamma})
\equiv \langle \zeta_{{\bf q}} | \zeta_{\bf p} \rangle
= g_{\bf p}({\bf q})^*.
\end{equation}

It will be recalled that the plane wave and the Gaussian basis functions
are the product of single particle functions.
Similarly, the expansion coefficient
is the product of single particle factors,
$g_{\bf p}({\bf q}) = \prod_{j=1}^N g_{{\bf p}_j}({\bf q}_j)$,
with
\begin{eqnarray}
g_{{\bf p}_j}({\bf q}_j)
& = &
\langle \zeta_{{\bf p}_j}({\bf r}_j) | \zeta_{{\bf q}_j}({\bf r}_j) \rangle
\nonumber \\ & = &
\frac{ (8\pi\xi^2)^{3/4}}{V^{1/2}}
e^{- \xi^2 p_j^2/\hbar^2} e^{ {\bf q}_j \cdot {\bf p}_j/i\hbar }.
\end{eqnarray}
It follows that the monomer overlap factor can be written
\begin{equation}
\chi^{(1)}({\bf \Gamma})
=
\prod_{j=1}^N
\langle \zeta_{{\bf q}_j}({\bf r}_j) | \zeta_{{\bf p}_j}({\bf r}_j) \rangle
=  \prod_{j=1}^N g_{{\bf p}_j}({\bf q}_j)^*.
\end{equation}

The dimer overlap factor in the microstate ${\bf \Gamma}$
for particles $j$ and $k$ is
\begin{eqnarray}
\chi^{(2)}_{jk}({\bf \Gamma})
& = &
\pm \langle \zeta_{\hat{\mathrm P}_{jk}{\bf q}}( {\bf r})
| \zeta_{\bf p}({\bf r}) \rangle
\nonumber \\ & = &
\pm
\langle \zeta_{{\bf q}_k}({\bf r}_j) | \zeta_{{\bf p}_j}({\bf r}_j) \rangle
\langle \zeta_{{\bf q}_j}({\bf r}_k) | \zeta_{{\bf p}_k}({\bf r}_k) \rangle
\nonumber \\ & & \mbox{ } \times
\prod_{m=1}^N \!^{ m \ne j,k}
\langle \zeta_{{\bf q}_j}({\bf r}_j) | \zeta_{{\bf p}_j}({\bf r}_j) \rangle
\nonumber \\ & = &
\pm
\frac{ g_{{\bf p}_j}({\bf q}_k)^* g_{{\bf p}_k}({\bf q}_j)^*
}{
g_{{\bf p}_j}({\bf q}_j)^* g_{{\bf p}_k}({\bf q}_k)^*
}
g_{\bf p}({\bf q})^*
\nonumber \\ & = &
\pm
e^{ ({\bf q}_j-{\bf q}_k) \cdot {\bf p}_j /i\hbar }
e^{ ({\bf q}_k-{\bf q}_j) \cdot {\bf p}_k /i\hbar }
g_{\bf p}({\bf q})^*
\nonumber \\ & \equiv &
\tilde \chi^{(2)}_{jk}({\bf \Gamma}) \; g_{\bf p}({\bf q})^*.
\end{eqnarray}
The coordinate argument within the expectation value is
just a dummy variable of integration
and is henceforth dropped.
The important point is the formally exact factorization
of the expectation value,
with the non-trivial part involving the permuted particles alone.

The overlap factors with a tilde here and below are localized
in the sense that they are only non-zero when all the particles
are close together.
(More precisely, localization means
that the separations between consecutive neighbors around the loop
are all small.)
This will be shown explicitly below,
but here it can be noted that the explicit exponents give highly oscillatory
and therefore canceling behavior
unless the differences in  configuration positions are all close to zero.

Similarly the trimer overlap factor for particles $j$, $k$, and $ l$ is
\begin{eqnarray}
\chi^{(3)}_{jkl}({\bf \Gamma})
& = &
\langle \zeta_{\hat{\mathrm P}_{jk}\hat{\mathrm P}_{kl}{\bf q}}
| \zeta_{\bf p}  \rangle
\nonumber \\ & = &
\frac{ g_{{\bf p}_j}({\bf q}_k)^* g_{{\bf p}_k}({\bf q}_l)^*
g_{{\bf p}_l}({\bf q}_j)^*
}{
g_{{\bf p}_j}({\bf q}_j)^* g_{{\bf p}_k}({\bf q}_k)^*
 g_{{\bf p}_l}({\bf q}_l)^*
}
g_{\bf p}({\bf q})^*
\nonumber \\ & = &
e^{ ({\bf q}_j-{\bf q}_k) \cdot {\bf p}_j /i\hbar }
e^{ ({\bf q}_k-{\bf q}_l) \cdot {\bf p}_k /i\hbar }
e^{ ({\bf q}_l-{\bf q}_j) \cdot {\bf p}_l /i\hbar }
g_{\bf p}({\bf q})^*
\nonumber \\ & \equiv &
\tilde \chi^{(3)}_{jkl}({\bf \Gamma}) \; g_{\bf p}({\bf q})^*.
\end{eqnarray}

Because of the factorization property,
the overlap factor for the product of dimer loops shown explicitly above
is just the product of dimer overlap factors,
\begin{eqnarray}
\chi^{(2,2)}_{ij,kl}({\bf \Gamma})
& = &
\langle \zeta_{\hat{\mathrm P}_{ij}\hat{\mathrm P}_{kl}{\bf q}}
| \zeta_{\bf p} \rangle
\nonumber \\ & = &
\tilde \chi^{(2)}_{ij}({\bf \Gamma}) \, \tilde \chi^{(2)}_{kl}({\bf \Gamma})
\; g_{\bf p}({\bf q})^*  .
\end{eqnarray}
By definition of distinct permutations,
the $i,\, j,\, k, $ and $l$ must all be different here.

Because the overlap factor
$\chi^\pm({\bf \Gamma})
\equiv \chi^\pm(\zeta_{\bf q},\zeta_{\bf p})$
is the sum over all permutations,
it can be rewritten as the sum over all possible monomers and loops.
This gives the loop expansion for the overlap factor
for use in the partition function as
\begin{eqnarray}
\chi^\pm({\bf \Gamma})
&=&
\chi^{(1)}({\bf \Gamma})
+ \sum_{ij}\!' \chi_{ij}^{(2)}({\bf \Gamma})
+ \sum_{ijk}\!' \chi_{ijk}^{(3)}({\bf \Gamma})
\nonumber \\ && \mbox{ }
+ \sum_{ijkl}\!'  \chi_{ij}^{(2)}({\bf \Gamma}) \chi_{kl}^{(2)}({\bf \Gamma})
+ \ldots
\end{eqnarray}
It is useful to define
$\tilde \chi^\pm({\bf \Gamma})  \equiv
\chi^\pm({\bf \Gamma}) /g_{\bf p}({\bf q})^* $,
and to write this as
\begin{eqnarray} \label{Eq:chi_n}
\tilde \chi^\pm({\bf \Gamma})
&=&
1
+ \sum_{ij}\!' \tilde \chi_{ij}^{(2)}({\bf \Gamma})
+ \sum_{ijk}\!' \tilde \chi_{ijk}^{(3)}({\bf \Gamma})
\nonumber \\ &&  \mbox{ }
+ \sum_{ijkl}\!' \tilde\chi_{ij}^{(2)}({\bf \Gamma})
\tilde\chi_{kl}^{(2)}({\bf \Gamma})
+ \ldots
\end{eqnarray}
Note that the parity factor for fermions and bosons, $(\pm 1)^{l-1}$,
has been incorporated into the definition
of the $\tilde \chi^{(l)}({\bf \Gamma})$.

%%%%%%%%%%%%%%%%%%%%%%%%%%%%%%%%%%%%%%%%%%%%%%%%%%%%%%%%%%%
\subsubsection{Expansion of the Grand Partition Function} \label{Sec:expXi}

With these results,
the grand canonical partition function in the continuum limit,
Eq.~(\ref{Eq:Xi-pm-cont}), becomes
\begin{eqnarray}
\lefteqn{
\Xi^\pm
} \nonumber \\ & = &
\sum_{N}\frac{z^N  }{N!( \Delta_q \Delta_p )^{3N}}
%\nonumber \\ && \mbox{ } \times
\int \mathrm{d}{\bf \Gamma} \;
 e^{ -\beta {\cal H}({\bf \Gamma})} W({\bf \Gamma})
g_{\bf p}({\bf q})
\chi^\pm({\bf \Gamma})
\nonumber \\ & = &
\sum_{N}\frac{z^N }{N!( \Delta_q \Delta_p )^{3N}}
\nonumber \\ && \mbox{ } \times
\int \mathrm{d}{\bf \Gamma} \;
 e^{ -\beta {\cal H}({\bf \Gamma})} W({\bf \Gamma})
g_{\bf p}({\bf q}) g_{\bf p}({\bf q})^*
\tilde \chi^\pm({\bf \Gamma})
\nonumber \\ & = &
\sum_{N}\frac{z^N }{N! h^{3N}  }
%\nonumber \\ && \mbox{ } \times
\int \mathrm{d}{\bf \Gamma} \;
 e^{ -\beta {\cal H}({\bf \Gamma})}W({\bf \Gamma})
e^{- 2\xi^2 p^2/\hbar^2}
\tilde \chi^\pm({\bf \Gamma})
\nonumber \\ & = &
\sum_{N}\frac{z^N }{N! h^{3N}  }
%\nonumber \\ && \mbox{ } \times
\int \mathrm{d}{\bf \Gamma} \;
 e^{ -\beta {\cal H}({\bf \Gamma})} W({\bf \Gamma})
\tilde \chi^\pm({\bf \Gamma}) .
\end{eqnarray}
In the third equality,
the volume element
$(\Delta_q\Delta_p)^{3N} =h^{3N}(8\pi\xi^2)^{3N/2}/V^{N}$
has canceled with the pre-exponential part of
$g_{\bf p}({\bf q}) g_{\bf p}({\bf q})^*$,
namely $[ (8\pi\xi^2)^{3N/4}/V^{N/2}]^2$,
Eq.~(\ref{Eq:zq=gqp}),
leaving a factor of $h^{-3N}$.
The final equality holds in the limit $\xi \rightarrow 0$,
since the $\tilde \chi^\pm({\bf \Gamma})$ are independent of $\xi$.
At this stage and henceforth,
the width $\xi$ has been entirely removed from the formulation.

In this there are two types of quantum corrections.
The quantum corrections due to non-commutativity of
(or lack of simultaneity in) positions and momenta
are embodied in
$ W({\bf \Gamma}) = \sum_{n=0}^\infty W_n({\bf \Gamma})\hbar^n  $,
with $ W_0({\bf \Gamma}) = 1$.
The quantum corrections due to the symmetrization of the wave function
are contained in the loop expansion of the overlap factor,
$\tilde \chi^\pm({\bf \Gamma})$,
which is the sum of products of the $\tilde \chi^{(l)}({\bf \Gamma})$.
The monomer term, $l=1$, is $\tilde \chi^{(1)}({\bf \Gamma}) =1$.

The leading order term in both series,
which may be designated $\{1,0\}$,
is obviously the classical term.
For the grand partition function it is
\begin{eqnarray} \label{Eq:Xi10}
\Xi_{1,0}^\pm
& = &
\sum_{N}\frac{z^N }{N! h^{3N}  }
%\nonumber \\ && \mbox{ } \times
\int \mathrm{d}{\bf \Gamma} \;
 e^{ -\beta {\cal H}({\bf \Gamma})} .
\end{eqnarray}
This is just
the  classical equilibrium grand canonical  partition function.
Since this is the same for bosons and for fermions,
the superscript $\pm$ is redundant.

One may keep the quantum corrections due to non-commutativity,
in which case the monomer grand potential is
\begin{eqnarray}
\Xi_{1}^\pm
& \equiv &
\sum_{N}\frac{z^N }{N! h^{3N}  }
%\nonumber \\ && \mbox{ } \times
\int \mathrm{d}{\bf \Gamma} \;
 e^{ -\beta {\cal H}({\bf \Gamma})} W({\bf \Gamma}).
\end{eqnarray}
Again the superscript $\pm$ is redundant.
In places below the left hand side will be written
$\Xi_{1,W}$ or as $\Xi_{1}(W)$
when it is desirable to emphasize that the
quantum corrections due to  non-commutativity are included
in the phase space weight.
Except for $\Xi_{1,0}$ or $\Xi_{l,0}$,
one should always assume that the quantum weight is included.

The ratio of the total grand partition function to the monomer
grand partition function is clearly just an equilibrium average,
\begin{equation}
\frac{\Xi^\pm}{\Xi_{1}}
=
\left< \tilde \chi^\pm({\bf \Gamma})  \right>_{1,W} .
\end{equation}
The phase space weight factor signified by the subscript $1,W$ is
$ e^{ -\beta {\cal H}({\bf \Gamma})} W({\bf \Gamma})$,
which is to say that it is an equilibrium average,
not a classical equilibrium average.

For the single dimer term one has
\begin{eqnarray}
\frac{\Xi^\pm_2}{\Xi_{1}}
& = &
\left<
\sum_{ij}\!' \tilde\chi_{ij}^{(2)}({\bf \Gamma}) \right>_{1,W}
\nonumber \\ & = &
\left< \frac{N(N-1)}{2} \tilde\chi_{12}^{(2)}({\bf \Gamma}) \right>_{1,W} .
\end{eqnarray}
Note that the the particle symmetry factor $\pm 1$
is included in the definition of $ \tilde\chi^{(2)}({\bf \Gamma})$.
The second equality follows because the sum over all states
makes all dimer pairs equivalent.
Because of this the subscript 12 on  $\tilde \chi_{12}^{(2)}({\bf \Gamma})$
is redundant and will be dropped for this and the following overlap factors.
(One can replace $N(N-1)$ by $N^2$ here and below.)
This scales with volume because the overlap factor is only non-zero
when the two particles are close together.
Here and below, the $N$ that appears explicitly in the grand canonical average
is the total number of particles in each term,
which is to say that there are $l$ loop particles and $N-l$ monomers,
all able to interact  depending upon their proximity in each microstate.

In the same way the single trimer term gives
\begin{eqnarray}
\frac{\Xi^\pm_3}{\Xi_{1}}
& = &
\left<
\sum_{ijk}\!' \tilde\chi_{ijk}^{(3)}({\bf \Gamma}) \right>_{1,W}
\nonumber \\ & = &
\left< \frac{N!}{3(N-3)!} \tilde\chi^{(3)}({\bf \Gamma}) \right>_{1,W} .
\end{eqnarray}
The numerator in the combinatorial pre-factor
is the total number of re-arrangements of the particles.
The denominator counts the number of re-arrangements that leave the system
topologically unchanged.
In general for a loop of $l$ particles,
There are $(N-l)!$ such re-arrangements of the monomers,
and there are $l$ discrete rotations of the loop labels
that leave the neighbors unchanged.
The loop particles are distinguished by their relative positions
and cannot be interchanged except by a rigid rotation of all the labels.
This factor can be written $N!/l(N-l)! = (l-1)! \, ^N\!C_l$.
In the thermodynamic limit this is $N^l/l$
Obviously  $l=3$ for this trimer,
and one can write ${N!}/{3(N-3)!} =N^3/3$ here.
This average also scales with the volume.

The double dimer product term has average
\begin{eqnarray} \label{Eq:Xi22}
\frac{\Xi_{22}^\pm}{\Xi_{1}}
& = &
\left< \sum_{ijkl}\!' \tilde\chi^{(2,2)}_{ij,kl}({\bf \Gamma}) \right>_{1,W}
\nonumber \\ & = &
\left< \sum_{ijkl}\!'
\tilde\chi^{(2)}_{ij}({\bf \Gamma})\, \tilde\chi^{(2)}_{kl}({\bf \Gamma})
 \right>_{1,W}
\nonumber \\ & = &
\left<  \frac{N!}{2^3(N-4)!}
\tilde\chi^{(2)}_{12}({\bf \Gamma})\, \tilde\chi^{(2)}_{34}({\bf \Gamma})
 \right>_{1,W}
\nonumber \\ & = &
\frac{1}{2^3  }
\left<  N^4
\tilde\chi^{(2)}_{12}({\bf \Gamma})\, \tilde\chi^{(2)}_{34}({\bf \Gamma})
 \right>_{1,W}
\nonumber \\ & = &
\frac{1}{2!}
\left[ \frac{1}{2!} \left< N^2 \tilde\chi^{(2)}({\bf \Gamma})  \right>_{1,W}
\right]^2 .
\end{eqnarray}
In the final equality the average of the product has been written
as the product of the averages.
This is justified because
there are many more microstates in which the two dimers are far apart
and independent than there are those in which they are close together
and influencing each other.

This particular factorization appears to be formally exact
at all densities in the thermodynamic limit,
$N \rightarrow \infty$, $V \rightarrow \infty$, $N/V =\mbox{const}$.
(Equivalently, $V \rightarrow \infty$ with $N = \overline N(\mu,T,V)$.)
The reason is that the contribution
from two independent dimers scales with $V^2$,
whereas that from two  dimers close enough to interact scales with $V$,
which is relatively negligible.
The analogous argument holds for all the powers of loop overlap factors
that occur next.

Continuing in this fashion it is clear that
\begin{eqnarray} \label{Eq:Xipm-resum}
\lefteqn{
\Xi^\pm
} \nonumber \\
& = &
\Xi_{1}
\left\{ 1 +
\frac{1}{2} \left< N^2 \tilde\chi^{(2)}({\bf \Gamma}) \right>_{1,W}
+ \frac{1}{3} \left< N^3 \tilde\chi^{(3)}({\bf \Gamma})\right>_{1,W}
\right. \nonumber \\ && \left. \mbox{ }
+ \frac{1}{2!}
\left[ \frac{1}{2} \left< N^2
\tilde\chi^{(2)}({\bf \Gamma})  \right>_{1,W} \right]^2
+ \ldots
\right\}
\nonumber \\ & = &
\Xi_{1} \sum_{\{m_l\}} \prod_{l=2}^\infty \frac{1}{m_l!}
\left[ \frac{1}{l}
\left< \frac{N!}{(N-l)!}
\tilde\chi^{(l)}({\bf \Gamma}) \right>_{1,W} \right]^{m_l}
\nonumber \\ & = &
\Xi_{1} \prod_{l=2}^\infty \sum_{m_l=0}^\infty
\frac{1}{m_l!}
\left[ \frac{1}{l}
\left< \frac{N!}{(N-l)!}
\tilde\chi^{(l)}({\bf \Gamma})  \right>_{1,W} \right]^{m_l}
\nonumber \\ & = &
\Xi_{1} \prod_{l=2}^\infty \exp
\left[ \left< \frac{N!}{l(N-l)!}
\tilde\chi^{(l)}({\bf \Gamma}) \right>_{1,W} \right] .
\end{eqnarray}
Here $m_l$ is the number of loops of $l$ particles.
Here  $N!/(N-l)!$ has been written in place of $N^l$,
although in truth either is justified.

The combinatorial and loop overlap factors being averaged here
will occur frequently below
and it will save space to define
\begin{equation} \label{Eq:defX}
X^{(l)}({\bf \Gamma})
\equiv
\frac{N!}{l(N-l)!}
\tilde\chi^{(l)}({\bf \Gamma}) .
\end{equation}

%%%%%%%%%%%%%%%%%%%%%%%%%%%%%%%%%%%%%%%%%%%%%%%%%%%%%%%%%%%
\subsubsection{Expansion of the Grand Potential} \label{Sec:expand-Omega}

This expansion and re-summation
of the grand partition function
appears the same as that given in Ref.~\onlinecite{Attard16}.
The differences in detail are that the factorizations that were regarded
as approximations there (the so-called localization approximation)
have here been derived as theorems that are formally exact
in the limit $\xi \rightarrow 0$.
Also the $\tilde \chi$ here are for the overlap
of a Gaussian and a plane wave basis function,
whereas in Ref.~\onlinecite{Attard16} they were for
the overlap of two minimum uncertainty wave packets.
Finally, the quantum corrections due to non-commutativity,
the $W({\bf \Gamma})$, were not included originally.

Since the grand partition function is formulated above
as the product of exponentials,
the grand potential is essentially just the sum of the exponents
\begin{eqnarray}
\Omega^\pm(\mu,V,T)  & = &
- k_\mathrm{B}T \ln \Xi^\pm(z,V,T)
\nonumber \\ & = &
\sum_{l=1}^\infty \Omega_l^\pm .
\end{eqnarray}

The monomer grand potential is
\begin{equation}
\Omega_1 = - k_\mathrm{B}T \ln \Xi_1,
\end{equation}
with the monomer grand partition function being
\begin{eqnarray} \label{Eq:Xi1}
\Xi_1
& = &
\sum_{N}\frac{z^N }{N! h^{3N}  }
%\nonumber \\ && \mbox{ } \times
\int \mathrm{d}{\bf \Gamma} \;
 e^{ -\beta {\cal H}({\bf \Gamma})} W({\bf \Gamma}) .
\end{eqnarray}
This does not depend upon the wave particle symmetrization,
since it arises from the identity permutation,
$\hat{\mathrm P}^{(1)} = \hat{\mathrm I}$.
This will also be written below as $\Xi_{1,W}$ or as $\Xi_1(W)$.

The $l$-loop grand potential for $l \ge 2$ is
\begin{eqnarray} \label{Eq:Omega_l=<chi^l>}
\Omega_l
& = &
- k_\mathrm{B}T \left< \frac{N!}{l(N-l)!}
\tilde\chi^{(l)}({\bf \Gamma})  \right>_{1,W}
\nonumber \\ & = &
\frac{- k_\mathrm{B}T}{ \Xi_1}
\sum_{N=l}^\infty  \frac{z^N}{ N!h^{3N}} \frac{N!}{l(N-l)!}
\nonumber \\ && \mbox{ } \times
\int \mathrm{d}{\bf \Gamma} \;  e^{ -\beta {\cal H}({\bf \Gamma})}
W({\bf \Gamma})
\tilde\chi^{(l)}({\bf \Gamma}) .
\end{eqnarray}
Using the above definition, Eq.~(\ref{Eq:defX}),
%$X^{(l)}({\bf \Gamma}) \equiv {N!} \tilde\chi^{(l)}({\bf \Gamma})/{l(N-l)!}$,
this is more succinctly written as
\begin{eqnarray}
\Omega_l
& = &
- k_\mathrm{B}T \left< X^{(l)}({\bf \Gamma})  \right>_{1,W}
 \\ & = &
\frac{- k_\mathrm{B}T}{ \Xi_1}
\sum_{N=l}^\infty  \frac{z^N}{ N!h^{3N}}
%\nonumber \\ && \mbox{ } \times
\int \mathrm{d}{\bf \Gamma} \;  e^{ -\beta {\cal H}({\bf \Gamma})}
W({\bf \Gamma})
X^{(l)}({\bf \Gamma}) .\nonumber
\end{eqnarray}

In this expression,
the $l$-loop overlap factor is
\begin{eqnarray} \label{Eq:tilde-chi-l}
\tilde \chi^{(l)}({\bf \Gamma})
& = &
(\pm 1)^{l-1} \prod_{j=1}^{l}
\frac{ g_{{\bf p}_j}({\bf q}_{j+1})^* }{ g_{{\bf p}_j}({\bf q}_{j})^* }
\nonumber \\ & = &
(\pm 1)^{l-1} \prod_{j=1}^{l}
e^{ ({\bf q}_j-{\bf q}_{j+1}) \cdot {\bf p}_j /i\hbar }
\nonumber \\ & \equiv &
(\pm 1)^{l-1}
e^{ ({\bf q}-{\bf q}') \cdot {\bf p} /i\hbar }.
\end{eqnarray}
Here
${\bf q}_{l+1} \equiv {\bf q}_{1}$,
${\bf q}_{0} \equiv {\bf q}_{l}$,
and $ {\bf q}'_j \equiv {\bf q}_{j-1} $,
in this and similar expressions.
This convention is needed because one is dealing with an $l$-loop.
Note that the symmetry factor for bosons and fermions,
$(\pm 1)^{l-1}$, is included in the definition of the loop overlap factor.

The classical grand partition function
is just the zeroth term, $W_0({\bf \Gamma}) = 1$,
of the monomer grand partition function,
\begin{eqnarray} %\label{Eq:Xi10}
\Xi_{1,0}
& \equiv &
\sum_{N}\frac{z^N }{N! h^{3N}  }
%\nonumber \\ && \mbox{ } \times
\int \mathrm{d}{\bf \Gamma} \;
 e^{ -\beta {\cal H}({\bf \Gamma})} .
\end{eqnarray}
Obviously the classical grand potential
is the logarithm of this, $\Omega_{1,0} = - k_\mathrm{B}T \ln \Xi_{1,0}$.
The expression for the loop grand potential
can be written as the ratio of classical averages
by multiplying and dividing by the  classical grand partition function,
\begin{eqnarray} \label{Eq:Omega_l=<X^lW>10}
\Omega_l
& = &
\frac{ - k_\mathrm{B}T \Xi_{1,0} }{\Xi_1}
 \left<  X^{(l)}({\bf \Gamma}) W({\bf \Gamma}) \right>_{1,0}
\nonumber \\ & = &
- k_\mathrm{B}T
\frac{
 \left<  X^{(l)}({\bf \Gamma})  W({\bf \Gamma}) \right>_{1,0}
}{
\left<  W({\bf \Gamma}) \right>_{1,0}
} .
\end{eqnarray}
For what it is worth,
the denominator can also be written
as the exponential of grand potential differences,
\begin{equation}
\left<  W({\bf \Gamma}) \right>_{1,0}
=
\frac{\Xi_1}{\Xi_{1,0} }
=
e^{-\beta [\Omega_1 -\Omega_{1,0}] }.
\end{equation}

\comment{ %%%%%%%%%%%%%%%%%%%%%%%%%%%%%%%%%%%%%%%%%%%%%%%%%%%%%%%%%
It is clear that the quantum grand potential has been formulated
as a double series of quantum corrections,
one the sum over symmetrization loops $\Omega_l$,
and the other the sum over non-commuting corrections $W_j$.
There are many different ways of developing approximations
based on terminating the sums.
One convention would be to \emph{define} the $lj$-th approximation
to the grand potential
as including all loops up to $l$ and all non-commuting corrections up to $j$,
\begin{equation}
\Omega^{(l,j)}
\equiv
\frac{ - k_\mathrm{B}T }{\sum_{j'=0}^j \left<  W_{j'} \right>_{1,0} }
%\nonumber \\ & & \mbox{ } \times
\sum_{l'=1}^l \sum_{j'=0}^j
 \left<  X^{(l')}({\bf \Gamma})  W_{j'} \right>_{1,0}
\end{equation}
There are different possible variations on this theme.

One typical approach would expand $W({\bf \Gamma})$
in the denominator up to a certain order,
and then perform an harmonic expansion to bring these
additional terms up to the numerator.
For example,
from the analysis in \S \ref{Sec:W20},
to quadratic order in $\hbar$ one has
\begin{eqnarray}
\left<  W({\bf \Gamma}) \right>_{1,0}^{-1}
& = &
\left< 1 +  W_1 + W_2 \right>_{1,0}^{-1}
 \\ & = &
1 - \left< W_1 + W_2 \right>_{1,0} + \left< W_1  \right>_{1,0}^2
+{\cal O}(\hbar^3).\nonumber
\end{eqnarray}
Since  the classical phase space weight is an even function of momentum,
and the term in $W_1 $ linear in $\hbar$ is linear in momentum,
then $\left< W_1  \right>_{1,0}^2 = {\cal O}(\hbar^4)$
and can be neglected.
} % end comment %%%%%%%%%%%%%%%%%%%%%%%%%%%%%%%%%%%%%%%%%%%%%%%%

%%%%%%%%%%%%%%%%%%%%%%%%%%%%%%%%%%%%%%%%%%%%%%
\subsubsection{Commuting Part of the Expansion} \label{Sec:commute}

Of particular interest is what might be called the commuting part of the
expansion, in which case one sets $W_j({\bf r}) = 0$, $j \ge1$.
In this case the commuting part of the loop potential is
\begin{equation}
\Omega_{l,0}
=
- k_\mathrm{B}T
 \left<  X^{(l)}({\bf \Gamma})   \right>_{1,0} .
\end{equation}
Recall Eq.~(\ref{Eq:defX}),
$X^{(l)}({\bf \Gamma}) \equiv {N!} \tilde\chi^{(l)}({\bf \Gamma})/{l(N-l)!}$.

Since this $l$-loop overlap factor
depends only on the first $l$ particles,
its classical equilibrium average can be written
as an integral over  the configuration momenta and positions of these
weighted with the $l$-particle density
(and the Maxwell-Boltzmann factor of the kinetic energy).
The standard definition in classical equilibrium statistical mechanics
of the $l$-particle density is\cite{TDSM}
\begin{eqnarray}
\lefteqn{
\rho^{(l)}({\bf q}^l)
}  \\
& = &
\frac{1}{\Xi_{1,0}}
\sum_{N=l}^\infty \frac{\Lambda^{-3N} z^N}{(N-l)!}
%\nonumber \\ && \mbox{ } \times
 \int \mathrm{d}{\bf q}_{l+1} \ldots \mathrm{d}{\bf q}_{N} \;
 e^{ -\beta U({\bf q}^N) }
 \nonumber \\ & = &
 \frac{1}{\Xi_{1,0}}
\sum_{N=l}^\infty
\frac{ z^N \Lambda^{-3l} e^{\beta {\cal K}({\bf p}^l)} }
{ (N-l)!h^{3(N-l)} }
%\nonumber \\ && \mbox{ } \times
 \int \mathrm{d}{\bf \Gamma}_{l+1} \ldots \mathrm{d}{\bf \Gamma}_{N} \;
 e^{ -\beta {\cal H}({\bf \Gamma}^N) }   .\nonumber
\end{eqnarray}
With this  the commuting part of the $l$-loop  grand potential can be written
\begin{eqnarray}
-\beta \Omega_{l,0}
& = &
\frac{1}{ \Xi_{1,0}}
\sum_{N=l}^\infty  \frac{z^N}{ l(N-l)!h^{3N}}
%\nonumber \\ && \mbox{ } \times
\int \mathrm{d}{\bf \Gamma} \;  e^{ -\beta {\cal H}({\bf \Gamma})}
\tilde\chi^{(l)}({\bf \Gamma})
 \nonumber \\ & = &
\frac{\Lambda^{3l}}{ l h^{3l}}
%\nonumber \\ && \mbox{ } \times
\int \mathrm{d}{\bf \Gamma}^l \;  e^{ -\beta {\cal K}({\bf p}^l)}
\rho^{(l)}({\bf q}^l)
\tilde\chi^{(l)}({\bf \Gamma}) .
\end{eqnarray}
The factor of $\Lambda^{3l} e^{-\beta {\cal K}({\bf p}^l)}$
in the second equality here came from its reciprocal
in the second equality of the previous equation.

In view of the overlap factor given above,
the momentum exponent is
\begin{eqnarray}
\frac{-\Lambda^2 p^2}{4\pi\hbar^2}
+ \frac{{\bf p}\cdot( {\bf q} - {\bf q}')}{i\hbar}
& = &
\frac{-\Lambda^2 }{4\pi\hbar^2}
\left[
{\bf p} -  \frac{2\pi\hbar^2}{\Lambda^2 i\hbar}( {\bf q} - {\bf q}')
\right]^2
\nonumber \\ && \mbox{ }
- \frac{\pi}{\Lambda^2}( {\bf q} -  {\bf q}')^2 .
\end{eqnarray}
Hence the momentum integral can be performed analytically
and one finally obtains
\begin{eqnarray} \label{Eq:bWpk-int-q}
\lefteqn{
-\beta \Omega_{l,0}^\pm
} \nonumber \\
 & = &
\frac{ (\pm 1)^{l-1} \Lambda^{3l} }{ l h^{3l}   }
(4\pi^2\hbar^2/\Lambda^2)^{3l/2}
\nonumber \\ && \mbox{ } \times
\int \mathrm{d}{\bf q}^l\;
e^{- \pi ( {\bf q} -  {\bf q}')^2/{\Lambda^2} }
\rho^{(l)}({\bf q})
\nonumber \\ & = &
\frac{ (\pm 1)^{l-1}  }{ l   }
\int \mathrm{d}{\bf q}^l\;
e^{- \pi ( {\bf q} -  {\bf q}')^2/{\Lambda^2} }
\rho^{(l)}({\bf q})
\nonumber \\ & = &
\frac{ (\pm 1)^{l-1} V }{ l   }
\int \mathrm{d}{\bf q}^{l-1}\;
e^{- \pi ( {\bf q} -  {\bf q}')^2/{\Lambda^2} }
\rho^{(l)}({\bf q}).
\end{eqnarray}
Since this is homogeneous in the volume,
in the final equality the final coordinate has been fixed,
${\bf q}_l = {\bf 0}$,
and  a factor of $V$ has replaced the integration over this coordinate.
Here
${\bf q} = \{{\bf q}_1,{\bf q}_2,\ldots,{\bf q}_{l}\}$
and
${\bf q}' \equiv \hat{\mathrm P}^{(l)} {\bf q} =
\{{\bf q}_l,{\bf q}_1,{\bf q}_2,\ldots,{\bf q}_{l-1}\} $.

On a technical note,
in this expression the multi-particle density appears explicitly,
whereas the fugacity (equivalently, chemical potential)
is the independent variable of the grand potential.
The density that appears here can be regarded
as an implicit classical function of the given fugacity.
This means that this particular term
written in this form contributes to the second order
and to the higher orders in the fugacity expansion
of the quantum grand potential.

It is worth pointing out
that the Gaussian exponent can be written as a quadratic form.
With ${\bf q}_l = {\bf 0}$, for each direction  $\alpha=x,y,z$
(three dimensions are assumed)
one simply has
\begin{eqnarray}
\lefteqn{
( {\bf q} -  {\bf q}')_\alpha^2
} \nonumber \\
& = &
q_{1;\alpha}^2
+ q_{12;\alpha}^2 + q_{23;\alpha}^2  + \ldots + q_{l-2,l-1;\alpha}^2
 + q_{l-1;\alpha}^2
 \nonumber \\ & = &
 \underline{\underline A}^{(l-1)} :
\underline { q}_\alpha^{l-1} \underline { q}_\alpha^{l-1} ,
\end{eqnarray}
where $ q_{jk,\alpha}^2 = ( q_{j,\alpha}-q_{k,\alpha})^2$.
Here $ \underline{\underline A}^{(l-1)} $ is an
$(l-1)\times(l-1)$ tridiagonal matrix
with $2$ on the main diagonal and $-1$ immediately above and below
the main diagonal, and all other entries 0.
It is readily shown that this has determinant
\begin{eqnarray} \label{Eq:Det(A)}
\left| \underline{\underline A}^{(l-1)} \right|
& = & 2 \left| \underline{\underline A}^{(l-2)} \right|
- \left| \underline{\underline A}^{(l-3)} \right|
 \nonumber \\ & = &
 l.
\end{eqnarray}
This result will be used in \S \ref{Sec:Ideal} below.

%%%%%%%%%%%%%%%%%%%%%%%%%%%%%%%%%%%%%%%%%%%%%%%%%%%%%%%%%%%%%
\subsection{Statistical Average} \label{Sec:Stat-Av}

\subsubsection{General Expressions}

The statistical average of an operator can be written
\begin{eqnarray}
\lefteqn{
\left< \hat O\right>_{\mu,V,T}
}  \\
& = &
\frac{1}{\Xi} \mbox{TR } z^N \hat O e^{-\beta \hat{\cal H} }
\nonumber \\ & = &
\frac{1}{\Xi} \sum_N \frac{z^N}{N!}
\sum_{\bf m} \chi(\phi_{\bf m})
\left< \phi_{\bf m} \right|
\hat O e^{-\beta \hat{\cal H} } \left| \phi_{\bf m} \right>
\nonumber \\ & = &
\frac{1}{\Xi} \sum_N \frac{z^N}{N!} \sum_{\hat{\mathrm P}} (\pm 1)^p
\sum_{\bf m}
\left< \phi_{\bf m} \right| \hat O e^{-\beta \hat{\cal H} }
\left| \phi_{\hat{\mathrm P}{\bf m}} \right>
\nonumber \\ & = &
\frac{1}{\Xi} \sum_N \frac{z^N}{N!}
 \sum_{\bf \Gamma} g_{\bf p}({\bf q})
 \sum_{\hat{\mathrm P}} (\pm 1)^p
\left< \zeta_{\bf q} \left|
\hat O e^{-\beta \hat{\cal H} }
\right| \zeta_{\hat{\mathrm P}{\bf p}} \right> .\nonumber
\end{eqnarray}
The manipulations leading to the final equality
are the same as in Eq.~(\ref{Eq:Xipm-1}).

The derivation of the expression given in \S \ref{Sec:MBtilde}
for the Maxwell-Boltzmann operator acting a plane wave basis function
can also be carried out here.
One has
\begin{eqnarray} \label{Eq:hat-O-eBH-zp}
\hat O e^{-\beta \hat{\cal H} }
\zeta_{{\bf p}}({\bf r})
& = &
\zeta_{{\bf p}}({\bf r})
\hat {\tilde O} e^{-\beta \hat{\tilde {\cal H}} } 1
\nonumber \\ & = &
\zeta_{{\bf p}}({\bf r})
\hat {\tilde O}
\left\{
e^{-\beta {\cal H}({\bf r},{\bf p}) } W({\bf r},{\bf p}) \right\} .
\end{eqnarray}
The transformed operator, which acts on everything to its right, is
\begin{equation}
\hat {\tilde O} \equiv
e^{{\bf p} \cdot{\bf r}/i\hbar}
\hat {O}
e^{-{\bf p} \cdot{\bf r}/i\hbar} .
\end{equation}

In general, once written in this form where the plane wave basis function
has commuted with the operator,
the expectation value can be applied,
and, because the Gaussian wave packet becomes a Dirac-$\delta$ function
in the limit $\xi \rightarrow 0$,
the modified operator and Maxwell-Boltzmann weight can be
evaluated at ${\bf r} = {\bf q}$ and taken outside of
the expectation value.
For a general operator that is a function of both position
and the momentum operator,
$\hat O = O({\bf r}, -i\hbar \nabla_{\bf r})$,
\begin{eqnarray}
\lefteqn{
\langle \zeta_{\hat{\mathrm P} {\bf q}}({\bf r})
| O({\bf r}, -i\hbar \nabla_{\bf r}) e^{-\beta \hat{\cal H}({\bf r}) } |
\zeta_{{\bf p}}({\bf r}) \rangle
} \nonumber \\
& = &
\langle \zeta_{\hat{\mathrm P} {\bf q}}({\bf r}) |
\zeta_{{\bf p}}({\bf r})
{\tilde O}({\bf r}, -i\hbar \nabla_{\bf r})
e^{-\beta {\hat{\tilde{{\cal H}}}({\bf r}) } } 1  \rangle
\nonumber \\ & = &
\langle \zeta_{\hat{\mathrm P} {\bf q}}({\bf r}) |
\zeta_{{\bf p}}({\bf r}) \rangle
{\tilde O}(\hat{\mathrm P} {\bf q}, -i\hbar \nabla_{\hat{\mathrm P}{\bf q}})
\left\{ e^{-\beta {\cal H}(\hat{\mathrm P}{\bf q},{\bf p}) }
W(\hat{\mathrm P}{\bf q},{\bf p})  \right\}
\nonumber \\ & = &
\langle \zeta_{\hat{\mathrm P} {\bf q}}({\bf r}) |
\zeta_{{\bf p}}({\bf r}) \rangle
{\tilde O}({\bf q},-i\hbar \nabla_{\bf q})
\left\{ e^{-\beta {\cal H}({\bf \Gamma}) }
W({\bf \Gamma})  \right\}.
\end{eqnarray}
The third equality assumes that the operator is symmetric
with respect to permutations of the particles.
Notice how the gradient operator now acts on the configuration positions;
henceforth this will simply be written
$\nabla \equiv \partial/\partial{\bf q}$.
These gradient operators act on everything to their right.

The part outside of the expectation value is a phase space
function used to obtain the statistical average.
For brevity it is useful to define
\begin{equation}
W_O({\bf \Gamma}) \equiv
e^{\beta {\cal H}({\bf \Gamma}) }
{\tilde O}({\bf q},-i\hbar \nabla_{\bf q})
\left\{ e^{-\beta {\cal H}({\bf \Gamma}) }
W({\bf \Gamma})  \right\} .
\end{equation}
The classical equilibrium average of this
is essentially the phase space integral
of the part outside of the expectation value,
since the  the Maxwell-Boltzmann factor will cancel with
the leading factor here.
With this and in view of Eq.~(\ref{Eq:Xi-pm-cont})
and of \S \ref{Sec:expXi},
the average can therefore be written
\begin{eqnarray}
\lefteqn{
\left< O({\bf r}, -i\hbar \nabla_{\bf r}) \right>_{\mu,V,T}
} \nonumber \\
& = &
\frac{1}{\Xi^\pm(W)}
\sum_{N}\frac{z^N  }{N!( \Delta_q \Delta_p )^{3N}}
\int \mathrm{d}{\bf \Gamma} \;
\nonumber \\ && \mbox{ } \times
g_{\bf p}({\bf q})
\chi^\pm(\zeta_{\bf q},\zeta_{\bf p})
{\tilde O}({\bf q},-i\hbar \nabla_{\bf q})
e^{-\beta {\cal H}({\bf \Gamma}) }
W({\bf \Gamma})
\nonumber   \\ & = &
\frac{1}{\Xi^\pm(W)}
\sum_{N}\frac{z^N  }{N!( \Delta_q \Delta_p )^{3N}}
\int \mathrm{d}{\bf \Gamma} \;
\nonumber \\ && \mbox{ } \times
g_{\bf p}({\bf q})
\chi^\pm(\zeta_{\bf q},\zeta_{\bf p})
e^{-\beta {\cal H}({\bf \Gamma}) }
W_O({\bf \Gamma})
\nonumber   \\ & = &
\frac{1}{\Xi^\pm(W)}
\sum_{N}\frac{z^N  }{N!h^{3N}}
%\nonumber \\ && \mbox{ } \times
\int \mathrm{d}{\bf \Gamma} \;
 e^{ -\beta {\cal H}({\bf \Gamma})}
\tilde \chi^\pm(\zeta_{\bf q},\zeta_{\bf p})
W_O({\bf \Gamma})
\nonumber   \\ & = &
\frac{\Xi^\pm(W_O)}{\Xi^\pm(W)} .
\end{eqnarray}
In the penultimate equality, the limit $\xi \rightarrow 0$ has been taken,
and there is now no $\xi$ dependence.

The loop resummation leading to the product form
of the grand partition function,
Eq.~(\ref{Eq:Xipm-resum})
may now be applied with the formal change
$W \Rightarrow W_O$.
That is
\begin{eqnarray}
\lefteqn{
\left<   O({\bf r}, -i\hbar \nabla_{\bf r}) \right>_{\mu,V,T}
}  \\
& = &
\frac{\Xi^\pm(W_O)}{\Xi^\pm(W)}
\nonumber   \\ & = &
\frac{\Xi_1(W_O)}{\Xi^\pm(W)}
\prod_{l=2}^\infty \exp
\left[ \left< X^{(l)}({\bf \Gamma}) \right>_{1,W_O} \right]
\nonumber   \\ & = &
\frac{\Xi_1(W_O)}{\Xi_1(W)}
\prod_{l=2}^\infty \exp
\left[ \left< X^{(l)}({\bf \Gamma}) \right>_{1,W_O}
%\right. \nonumber   \\ &  & \left. \mbox{ }
- \left< X^{(l)}({\bf \Gamma}) \right>_{1,W} \right] .\nonumber
\end{eqnarray}
Here $\Xi^\pm(W)$ is the full grand partition function,
Eq.~(\ref{Eq:Xipm-resum}),
and $\Xi_1(W)$ is the monomer  grand partition function, Eq.~(\ref{Eq:Xi1}).
Recall Eq.~(\ref{Eq:defX}),
$X^{(l)}({\bf \Gamma}) \equiv {N!} \tilde\chi^{(l)}({\bf \Gamma})/{l(N-l)!}$.

One has
\begin{eqnarray}
%\lefteqn{
\frac{\Xi_1(W_O)}{\Xi_1(W)}
%} \nonumber \\
& = &
\frac{1}{\Xi_1(W)}
\sum_{N}\frac{z^N }{N! h^{3N}  }
%\nonumber \\ && \mbox{ } \times
\int \mathrm{d}{\bf \Gamma} \;
 e^{ -\beta {\cal H}({\bf \Gamma})} W_O({\bf \Gamma})
 \nonumber   \\ & = &
\frac{ \left<  W_O  \right>_{1,0}}{\left<  W  \right>_{1,0} } ,
\end{eqnarray}
which gives
\begin{eqnarray} \label{Eq:<O(r,p)>-W}
\lefteqn{
\left<  O({\bf r}, -i\hbar \nabla_{\bf r}) \right>_{\mu,V,T}
}  \\
& = &
\frac{\langle W_O \rangle_{1,0}}{\langle W \rangle_{1,0}}
\prod_{l=2}^\infty \exp
%\right. \nonumber \\ && \left. \mbox{ } \times
\left< X^{(l)}
\left[ \frac{W_O}{ \langle W_O \rangle_{1,0} }
 -  \frac{ W }{\langle W \rangle_{1,0}}
 \right] \right>_{1,0}
\nonumber \\ & \approx &
\frac{\langle W_O \rangle_{1,0}}{\langle W \rangle_{1,0}}
%\nonumber \\ && \mbox{ }
+
\sum_{l=2}^\infty
\left< X^{(l)}
\left[ \frac{W_O}{ \langle W_O \rangle_{1,0} }
 -  \frac{ W }{\langle W \rangle_{1,0}}
 \right] \right>_{1,0}  .\nonumber
\end{eqnarray}
Of course one can write these classical equilibrium averages
as weighted averages,
$  \langle W_O \rangle_{1,0} =
 \langle W \rangle_{1,0} \langle W^{-1} W_O \rangle_{1,W}$.
One can linearize the exponential,
as in the final equality,
when the loop contribution is small.
This is no real advantage computationally.

The present results for the average of an operator
that is a general function of position and momentum operators
appears to preclude the existence of a universal
classical phase space probability density for quantum averages.
The fact that the result is written as a classical average
means that there does exist a phase space probability density
that is particular for each function being averaged,
namely $ e^{ -\beta {\cal H}({\bf \Gamma})} W_O({\bf \Gamma})$,
but this is not universal and independent of the operator function.
In contrast both Wigner\cite{Wigner32} and Kirkwood\cite{Kirkwood33}
give such a universal phase space probability density,
albeit a different one for each author.

If the function being averaged is solely
an ordinary function of position alone,
or a function of the momentum operator alone,
then there is reason to believe that the present results
will yield a universal probability density for configuration positions,
or for configuration momenta, respectively
and that these will agree with the respective quantities
of  both Wigner\cite{Wigner32} and Kirkwood\cite{Kirkwood33}
(see \S\S \ref{Sec:negl-sym} and \ref{Sec:<KE>})
when wave function symmetrization is neglected.

For the case that the operator being averaged is an ordinary function
of position alone,
$\hat O = O({\bf r})$,
then $W_O = O({\bf q}) W({\bf \Gamma})$ and this becomes
\begin{eqnarray} \label{Eq:<O(r)>-W}
\lefteqn{
\left<  O({\bf r}) \right>_{\mu,V,T}
}  \\
& = &
\frac{\langle OW \rangle_{1,0}}{\langle W \rangle_{1,0}}
\prod_{l=2}^\infty \exp
% \nonumber \\ &&   \mbox{ } \times
\left<X^{(l)}
\left[
\frac{OW}{\langle OW \rangle_{1,0}}
 -  \frac{ W }{\langle W \rangle_{1,0}}
 \right] \right>_{1,0}
 \nonumber \\ & = &
\langle O \rangle_{1,W}
\prod_{l=2}^\infty \exp
\left< X^{(l)}
\left[ \frac{O}{\langle O \rangle_{1,W}} -  1
 \right] \right>_{1,W}
\nonumber \\ & \approx &
\langle O \rangle_{1,W}
%\nonumber \\ && \mbox{ }
+  \sum_{l=2}^\infty
\left< X^{(l)}({\bf \Gamma})
\left[ O({\bf q}) -  \langle O({\bf q})  \rangle_{1,W}  \right] \right>_{1,W}
 .\nonumber
\end{eqnarray}
As above the monomer averages can be recast as
the ratio of classical averages
$ \left<  f \right>_{1,W} =
{ \left<  W f \right>_{1,0}}/{ \left<  W \right>_{1,0}}  $.

%%%%%%%%%%%%%%%%%%%%%%%%%%%%%%%%%%%%%%%%%%%
\subsubsection{Thermodynamic Derivatives}

Making the usual assumption that most likely values equal average values,
\cite{TDSM}
in classical thermodynamics and statistical mechanics
the average number is
\begin{eqnarray}
\left< N \right>
& = &
\frac{-\partial  \Omega}{\partial  \mu} .
\end{eqnarray}
(It saves space to drop the subscript $\mu,V,T$ from the average.)
For quantum systems,
the right hand side is a sum of monomer and loop potentials,
so that one has
\begin{eqnarray}
\left< N \right>
& = &
\sum_{l=1}^\infty \left< N \right>^{(l)}
\equiv
\sum_{l=1}^\infty \frac{-\partial  \Omega_l}{\partial  \mu} .
\end{eqnarray}

The monomer term is
$\left< N \right>^{(1)} \equiv \left< N \right>_{1,W}$, or
\begin{eqnarray}
\lefteqn{
\left< N \right>_{1,W}
}  \\
& = &
\frac{-\partial  \Omega_1}{\partial  \mu}
\nonumber \\ & = &
\frac{k_\mathrm{B}T}{\Xi_1(W)}
\frac{\partial  \Xi_1(W)}{\partial  \mu}
\nonumber \\ & = &
\frac{1}{ \Xi_1(W)}
\sum_{N=1}^\infty  \frac{N z^N}{ N!h^{3N}}
%\nonumber \\ && \mbox{ } \times
\int \mathrm{d}{\bf \Gamma} \;  e^{ -\beta {\cal H}({\bf \Gamma})}
W({\bf \Gamma}) .\nonumber
\end{eqnarray}
The symmetrization quantum corrections to the average number
are given by the terms  $l\ge 2$,
\begin{eqnarray}
\lefteqn{
\left<  N \right>^{(l)}
} \nonumber \\
& = &
\frac{-\partial \Omega_l}{\partial  \mu}
\nonumber \\ & = &
\frac{ \Omega_l }{\Xi_{1,W}}\frac{\partial  \Xi_{1,W}}{\partial  \mu}
%\nonumber \\ &  & \mbox{ }
+ \frac{1}{ \Xi_{1,W} }
\sum_{N=l}^\infty  \frac{N z^N}{ l(N-l)!h^{3N}}
\nonumber \\ && \mbox{ } \times
\int \mathrm{d}{\bf \Gamma} \;  e^{ -\beta {\cal H}({\bf \Gamma})}
W({\bf \Gamma})
\tilde\chi^{(l)}({\bf \Gamma})
\nonumber \\ & = &
 \beta \Omega_l \left<  N \right>_{1,W}
+  \left< X^{(l)}({\bf \Gamma})  N  \right>_{1,W}
\nonumber \\ & = &
\left< X^{(l)}({\bf \Gamma})
\left[ N -\left<  N \right>_{1,W} \right]
\right>_{1,W} .
\end{eqnarray}
In the penultimate equality both terms are ${\cal O}(N^2)$.
The final equality shows the cancelation between them,
with the residue quantum correction presumably ${\cal O}(N)$.

Interestingly enough, this thermodynamic derivative
gives the same answer as the linearized form of the average
given above.

The temperature derivative of the grand potential
gives, in essence, the average energy,\cite{TDSM}
\begin{eqnarray}
\left< E - \mu N \right>
& = &
\frac{\partial  \beta \Omega}{\partial  \beta} .
\end{eqnarray}
This is the thermodynamic expression.

But the present theory for quantum statistical mechanics
gives a slightly different result.
The monomer result is
%\left< E - \mu N \right>_{1,W}
\begin{eqnarray}
\lefteqn{
\frac{\partial  \beta \Omega_1}{\partial  \beta}
} \nonumber \\
& = &
\frac{-1}{\Xi_{1,W}}
\frac{\partial  \Xi_{1,W}}{\partial  \beta}
\nonumber \\ & = &
\frac{1}{\Xi_{1,W}}
\sum_{N=1}^\infty  \frac{z^N}{ N!h^{3N}}
%\nonumber \\ && \mbox{ } \times
\int \mathrm{d}{\bf \Gamma} \;
e^{ -\beta {\cal H}({\bf \Gamma})} W({\bf \Gamma})
\nonumber \\ && \mbox{ } \times
\left[ {\cal H}({\bf \Gamma}) - \mu N
- \frac{\partial  \ln W({\bf \Gamma})}{\partial  \beta} \right]
\nonumber \\ & = &
\left<
{\cal H}({\bf \Gamma}) - \mu N
%\right. \nonumber \\ && \left. \mbox{ }
- \frac{\partial  \ln W({\bf \Gamma})}{\partial  \beta}
\right>_{1,W} .
\end{eqnarray}
Because the quantum corrections for non-commutativity
are temperature-dependent,
one sees that it is only the classical term
$W({\bf \Gamma}) = W_0({\bf \Gamma}) = 1$
that has the form given by classical thermodynamics,
\begin{eqnarray}
\frac{\partial  \beta \Omega_{1,0}}{\partial  \beta}
=
\left< {\cal H}({\bf \Gamma}) - \mu N  \right>_{1,0} .
\end{eqnarray}

For the symmetrization quantum corrections, $l \ge 2$, one has
\begin{eqnarray}
\lefteqn{
\frac{\partial  \beta \Omega_l}{\partial  \beta}
}  \\
& = &
%\frac{\partial  \beta \Omega_1}{\partial  \beta} \beta\Omega_l
\frac{-\beta\Omega_l}{\Xi_{1,W}}
\frac{\partial  \Xi_{1,W}}{\partial  \beta}
%\nonumber \\ &  & \mbox{ }
+ \frac{1}{ \Xi_{1,W}}
\sum_{N=l}^\infty  \frac{z^N}{ N!h^{3N}}
\nonumber \\ && \mbox{ } \times
\int \mathrm{d}{\bf \Gamma} \;
e^{ -\beta {\cal H}({\bf \Gamma})}  W({\bf \Gamma})
X^{(l)}({\bf \Gamma})
\nonumber \\ && \mbox{ } \times
\left[ {\cal H}({\bf \Gamma}) -\mu N
- \frac{\partial  \ln W({\bf \Gamma})}{\partial  \beta} \right]
\nonumber \\ & = &
\left< X^{(l)}({\bf \Gamma})
\left\{  \rule{0cm}{0.4cm}
 {\cal H}({\bf \Gamma}) -\mu N
 - \frac{\partial  \ln W({\bf \Gamma})}{\partial  \beta}
 \right. \right. \nonumber \\ && \left. \left. \mbox{ }
-\left<  {\cal H}({\bf \Gamma}) -\mu N
- \frac{\partial  \ln W({\bf \Gamma})}{\partial  \beta} \right>_{1,W} \right\}
\right>_{1,W} .\nonumber
\end{eqnarray}

%%%%%%%%%%%%%%%%%%%%%%%%%%%%%%%%%%%%%%%%%%%%%%%%%%%%%%%%%%%%%%%%%%%%%%%%%%
%
\section{Comparison with Known Results} \label{Sec:compare}
\setcounter{equation}{0} \setcounter{subsubsection}{0}
%
%%%%%%%%%%%%%%%%%%%%%%%%%%%%%%%%%%%%%%%%%%%%%%%%%%%%%%%%%%%%%%%%%%%%%%%%%%

%%%%%%%%%%%%%%%%%%%%%%%%%%%%%%%%%%%%%%%%%%%%%%%%%%%%%%%%%%%%%
\subsection{Ideal Gas} \label{Sec:Ideal}

One test of the present formalism is to apply it to the
case of the quantum ideal gas, where the exact fugacity expansion is known.
For an ideal gas the potential energy vanishes,
$U({\bf r}) = 0$.
Since the quantum corrections due to non-commutativity
depend upon the gradient of the potential,
these vanish for the ideal gas,
$W^\mathrm{id}({\bf r}) = 1$,
and one need only retain the commuting part of the expansion,
\S \ref{Sec:commute},
$\Omega_l^{\pm,\mathrm{id}}\equiv\Omega_{l,0}^{\pm}$.

For the case of the ideal gas
the classical $l$-particle density is
\begin{equation}
\rho^{(l),\mathrm{id}}({\bf q}) = \Lambda^{-3l} z^l  .
\end{equation}
This assumes a homogeneous system.

With this and using Eq.~(\ref{Eq:Det(A)}),
the $l$-loop grand potential for the ideal gas is
\begin{eqnarray}
\lefteqn{
-\beta \Omega_l^{\pm,\mathrm{id}}
} \nonumber \\
 & = &
\frac{ (\pm 1)^{l-1} V \Lambda^{-3l} z^l  }{ l   }
\int \mathrm{d}{\bf q}^{l-1}\;
e^{- \pi ( {\bf q} -  {\bf q}')^2/{\Lambda^2} }
\nonumber \\ & = &
\frac{ (\pm 1)^{l-1} V \Lambda^{-3l} z^l  }{ l   }
\int \mathrm{d}{\bf q}^{l-1}\;
e^{- \pi  \sum_\alpha \underline{\underline A}^{(l-1)} :
\underline { q}_\alpha^{l-1} \underline { q}_\alpha^{l-1} /{\Lambda^2} }
\nonumber \\ & = &
\frac{ (\pm 1)^{l-1} V \Lambda^{-3l} z^l  }{ l   }
\left( \frac{ 2\pi \Lambda^2}{ 2\pi  } \right)^{3(l-1)/2}
\left| \underline{\underline A}^{(l-1)} \right|^{-3/2}
\nonumber \\ & = &
\frac{ (\pm 1)^{l-1} \Lambda^{-3} V  z^l }{ l^{5/2}  } .
\end{eqnarray}
The upper sign is for bosons, and the lower for fermions.
This holds for $l \ge 2$.
For the monomer case direct calculation shows that
$-\beta \Omega_1^{\mathrm{id}} = zV/\Lambda^3$,
which in fact is just this expression for $-\beta \Omega_l^{\pm,\mathrm{id}}$
with $l=1$.

The thermodynamic relation between the pressure and the grand potential is
$p = -\Omega/V$.\cite{TDSM}
The classical ideal gas pressure is
$p^\mathrm{cl,id} = z k_\mathrm{B}T/\Lambda^{3}
= -\Omega_1^{\mathrm{id}}/V$.
With these and the above result,
the pressure of the quantum ideal gas is given by
\begin{eqnarray}
\beta p^{\pm,\mathrm{id}} \Lambda^{3}
& = & \frac{-\beta\Lambda^{3}}{V}
\sum_{l=1}^\infty \Omega_l^{\pm,\mathrm{id}}
\nonumber \\ & = &
\sum_{l=1}^\infty
(\pm 1)^{l-1}  z^l l^{-5/2}  .
\end{eqnarray}
This is the known result.\cite{Pathria72}

%%%%%%%%%%%%%%%%%%%%%%%%%%%%%%%%%%%%%%%%
\subsection{Comparison with Wigner (1932) and Kirkwood (1933)}

%%%%%%%%%%%%%%%%%%%%%%%%%%%%%%
\subsubsection{Position Configuration Weight Density} \label{Sec:negl-sym}

Wigner's\cite{Wigner32} formulation of the quantum states
that represent phase space and the consequent phase space
quasi-probability function neglect the effects of wave function symmetrization.
This may be compared to the $l=1$ term in the above expressions.
The monomer grand potential is
$\Omega_1 = - k_\mathrm{B}T \ln \Xi_1$,
with the monomer grand partition function being
given by Eq.~(\ref{Eq:Xi1})
\begin{eqnarray}
\Xi_{1,W}
& = &
\sum_{N}\frac{z^N }{N! h^{3N}  }
%\nonumber \\ && \mbox{ } \times
\int \mathrm{d}{\bf \Gamma} \;
 e^{ -\beta {\cal H}({\bf \Gamma})} W({\bf \Gamma})
 \nonumber \\ &=&
\Xi_{1,0} \left<  W  \right>_{1,0} ,
\end{eqnarray}
where classical averages appear in the final equality.
Hence the grand potential in this approximation is
\begin{eqnarray}
\Omega_{1,W}
& = &
\Omega_{1,0} - k_\mathrm{B}T \ln \left<  W  \right>_{1,0}
\nonumber \\ & \approx &
\Omega_{1,0} - k_\mathrm{B}T  \left<  \hbar W_1 + \hbar^2 W_2 \right>_{1,0}
+{\cal O}(\hbar^4)
\nonumber \\ & = &
\Omega_{1,0} - k_\mathrm{B}T  \left<  \hbar^2 W_2 \right>_{1,0}.
\end{eqnarray}
The final equality follows because
the first quantum correction due to non-commutativity,
Eq.~(\ref{Eq:W1}),
is proportional to the momentum ${\bf p}$,
which averages to zero in an equilibrium system,
$ \left<   W_1 \right>_{1,0} = 0$.

The second order term is given by Eq.~(\ref{Eq:W2}),
\begin{eqnarray}
W_2
& = &
\frac{-\beta^4}{8m^2} ({\bf p} \cdot \nabla U)^2
% \nonumber \\ && \mbox{ }
+ \frac{\beta^3}{6m^2}  ( {\bf p} \cdot \nabla )^2 U
\nonumber \\ && \mbox{ }
- \frac{\beta^2}{4m} \nabla^2  U
+ \frac{\beta^3}{6m} \nabla U \cdot \nabla U .
\end{eqnarray}
The classical equilibrium phase space weight
is an even function of the momenta,
and,
by the equipartition theorem or directly,
$ \left< {\bf p} {\bf p} \right>_{1,0} =  m  k_\mathrm{B}T \, \mathrm{I}$ .
Hence to quadratic order
\begin{eqnarray} \label{Eq:<W>10}
%\lefteqn{
 \left<  W \right>_{1,0}
%} \nonumber \\
& = &
1 +  \hbar^2 \left<  W_2 \right>_{1,0} +{\cal O}(\hbar^4)
\nonumber \\ & = &
1 -
\hbar^2 \left<
\frac{\beta^2}{4m}\nabla^2 {U}
- \frac{\beta^3}{6m} \nabla U \cdot \nabla U
\right. \nonumber \\ && \left. \mbox{ }
+ \frac{\beta^3}{8m}  \nabla U \cdot \nabla U
- \frac{\beta^2}{6m}  \nabla^2 {U}
\right>_{1,0}
\nonumber \\ & = &
1 -
\hbar^2 \left<
\frac{\beta^2}{12m}\nabla^2 {U}
- \frac{\beta^3}{24m} \nabla U \cdot \nabla U
\right>_{1,0}
\nonumber \\ & = &
1 -
\frac{\hbar^2 \beta^2}{24m}
\left< \nabla^2 {U} \right>_{1,0} .
\end{eqnarray}
The final equality uses the fact that
$  \langle \nabla U  \cdot \nabla U \rangle_{1,0}
=  \beta^{-1} \langle \nabla^2 U  \rangle_{1,0}$,
which may be derived by an integration by parts,
assuming that either $e^{-\beta U({\bf q})}$ or $\nabla U$
vanish on the boundary of the system.
(Or else that boundary contributions are relatively negligible
in the thermodynamic limit.)
This gives the first quantum correction to the grand potential
due to the non-commutativity of the position and momentum operators.

The statistical average of an operator that
is an ordinary function of the position coordinates
is given in Eq.~(\ref{Eq:<O(r)>-W}).
Taking only the $l=1$ term of this gives
\begin{equation}
\left<  O({\bf r}) \right>_{\mu,V,T}
=
\left<  O  \right>_{1,W}
=
\frac{ \left<  W O \right>_{1,0}}{ \left<  W \right>_{1,0} } ,
\end{equation}

It follows that since the operator is a function only of the position
\begin{eqnarray}
\lefteqn{
 \left<  W({\bf \Gamma}) O({\bf q}) \right>_{1,0}
}  \\
& = &
\left< O \right>_{1,0} -
\hbar^2 \left<
\left[
\frac{\beta^2}{12m}\nabla^2 {U}
- \frac{\beta^3}{24m} \nabla U \cdot \nabla U
\right] O \right>_{1,0} \nonumber .
\end{eqnarray}
Hence
the leading non-commutativity correction
to an average of a function
of the position is
\begin{eqnarray}
\lefteqn{
\left<  O({\bf r}) \right>_{\mu,V,T}
}  \\
& = &
\left<  O({\bf q})  \right>_{1,W}
\nonumber \\ & = &
\frac{ \left<  W O({\bf q}) \right>_{1,0}}{ \left<  W \right>_{1,0} }
\nonumber \\ & = &
\left< O({\bf q}) \right>_{1,0} -
\hbar^2 \left<
\left[
\frac{\beta^2}{12m}\nabla^2 {U}
- \frac{\beta^3}{24m} \nabla U \cdot \nabla U
\right]
\right. \nonumber \\ &  & \left. \mbox{ } \times
\left[ O({\bf q}) - \left< O({\bf q}) \right>_{1,0} \right]\right>_{1,0}
+{\cal O}(\hbar^4).
\end{eqnarray}

From this
one can identify the probability of a configuration position
in classical phase space,
%taking into account non-commutativity of positions and momenta, but
neglecting wave function symmetrization, to quadratic order in $\hbar$
\begin{equation}
\wp({\bf q}) \propto
e^{-\beta U({\bf q})} \left\{
 1 - \frac{\hbar^2 \beta^2}{12m}\nabla^2 {U}
+ \frac{\hbar^2 \beta^3}{24m} \nabla U \cdot \nabla U
\right\} .
\end{equation}
This is identical
to  Wigner's Eq.~(28).\cite{Wigner32}
(Kirkwood says that his configuration position probability density
is also identical to Wigner's.)\cite{Kirkwood33}

One has to distinguish three quantities:
the  configuration position probability density,
the configuration momentum probability density,
and the phase space probability density.
As just mentioned,
the present theory agrees with Wigner and Kirkwood
for the  configuration position probability density
(in the absence of wave function symmetrization).
As will be shown next,
it gives the same result for the average kinetic energy,
which suggests that it also agrees with Wigner and Kirkwood
for the  configuration momentum probability density
(in the absence of wave function symmetrization).
However the present theory does not appear to give
a phase space probability density
(ie.\ one that is independent of the operator being averaged),
although it does give a classical average for an arbitrary function
of position and momentum operators.
This is in contrast to Wigner\cite{Wigner32} and Kirkwood\cite{Kirkwood33}
who each claim to give a phase space quasi-probability density.
The densities in this case do not agree with each other,
and Kirkwood specifically states that due to the lack of
simultaneity in position and momentum,
one should not really expect a phase space probability density
to exist.\cite{Kirkwood33}

\comment{ %%%%%%%%%%%%%%%%%%%%%%%%%%%%%%%%%%%%%%%%%%%%
It is worth mentioning
that the equilibrium average of the momentum operator vanishes,
$\left< - i\hbar \nabla \right>_{\mu,V,T} = 0$.
This is obviously a necessary condition for any exact theory.
In the present case it follows because
the average of a Hermitian operator must be real,
and therefore this must couple to odd powers of $i$ in $W_\nabla$,
which themselves are multiplied by the same power of ${\bf p}$,
which therefore averages to zero.
This can be shown explicitly for the leading order term.
} % end comment %%%%%%%%%%%%%%%%%%%%%%%%%%%%%%%%%%%%%%

%%%%%%%%%%%%%%%%%%%%%%%%%%%%%%
\subsubsection{Average Kinetic Energy} \label{Sec:<KE>}

Now as an example of an operator that is a function of the momentum operator,
consider the kinetic energy operator,
$\hat{\cal K} =(-\hbar^2/2m)\nabla^2$.
This is transformed into
\begin{eqnarray}
\hat{\tilde{\cal K}}
& = &
e^{{\bf p} \cdot{\bf r}/i\hbar}
\hat{\cal K}
e^{-{\bf p} \cdot{\bf r}/i\hbar}
\nonumber \\ & = &
\left\{
\frac{  p^2 }{2m}
- \frac{ i\hbar}{m} {\bf p} \cdot \nabla
 -\frac{\hbar^2}{2m} \nabla^2
\right\}  .
\end{eqnarray}
Henceforth $\nabla = \partial/\partial{\bf q}$.

Keeping only the leading quantum correction to quadratic order in $\hbar$,
$W_0({\bf q}) = 1$,
$ W_1({\bf \Gamma})$ was given in Eq.~(\ref{Eq:W1})
and
$ W_2({\bf \Gamma})$ was given in Eq.~(\ref{Eq:W2}),
\begin{eqnarray}
\lefteqn{
W_{\cal K}({\bf \Gamma})
} \nonumber \\
& \equiv &
e^{\beta {\cal H}({\bf \Gamma}) }
\hat{\tilde{{\cal K}}}
\left\{
e^{-\beta {\cal H}({\bf \Gamma}) } W({\bf \Gamma}) \right\}
\nonumber \\ & = &
e^{\beta {\cal H}({\bf \Gamma}) }
\left\{
\frac{  p^2 }{2m}
- \frac{ i\hbar}{m} {\bf p} \cdot \nabla
 -\frac{\hbar^2}{2m} \nabla^2
\right\}
\nonumber \\ &  & \mbox{ } \times
\left\{
e^{-\beta {\cal H}({\bf \Gamma}) }
\left[ 1 + \hbar W_1({\bf \Gamma}) + \hbar^2 W_2({\bf \Gamma}) \right]  \right\}
\nonumber \\ & = &  %%%%%%%%%%%%%%%%%%%%%%%%%%%%%%%%%%%%%%%%%%%%%
e^{\beta {\cal H}({\bf \Gamma}) }
\frac{  p^2 }{2m} e^{-\beta {\cal H}({\bf  \Gamma}) }
\left[ 1 + \hbar W_1({\bf \Gamma}) + \hbar^2 W_2({\bf \Gamma}) \right]
\nonumber \\ &  & \mbox{ }
-
e^{\beta {\cal H}({\bf \Gamma}) }
\frac{ i\hbar}{m} {\bf p} \cdot \nabla
\left\{ e^{-\beta {\cal H}({\bf \Gamma}) }
 \frac{(-i\hbar \beta^2)}{2m}  {\bf p} \cdot \nabla U
\right\}
\nonumber \\ &  & \mbox{ }
 -
e^{\beta {\cal H}({\bf \Gamma}) }
\frac{\hbar^2}{2m} \nabla^2  e^{-\beta {\cal H}({\bf \Gamma}) }
+ {\cal O}(\hbar^4)
\nonumber \\ & = &   %%%%%%%%%%%%%%%%%%%%%%%%%%%%%%%%%%%%%
\frac{  p^2 }{2m}
-\frac{\hbar^2\beta^2 p^2}{8m^2}   \nabla^2 {U}
+ \frac{\hbar^2\beta^3p^2}{12m^2}  \nabla U  \cdot \nabla U
\nonumber \\ &  &  \mbox{ }
- \frac{\hbar^2\beta^4p^2}{16m^3}  ({\bf p}\cdot \nabla U)^2
+ \frac{\hbar^2\beta^3p^2}{12m^3}   {\bf p} {\bf p} : \nabla\nabla U
 \nonumber \\ &  &  \mbox{ }
+
\frac{\hbar^2\beta^3}{2m^2} [ {\bf p}\cdot \nabla U ]^2
%\right. \nonumber \\ &  & \left. \mbox{ }
-  \frac{ \hbar^2\beta^2}{2m^2} {\bf p} {\bf p} : \nabla \nabla U
 \nonumber \\ &  &  \mbox{ }
+ \frac{\hbar^2\beta }{2m} \nabla^2  U
 -\frac{\hbar^2\beta^2 }{2m} \nabla U \cdot \nabla U .
\end{eqnarray}
Terms linear in ${\bf p}$ have been neglected
because these will average to zero.
Terms of higher order than ${\cal O}(\hbar^2)$ have also been neglected.

Since an expansion to leading order in $\hbar$ has been performed,
the exponential can be linearized
and the corrections due symmetrization can be neglected.
Hence the statistical average of the kinetic energy operator is
given by Eq.~(\ref{Eq:<O(r,p)>-W}) with $X^{(l)} = 0$.
That is
\begin{equation}
\left< {\cal K} \right>_{\mu,V,T}
=
\frac{\langle W_{\cal K}({\bf \Gamma}) \rangle_{1,0}}{\langle W \rangle_{1,0}}.
\end{equation}
These are classical equilibrium averages.

The momentum terms may be averaged first.
One has the usual results
\begin{eqnarray}
\langle p^2 \rangle_{1,0} & = & 3N m k_\mathrm{B}T ,
\nonumber \\
\langle  {\bf p} {\bf p} \rangle_{1,0}
& = &
m k_\mathrm{B}T  {\mathrm I} ,
\nonumber \\
\langle p^2  {\bf p} {\bf p} \rangle_{1,0}
& = &
[ 3N-1 + 3 ] (m k_\mathrm{B}T)^2  {\mathrm I} .
\end{eqnarray}
The leading order with respect to $N$  in the thermodynamic limit  will cancel,
and so it will prove necessary to also keep the second order in the following.
Using these and the fact that
$  \langle \nabla U  \cdot \nabla U \rangle_{1,0}
=  \beta^{-1} \langle \nabla^2 U  \rangle_{1,0}$,
the average of the kinetic energy function becomes
\begin{eqnarray}
\lefteqn{
\langle W_{\cal K}({\bf \Gamma}) \rangle_{1,0}
} \nonumber \\
& = &
\frac{  3N k_\mathrm{B}T }{2}
\nonumber \\ &  &  \mbox{ }
- 3N \frac{\hbar^2\beta }{8m}   \langle \nabla^2 {U} \rangle_{1,0}
+ 3N \frac{\hbar^2\beta^2 }{12m}  \langle \nabla U  \cdot \nabla U \rangle_{1,0}
\nonumber \\ &  &  \mbox{ }
- (3N+2) \frac{\hbar^2\beta^2}{16m}
\langle \nabla U  \cdot \nabla U \rangle_{1,0}
\nonumber \\ &  &  \mbox{ }
+ (3N+2)  \frac{\hbar^2\beta}{12m}  \langle \nabla^2 {U} \rangle_{1,0}
 \nonumber \\ &  &  \mbox{ }
+
\frac{\hbar^2\beta^2}{2m} \langle \nabla U  \cdot \nabla U \rangle_{1,0}
%\right. \nonumber \\ &  & \left. \mbox{ }
-  \frac{ \hbar^2\beta}{2m} \langle \nabla^2 {U} \rangle_{1,0}
 \nonumber \\ &  &  \mbox{ }
+ \frac{\hbar^2\beta }{2m} \langle \nabla^2 {U} \rangle_{1,0}
 -\frac{\hbar^2\beta^2 }{2m} \langle \nabla U  \cdot \nabla U \rangle_{1,0}
\nonumber \\ & = &   %%%%%%%%%%%%%%%%%%%%%%%%%%%%%%%%%%%%%
\frac{  3N k_\mathrm{B}T }{2}
+ \frac{ \hbar^2\beta }{m}  \langle \nabla^2 {U} \rangle_{1,0}
\nonumber \\ &  &  \mbox{ } \times
\left\{ \frac{-3N}{8} + \frac{3N}{12} - \frac{3N+2}{16} + \frac{3N+2}{12}
\right\}
%\nonumber \\ & =  &
%\frac{  3N k_\mathrm{B}T }{2}
%+ \frac{ \hbar^2\beta }{m}  \langle \nabla^2 {U} \rangle_{1,0}
%\nonumber \\ &  &  \mbox{ } \times
%\frac{-18N + 12N -9N - 6 + 12N + 8 }{48}
\nonumber \\ & =  &
\frac{  3N k_\mathrm{B}T }{2}
- \frac{ (3N-2) \hbar^2\beta }{48 m}  \langle \nabla^2 {U} \rangle_{1,0}.
\end{eqnarray}
Using
$\langle W \rangle_{1,0} =
1 - { \hbar^2\beta^2 }  \langle \nabla^2 {U} \rangle_{1,0}/{24m}
+{\cal O}(\hbar^4)$,
which was shown in Eq.~(\ref{Eq:<W>10}),
the average of the kinetic energy operator is
\begin{eqnarray}
\left< {\cal K} \right>_{\mu,V,T}
& = &
\frac{\langle W_{\cal K}({\bf \Gamma}) \rangle_{1,0}
}{
\langle W({\bf \Gamma}) \rangle_{1,0} }
\nonumber \\ & = &
\frac{  3N k_\mathrm{B}T }{2}
+ \frac{ \hbar^2\beta }{24 m}  \langle \nabla^2 {U} \rangle_{1,0}
\nonumber \\ &  &  \mbox{ }
- \frac{ 3N \hbar^2\beta }{48 m}  \langle \nabla^2 {U} \rangle_{1,0}
+ \frac{ 3N \hbar^2\beta }{48m}  \langle \nabla^2 {U} \rangle_{1,0}
%\nonumber \\ &  &  \mbox{ }
%+{\cal O}(\hbar^4)
\nonumber \\ & = &
\frac{  3N k_\mathrm{B}T }{2}
+ \frac{ \hbar^2\beta }{24 m}  \langle \nabla^2 {U} \rangle_{1,0}
%\nonumber \\ &  &  \mbox{ }
+{\cal O}(\hbar^4).
\end{eqnarray}
This agrees with Wigner's Eq.~(30).\cite{Wigner32}

%%%%%%%%%%%%%%%%%%%%%%%%%%%%%%
\subsubsection{Dimer Grand Potential
with First Correction for Non-Commutativity} \label{Sec:Omega-21}

The loop potential is given in Eq.~(\ref{Eq:Omega_l=<X^lW>10}),
\begin{eqnarray}
\Omega_l
& = &
\frac{ - k_\mathrm{B}T \Xi_{1,0} }{\Xi_1}
 \left<  X^{(l)}({\bf \Gamma}) W({\bf \Gamma}) \right>_{1,0}
\nonumber \\ & = &
- k_\mathrm{B}T
\frac{
 \left<  X^{(l)}({\bf \Gamma})  W({\bf \Gamma}) \right>_{1,0}
}{
\left<  W({\bf \Gamma}) \right>_{1,0}
} .
\end{eqnarray}
With only the first order non-commuting correction,
this may be labeled and written
\begin{eqnarray}
-\beta \Omega_{l,1}
& = &
\beta \Omega_{l,0} +
\frac{
 \left<  X^{(l)}  \right>_{1,0}  +  \hbar \left<  X^{(l)}  W_1 \right>_{1,0}
}{
1 + \hbar \left<  W_1 \right>_{1,0}
}
\nonumber \\ & = &
\hbar \left<   X^{(l)} W_1 \right>_{1,0} ,
\end{eqnarray}
since $ \left<  W_1 \right>_{1,0}=0$.
For the dimer $l=2$,
\begin{equation}
X^{(2)} = \frac{ \pm N(N-1)}{2} e^{ {\bf q}_{12}\cdot {\bf p}_{12}/i\hbar}
\end{equation}
and
\begin{eqnarray}
-\beta \Omega_{2,1}
& = &
\frac{-i\hbar\beta^2}{2m}
\left<
\frac{ \pm N(N-1)}{2} e^{ {\bf q}_{12}\cdot {\bf p}_{12}/i\hbar}
 {\bf p} \cdot \nabla U  \right>_{1,0}
\nonumber \\ & = &
\frac{\mp i\hbar\beta^2}{2m \Xi_{1,0}}
\sum_{N=2}^\infty \frac{ z^N N(N-1) }{ 2h^{3N} }
\int \mathrm{d}{\bf \Gamma} \;
\nonumber \\ & & \mbox{ } \times
e^{-\beta p^2/2m} e^{-\beta U({\bf q})}
e^{ {\bf q}_{12}\cdot {\bf p}_{12}/i\hbar}
 {\bf p} \cdot \nabla U({\bf q})
\nonumber \\ & = &
\frac{\mp i\hbar\beta^2}{4m \Xi_{1,0}}
\sum_{N=2}^\infty \frac{ z^N N(N-1) }{ h^6 \Lambda^{3(N-2)} }
%\nonumber \\ & & \mbox{ } \times
\int \mathrm{d}{\bf p}_1 \,  \mathrm{d}{\bf p}_2 \, \mathrm{d}{\bf q}\;
\nonumber \\ & & \mbox{ } \times
e^{-\beta U({\bf q})} e^{-\Lambda^2 [p_1^2 + p_2^2]/4\pi\hbar^2}
e^{ {\bf q}_{12}\cdot {\bf p}_{12}/i\hbar}
\nonumber \\ & & \mbox{ } \times
[ {\bf p}_1 \cdot \nabla_1 + {\bf p}_2 \cdot \nabla_2 ] U .
\end{eqnarray}
Now
\begin{eqnarray}
\lefteqn{
\frac{ -\Lambda^2 [p_1^2 + p_2^2]}{4\pi\hbar^2}
+ \frac{ {\bf q}_{12}\cdot {\bf p}_{12} }{ i\hbar }
} \nonumber \\
&=&
\frac{ -\Lambda^2 }{4\pi\hbar^2}
\left\{
\left[ {\bf p}_1 - \frac{4\pi\hbar^2}{ 2i\hbar \Lambda^2 }{\bf q}_{12} \right]^2
%\right. \nonumber \\ & & \left. \mbox{ }
+
\left[ {\bf p}_2 - \frac{4\pi\hbar^2}{ 2i\hbar \Lambda^2 }{\bf q}_{21} \right]^2
\right\}
\nonumber \\ & & \mbox{ }
-
\frac{2\pi }{\Lambda^2 } {\bf q}_{12} \cdot {\bf q}_{12} .
\end{eqnarray}
Hence the correction becomes
\begin{eqnarray} \label{Eq:Omega-21}
\lefteqn{
-\beta \Omega_{2,1}
} \nonumber \\
& = &
\frac{\mp i\hbar\beta^2}{4m \Xi_{1,0}}
\frac{ (2\pi)^{3} (2\pi\hbar^2)^{3}}{\Lambda^6}
\sum_{N=2}^\infty \frac{ z^N N(N-1) }{ h^6 \Lambda^{3(N-2)} }
\nonumber \\ & & \mbox{ } \times
\frac{4\pi\hbar^2}{ 2i\hbar \Lambda^2 }
\int \mathrm{d}{\bf q}\;
%\nonumber \\ & & \mbox{ } \times
e^{-\beta U({\bf q})}
e^{- {2\pi }{\bf q}_{12} \cdot {\bf q}_{12} /{\Lambda^2 } }
\nonumber \\ & & \mbox{ } \times
[ {\bf q}_{12} \cdot \nabla_1 + {\bf q}_{21}\cdot \nabla_2 ] U
\nonumber \\ & = &
\frac{\mp \beta}{4 \Xi_{1,0}}
\sum_{N=2}^\infty \frac{ z^N N(N-1) }{  \Lambda^{3N} }
\int \mathrm{d}{\bf q}\;
e^{-\beta U({\bf q})}
\nonumber \\ & & \mbox{ } \times
e^{- {2\pi }{q}_{12}^2 /{\Lambda^2 } }
%\nonumber \\ & & \mbox{ } \times
[ {\bf q}_{12} \cdot \nabla_1 + {\bf q}_{21}\cdot \nabla_2 ] U
\nonumber \\ & = &
\mp \beta
\left< \sum_{j<k}
e^{- {2\pi }{q}_{jk}^2 /{\Lambda^2 } }
{q}_{jk} u'(q_{jk})
\right>_{1,0} .
\end{eqnarray}
The final equality holds for a pair-wise additive potential,
\begin{eqnarray}
\lefteqn{
\frac{N(N-1)}{2} [ {\bf q}_{12} \cdot \nabla_1 + {\bf q}_{21}\cdot \nabla_2 ] U
} \nonumber \\
& = &
\sum_{j<k} [ {\bf q}_{jk} \cdot \nabla_j + {\bf q}_{kj}\cdot \nabla_k ]
u(q_{jk})
%\nonumber \\ & = &
%\sum_{j<k}
%\left[ {\bf q}_{jk} \cdot u'(q_{jk}) \frac{{\bf q}_{jk}}{{q}_{jk}}
%+ {\bf q}_{kj}\cdot u'(q_{jk}) \frac{{\bf q}_{kj}}{{q}_{jk}} \right]
\nonumber \\ & = &
2 \sum_{j<k}  {q}_{jk} u'(q_{jk}) .
\end{eqnarray}

This result for the first quantum correction for both
non-commutativity and  symmetrization,
$\Omega_{2,1}$,
is a factor of two smaller than the corresponding result
given by Kirkwood, Eq.~(21).\cite{Kirkwood33}
(The difference between the canonical
and the grand canonical case is immaterial.
The minus sign arises because Kirkwood treats the partition function,
whereas the present result is for the grand potential.)
This factor of 2 arises because
 Kirkwood counts the pair transposition $\hat{\mathrm P}_{jk}$ twice.
In the opinion of the present author, this is an error.
It is essentially the same error  as in Kirkwood's expression
for the partition function, his Eq.~(1),\cite{Kirkwood33}
which does not sum over unique states,
in contrast to the present Eq.~(\ref{Eq:Xi=sum'-n}).

%\newpage % $\;$ \newpage
%%%%%%%%%%%%%%%%%%%%%%%%%%%%%%%%%%%%%%%%%%%%%%%%%%%%%%%%%%%%%%%%%%%%%%%%%%
%
\section{Numerical Analysis and Results} \label{Sec:numerical}
\setcounter{equation}{0} \setcounter{subsubsection}{0}
%
%%%%%%%%%%%%%%%%%%%%%%%%%%%%%%%%%%%%%%%%%%%%%%%%%%%%%%%%%%%%%%%%%%%%%%%%%%

%%%%%%%%%%%%%%%%%%%%%%%%%%%%%%
\subsection{Neglect Non-Commutativity}

An initial level of approximation is to neglect the quantum corrections
due to non-commutativity of the position and momentum operators.
That is $W({\bf \Gamma}) = W_0({\bf \Gamma}) = 1$.
In this case the grand potential is
$\Omega^\pm_0 = \sum_{l=1}^\infty \Omega_{l,0}^\pm$.
The monomer term is the classical grand potential,
$\Omega_{1,0} = - k_\mathrm{B}T \ln \Xi_{1,0}$,
with the monomer grand partition function being given by Eq.~(\ref{Eq:Xi10}),
\begin{eqnarray}
\Xi_{1,0}
& = &
\sum_{N}\frac{z^N }{N! h^{3N}  }
%\nonumber \\ && \mbox{ } \times
\int \mathrm{d}{\bf \Gamma} \;
 e^{ -\beta {\cal H}({\bf \Gamma})} .
\end{eqnarray}
The classical pressure is $p_\mathrm{cl} = -\Omega_{1,0}/V$.
The various loop corrections to this are
given in Eq.~(\ref{Eq:bWpk-int-q}), $l \ge 2$,
\begin{eqnarray}
%\lefteqn{
-\beta \Omega_{l,0}^\pm
%} \nonumber \\
 & = &
\frac{ (\pm 1)^{l-1} V }{ l   }
\nonumber \\ && \mbox{ } \times
\int \mathrm{d}{\bf q}^{l-1}\;
e^{- \pi ( {\bf q} -  {\bf q}')^2/{\Lambda^2} }
\rho^{(l)}({\bf q}).
\end{eqnarray}

%%%%%%%%%%%%%%%%%%%%%%%%%%%%%%
\subsection{Numerical and Computational Approximations}

All of the analysis and results in this section are for three dimensions.

%%%%%%%%%%%%%%%%%%%%%%%%%%
\subsubsection{Dimer Term Neglecting Commutativity} \label{Sec:Dimer}

For real particles,
the two particle density is proportional to the radial distribution function,
\begin{equation}
\rho^{(2)}({\bf q}_1,{\bf q}_2)
= \rho^2 g(q_{12}).
\end{equation}
With this the dimer grand potential is
\begin{equation}
\frac{ -\beta \Omega_{2,0}^{\pm}}{V}
=
\frac{ \pm 4\pi \rho^2  }{ 2   }
\int_0^\infty \mathrm{d}q \, q^2 e^{- 2\pi q^2/{\Lambda^2} } g(q) .
\end{equation}

As an approximation
that illustrates the role of the particle's impenetrable core,
at low densities, take
\begin{equation}
g(q)
=
\left\{ \begin{array}{ll}
0 , & q < d ,\\
1 , & q > d .
\end{array} \right.
\end{equation}
For most atoms and molecules,
the core diameter is larger or much larger than the thermal wave length,
$ d \agt \Lambda$.
(For the Lennard-Jones fluid treated in detail below,
one can take $d \approx \sigma$.
At the critical temperature,
helium has $\Lambda = 0.92 \sigma$,
and argon has $\Lambda = 0.062 \sigma$.)
With this one has
\begin{eqnarray}
\lefteqn{
\frac{ -\beta \Omega_{2,0}^{\pm}}{V}
} \nonumber \\
& = &
\frac{ \pm 4\pi \rho^2  }{ 2   }
\int_d^\infty \mathrm{d}q \, q^2 e^{- 2\pi q^2/\Lambda^2}
\nonumber \\ & = &
\frac{ \pm 4\pi \rho^2  }{ 2   }
\left\{
\frac{ \Lambda^2 }{4\pi} d e^{- 2\pi d^2/\Lambda^2}
%\right. \nonumber \\ && \left. \mbox{ }
- \frac{ \Lambda^2 }{4\pi}
\int_d^\infty \mathrm{d}q \, e^{- 2\pi q^2/\Lambda^2}
\right\}
%\nonumber \\ & = &
%\frac{ \pm \Lambda^2 \rho^2  }{ 2}
%\left\{ d e^{- 2\pi d^2/\Lambda^2}
%\right. \nonumber \\ && \left. \mbox{ }
%- \frac{ \Lambda }{\sqrt{2\pi}}
%\int_{\sqrt{2\pi} d/\Lambda}^\infty \mathrm{d}x \, e^{- x^2} \right\}
\nonumber \\ & = &
\frac{ \pm \Lambda^2 \rho^2  }{ 2}
\left\{
d e^{- 2\pi d^2/\Lambda^2}
%\right. \nonumber \\ && \left. \mbox{ }
- \frac{ \Lambda }{\sqrt{8}}
\mbox{erfc}\left( \frac{\sqrt{2\pi}d}{\Lambda}\right)
\right\}.
\end{eqnarray}
The asymptotic expansion for the complementary error function is,
Eq.~(AS7.1.23),\cite{Abramowitz65}
\begin{equation}
\mbox{erfc } z
=
\frac{e^{-z^2}}{\sqrt{\pi}z}
\left\{ 1 + \sum_{m=1}^\infty (-1)^m \frac{(2m)!}{m!(4z^2)^m} ,
\right\}
\end{equation}
valid for $z \rightarrow \infty$.
Hence keeping just the  leading order term,
which is valid if $d \gg \Lambda$,
one has
\begin{equation}
\frac{ -\beta \Omega_{2,0}^{\pm}}{V}
=
\frac{ \pm \Lambda^2 d \rho^2  }{ 2}
e^{- 2\pi d^2/\Lambda^2}
%\nonumber \\ &&  \mbox{ } \times
\left\{ 1 - \frac{ \Lambda^2 }{4\pi d^2}
%\right. \nonumber \\ && \left. \mbox{ }
+ {\cal O}\left(\frac{\Lambda}{d}\right)^4 \right\} .
\end{equation}
This is smaller than the ideal gas result
by a factor of about $e^{- 2\pi d^2/\Lambda^2} \alt 2 \times 10^{-3}$
if $d \agt \Lambda$.
One  concludes that the particle exclusion volume
and any inter-particle repulsion
reduce the effects of quantum symmetrization.

One could imagine a hypothetical very small particle
with $d \ll \Lambda$. In this case $g(r) \approx 1$ everywhere
and one would obtain the ideal gas result
(at low densities, neglecting any long range interaction potential).

%%%%%%%%%%%%%%%%%%%%%%%%%%
\subsubsection{Trimer Term Neglecting Commutativity}

For the trimer, $l=3$, one has
\begin{equation}
\rho^{(3)}({\bf q}_1,{\bf q}_2,{\bf q}_3)
= \rho^3 g^{(3)}(q_{13},q_{23},q_{12}) .
\end{equation}
With this the trimer grand potential is
\begin{eqnarray}
\lefteqn{
\frac{ -\beta \Omega_{3,0}^{\pm}}{V}
} \nonumber \\
& = &
\frac{  \rho^3  }{ 3   }
\int \mathrm{d}{\bf q}_1 \, \mathrm{d}{\bf q}_2 \;
\,  e^{- \pi [ q_1^2 + q_2^2  + q_{12}^2 ]/{\Lambda^2} }
g^{(3)}(q_{1},q_{2},q_{12})
\nonumber \\ & = &
\frac{ 8\pi^2 \rho^3  }{ 3   }
\int_0^\infty \mathrm{d}{q}_1 \int_0^\infty  \mathrm{d}{q}_2
\int_{-1}^1 \; \mathrm{d}x \;
\,  q_1^2 q_2^2
\nonumber \\ && \mbox{ } \times
e^{- \pi [ q_1^2 + q_2^2  + q_{12}^2 ]/{\Lambda^2} }
g^{(3)}(q_{1},q_{2},q_{12}) ,
\end{eqnarray}
where $q_{12} = \sqrt{ q_1^2 + q_2^2 - 2 {q}_1 q_2 x}$.

%%%%%%%%%%%%%%%%%%%%%%%%%%%%%%%%%%%%%%%%%%%%%%
\subsubsection{Calculation of the Distribution Functions}

The Lennard-Jones potential used here is
\begin{equation}
u(r) =
4 \epsilon \left[
\left( \frac{\sigma}{r} \right)^{12}
- \left( \frac{\sigma}{r}\right)^{6} \right] .
\end{equation}

For the Lennard-Jones  fluid,
the critical temperature, density, and pressure
in  dimensionless form are\cite{TDSM}
\begin{equation}
\frac{k_\mathrm{B}T_\mathrm{c}}{\epsilon} = 1.35, \;
\rho_\mathrm{c}\sigma^3 = 0.35 , \mbox{ and }
\frac{p_\mathrm{c}\sigma^3}{k_\mathrm{B}T_\mathrm{c}} = 0.142,
\end{equation}
respectively.

The radial distribution function, $g(r)$
can be obtained
by classical equilibrium computer simulations\cite{Allen87}
or by standard integral equation techniques.\cite{TDSM}
Here the latter approach is taken,
based on the Ornstein-Zernike equation,
\begin{equation}
h(r_{12}) = c(r_{12})
+ \rho \int  \mathrm{d} {\bf r}_3 \;
c(r_{13})h(r_{32}) .
\end{equation}
Here $h(r)=g(r)-1$ is the total correlation function and $c(r)$
is the direct correlation function.
This exact equation
is solved in conjunction with the hypernetted chain (HNC) closure approximation,
\begin{equation}
h(r) = e^{ - \beta u(r) + h(r) - c(r) } - 1 .
\end{equation}
The method of solution is the standard one
of iteration combined with the fast Fourier transformation.\cite{TDSM}
Convergence typically takes a few seconds on a personal computer.
The HNC approximation is known to be accurate
for reasonably long-ranged potentials,
at not too high densities and not too low temperatures.\cite{TDSM}

The triplet distribution function $g^{(3)}$,
can also be obtained by classical equilibrium computer simulation
\cite{Allen87,TDSM}
or by integral equation techniques.\cite{Attard89,Attard91a}
These are more demanding computationally
than are the programs for the pair distribution function
and they will not be pursued here.
Instead the well known Kirkwood superposition approximation will be invoked,
\begin{equation}
g^{(3)}({\bf q}_{1},{\bf q}_{2},{\bf q}_{3})
=
g^{(2)}(q_{12}) \, g^{(2)}(q_{23})  \, g^{(2)}(q_{31}) .
\end{equation}
The quantitative reliability of this approximation
in the regime in which it is applied below should be treated with caution.

%%%%%%%%%%%%%%%%%%%%%%%%%%
\subsubsection{Potential Gradients for Neglecting Symmetrization}

When symmetrization is neglected,
various potential gradients are required for $W_1$ and $W_2$.
The first of these is the Laplacian, namely
\begin{eqnarray}
\nabla^2 U
& = &
\sum_{j=1}^N \sum_{\alpha=x,y,z}
\frac{\partial^2 U}{\partial q_{j,\alpha}^2 }
 \nonumber \\ & = &
\sum_{j,\alpha}  \sum_{k}\!^{(k \ne j)}
\frac{\partial^2 u(q_{jk})}{\partial q_{j,\alpha}^2 }
 \nonumber \\ & = &
\sum_{j,\alpha}  \sum_{k}\!^{(k \ne j)}
\frac{\partial }{\partial q_{j,\alpha} }
\left[ u'(q_{jk}) \frac{q_{jk,\alpha}}{q_{jk}}\right]
 \nonumber \\ & = &
\sum_{j\ne k,\alpha}
\left[ u''(q_{jk}) \frac{ q_{jk,\alpha}^2 }{ q_{jk}^2 }
+ u'(q_{jk}) \frac{q_{jk}^2-q_{jk,\alpha}^2}{q_{jk}^3}
\right]
 \nonumber \\ & = &
\sum_{j \ne k}
\left[ u''(q_{jk})  +  \frac{ 2 }{q_{jk}} u'(q_{jk}) \right]
 \nonumber \\ & = &
N(N-1)
\left[ u''(q_{12})  +  \frac{ 2 }{q_{12}} u'(q_{12}) \right] .
\end{eqnarray}
The classical equilibrium average of this is
\begin{eqnarray}
\lefteqn{
\left<  \nabla^2 U \right>_{1,0}
} \nonumber \\
& = &
V \int \mathrm{d}{\bf q}_1 \;
\left[ u''(q_{1})  +  \frac{ 2 }{q_{1}} u'(q_{1}) \right]
\rho^{(2)}(q_{1})
\nonumber \\ & = &
4\pi \rho^2 V
\int_0^\infty \mathrm{d}{q} \, q^2
g(q)
\left[ u''(q)  +  \frac{ 2 }{q} u'(q) \right] .
\end{eqnarray}

The  second quantity required is square of the gradient, which is
\begin{eqnarray}
\lefteqn{
\nabla U \cdot \nabla U
} \nonumber \\
& = &
\sum_{j=1}^N \sum_{\alpha=x,y,z}
 \left( \frac{\partial U}{\partial q_{j,\alpha}} \right)^2
 \nonumber \\ & = &
\sum_{j,\alpha}
 \left(
 \sum_{k}\!^{(k \ne j)}
 u'(q_{jk}) \frac{q_{jk,\alpha}}{q_{jk}}
\right)^2
 \nonumber \\ & = &
\sum_{j,\alpha}
 \sum_{k}\!^{(k \ne j)}  \sum_{i}\!^{(i \ne j)}
 u'(q_{jk}) \frac{q_{jk,\alpha}}{q_{jk}}
 u'(q_{ji}) \frac{q_{ji,\alpha}}{q_{ji}}
 \nonumber \\ & = &
\sum_{j\ne k}
 [ u'(q_{jk}) ]^2
 \nonumber \\ & & \mbox{ } %\times
+
\sum_{j}
 \sum_{k}\!^{(k \ne j)}  \sum_{i}\!^{(i \ne j, k)}
 u'(q_{jk})  u'(q_{ji})
 \frac{ {\bf q}_{jk} \cdot {\bf q}_{ji}  }{q_{jk}q_{ji}}
 \nonumber \\ & = &
N^2  [ u'(q_{12}) ]^2
%\nonumber \\ & & \mbox{ }
+ N^3  u'(q_{12})  u'(q_{13})
 \frac{ {\bf q}_{12} \cdot {\bf q}_{13}  }{q_{12}q_{13}} .
\end{eqnarray}
In the final equality,
all particles have been treated as equivalent,
and $N^2$ has been written in place of $N(N-1)$, etc.
The classical equilibrium average of this is
\begin{eqnarray}
\lefteqn{
%\frac{\hbar^2\beta^3}{24m}
\left<  \nabla U \cdot \nabla U \right>_{1,0}
} \nonumber \\
& = &
%\frac{\hbar^2\beta^3 }{24m}
V \int \mathrm{d}{\bf q}_1 \;
 [ u'(q_{12}) ]^2 \rho^{(2)}(q_{12})
\nonumber \\ & & \mbox{ }
+
%\frac{\hbar^2\beta^3 }{24m}
V  \int \mathrm{d}{\bf q}_1 \, \mathrm{d}{\bf q}_2 \;
u'(q_{1})  u'(q_{2})
\frac{\bf{ q}_{1} \cdot \bf{ q}_{2}  }{q_{1}q_{2}}
\rho^{(3)}(q_{1},q_{2},q_{12})
\nonumber \\ & = &
4\pi \rho^2 V
\int_0^\infty \mathrm{d}{q} \, q^2
g(q) u'(q)^2
\nonumber \\ & & \mbox{ }
+ 8\pi^2 \rho^3 V
\int_0^\infty \mathrm{d}q_1
\int_0^\infty \mathrm{d}q_2
\int_{-1}^1 \mathrm{d} x \; q_1^2 q_2^2
\nonumber \\ & & \mbox{ } \times
u'(q_{1})  u'(q_{2}) x
g^{(3)}(q_{1},q_{2},q_{12}) ,
\end{eqnarray}
where, as above, $q_{12} = \sqrt{ q_1^2 + q_2^2 - 2 {q}_1 q_2 x}$.
As above, the Kirkwood superposition approximation
may be used to approximate the triplet distribution function.

Since $\Omega_{1,1}=0$,
the first non-zero quantum correction to the grand potential
for non-commutativity and neglecting symmetrization is
\begin{eqnarray} \label{Eq:<W1+W2>-10}
\lefteqn{
\frac{-\beta}{V} \Omega_{1,2}
} \nonumber \\
& = &
\frac{1}{V}
\left< \hbar^2 W_2 \right>_{1,0} + {\cal O}(\hbar^4)
%\nonumber \\ & = &
%\frac{1}{V} \left< \frac{ - \hbar^2 \beta^2}{12m}\nabla^2 {U}
%+ \frac{\hbar^2 \beta^3}{24m} \nabla U \cdot \nabla U \right>_{1,0}
\nonumber \\ & = &
 \frac{  -\hbar^2 \beta^2}{12mV} \left<\nabla^2 {U} \right>_{1,0}
+ \frac{\hbar^2 \beta^3}{24mV} \left< \nabla U \cdot \nabla U \right>_{1,0}
\nonumber \\ & = &
 \frac{  -\hbar^2 \beta^2}{24mV} \left<\nabla^2 {U} \right>_{1,0} .
\end{eqnarray}
The last equality avoids the triplet distribution function.
Numerical comparison of the final two equalities
tests the accuracy of the Kirkwood superposition approximation
(in combination with the hypernetted chain approximation).

%%%%%%%%%%%%%%%%%%%%%%%%%%%%%%%%%%%%%%%%%%%%%%
\subsubsection{Numerical Results}

%%%%%%%%%%%%%%%%%%%%%%%%%%%%%%%%%%%%%%%%%%%%%%%%%%%%%%%%%%%%%%%%%%
\begin{figure}[t!]
\centerline{
\resizebox{8.5cm}{!}{ \includegraphics*{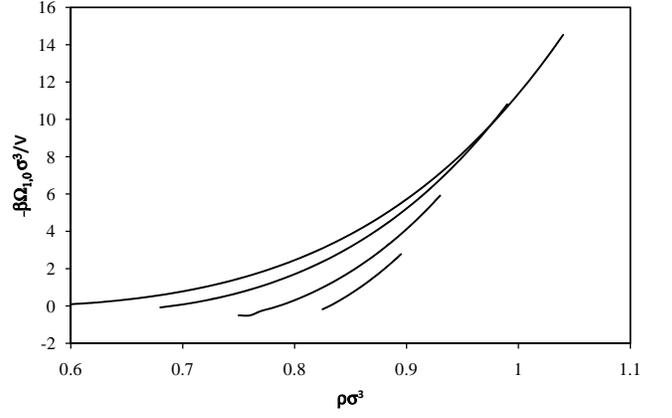} } }
% From My Documents\Projects\QSM-LJ16\LJ\LJxlsx:chart1
\caption{\label{Fig:Ptot}
Classical grand potential or pressure,
$-\beta \Omega_{1,0} \sigma^3/V = \beta p_\mathrm{cl} \sigma^3$,
as a function of density for a Lennard-Jones model
on liquid branch isotherms.
The temperatures are $k_\mathrm{B}T/\epsilon=$ 0.5, 0.6, 0.8, and 1.0
from bottom to top,  respectively.
} \end{figure}
%%%%%%%%%%%%%%%%%%%%%%%%%%%%%%%%%%%%%%%%%%%%%%%%%%%%%%%%%%%%%%%%%%

Figure~\ref{Fig:Ptot} shows the total classical grand potential
for a Lennard-Jones model calculated in hypernetted chain approximation.
The quantity plotted is equal to the pressure, $\beta p \sigma^3$.
(Actually in this figure the virial pressure has been calculated
rather than the grand potential in hypernetted chain approximation.)\cite{TDSM}
The curves represent the liquid branch of sub-critical isotherms.
They terminate at the last point of convergence of the hypernetted chain
algorithm, which correspond approximately to the gas and solid spinodal points.
Negative values of the pressure correspond to metastable states.
It can be seen that the pressure increases with increasing density,
and that it increases with increasing temperature.
The latter is a result of the reduction in weight of
the attractive tail of the Lennard-Jones potential
as the temperature is increased.
As a comparison, the ideal gas classical pressure
is equal to the density in these units
and would be a straight line.

%%%%%%%%%%%%%%%%%%%%%%%%%%%%%%%%%%%%%%%%%%%%%%%%%%%%%%%%%%%%%%%%%%
\begin{figure}[t!]
\centerline{
\resizebox{8.5cm}{!}{ \includegraphics*{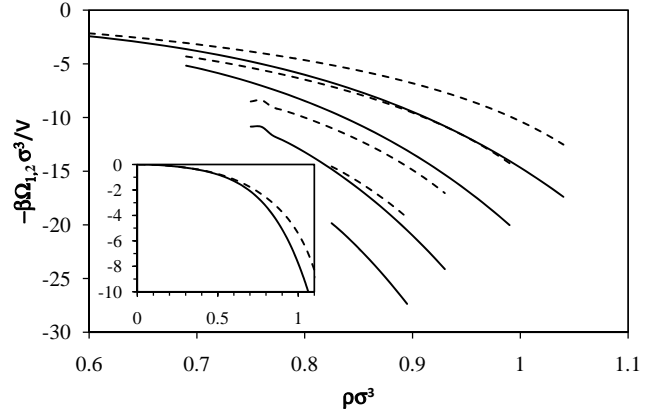} } }
% From My Documents\Projects\QSM-LJ16\LJ\LJxlsx:chart2
% corrected 10 Oct 2016
\caption{\label{Fig:W12}
Non-commuting quantum correction to order $\hbar^2$,
$-\beta \Omega_{1,2} \sigma^3/V$,
as a function of density for a Lennard-Jones model of helium
on liquid branch isotherms.
Both sets of curves represent Eq.~(\ref{Eq:<W1+W2>-10}):
the solid curves use only the pair distribution function
(the final equality),
and the dashed curves use the Kirkwood superposition approximation
for the triplet distribution function
(the penultimate equality).
The temperatures are $k_\mathrm{B}T/\epsilon=$ 0.5, 0.6, 0.8, and 1.0
from bottom to top,  respectively.
{\bf Inset.} Supercritical isotherm, $k_\mathrm{B}T/\epsilon= 1.5$
} \end{figure}
%%%%%%%%%%%%%%%%%%%%%%%%%%%%%%%%%%%%%%%%%%%%%%%%%%%%%%%%%%%%%%%%%%

Figure~\ref{Fig:W12} shows the leading quantum correction
due to the non-commutation of position and momentum.
The quantity plotted, $-\beta \Omega_{1,2} \sigma^3/V$,
which can be compared directly to the pressure,
gives the quantum correction to ${\cal O}(\hbar^2)$,
which in fact is proportional to the thermal wave length squared,
The calculations are for a Lennard-Jones model of helium.
with mass $m=4.003\,$amu,
potential well parameter $\epsilon/k_\mathrm{B}=10.22\,$K,
and core parameter $\sigma=0.2556\,$nm.\cite{Sciver12}
Again the hypernetted chain result for the radial distribution
function is used.
Classical simulations have previously been used to obtain
the first quantum correction
due to non-commutativity in argon and neon.\cite{Barker73,Singer84}

The magnitude of the quantum correction
shown in Fig.~\ref{Fig:W12}
increases with decreasing temperature.
This is the opposite trend to that for the total pressure,
Fig.~\ref{Fig:Ptot},
and so one sees that quantum effects become relatively more
important at low temperatures, as one would expect.
The fact that this quantum correction is negative means that
the quantum effects due to non-commutativity
of the position and momentum operators
decreases the pressure compared to an otherwise equivalent classical system.
For the lowest temperature and highest density here,
the magnitude of the first quantum correction due to non-commutativity
is larger than the total classical pressure itself.
This suggests that it is necessary to include more terms
in the series if the quantum correction due to non-commutativity
is to be reliably obtained for low temperature liquid helium.

Figure~\ref{Fig:W12} also tests the Kirkwood superposition approximation
for $g^{(3)}$  (cf.\ Eq.~(\ref{Eq:<W1+W2>-10})).
One can see that the superposition approximation
is reasonably accurate for this quantum correction
at the highest temperature and lowest densities shown.
As can be seen in the inset,
at the supercritical temperature of $k_\mathrm{B}T/\epsilon=1.5$,
the two calculations are within 10\% for $\rho\sigma^3 \alt 0.5$.
However  the superposition approximation significantly underestimates
the magnitude of the correction
at higher densities and at lower temperatures.

%%%%%%%%%%%%%%%%%%%%%%%%%%%%%%%%%%%%%%%%%%%%%%%%%%%%%%%%%%%%%%%%%%
\begin{figure}[t!]
\centerline{
\resizebox{8.5cm}{!}{ \includegraphics*{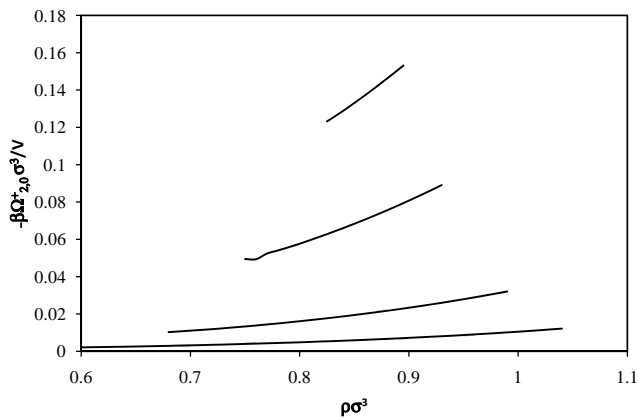} } }
% From My Documents\Projects\QSM-LJ16\LJ\LJxlsx:chart3
\caption{\label{Fig:W2}
Dimer grand potential for bosons, $-\beta \Omega^+_{2,0} \sigma^3/V$,
with all curves and parameters as in the preceding figure,
except that the temperature increases from top to bottom.
} \end{figure}
%%%%%%%%%%%%%%%%%%%%%%%%%%%%%%%%%%%%%%%%%%%%%%%%%%%%%%%%%%%%%%%%%%

Figure~\ref{Fig:W2} shows the first quantum correction
due to symmetrization, $-\beta \Omega^+_{2,0} \sigma^3/V$,
again for helium.
The data plotted is for bosons;
the results for fermions are equal in magnitude and opposite in sign.
The positive values in Fig.~\ref{Fig:W2}
indicate that particle symmetrization effects
increase the pressure for bosons
and decrease the pressure for fermions
compared to a classical liquid at the same density.
It can be seen that again the magnitude of the correction
increases with decreasing temperature.
However in this regime for helium,
the correction due to symmetrization
is about two orders of magnitude smaller
than the correction due to non-commutativity.
(At the highest density and lowest temperature shown,
$\Omega^+_{2,0} \approx \Omega^+_{1,2}/200$.)
Nevertheless the effect of quantum symmetrization
is on the order of 10--20\% of the total classical pressure,
with it being relatively larger as the density is decreased
or as the temperature is lowered.

This first quantum correction due to symmetrization
is for a Lennard-Jones liquid about 3--4 times smaller
than the same correction for an ideal gas.
This reduction
is due to the short-range repulsion between the Lennard-Jones particles,
as discussed in \S \ref{Sec:Dimer}.

For helium, at  $k_\mathrm{B}T/\epsilon=$ 0.5,
the thermal wavelength is $\Lambda=1.51 \sigma$,
and at  $k_\mathrm{B}T/\epsilon=$ 1.0,
it is $\Lambda=1.07 \sigma$.
In these cases there is enough overlap between
the Gaussian symmetrization factor
and the non-zero part of the radial distribution function
for the symmetrization quantum correction to be non-negligible.

For the case of argon,
$m=39.948\,$amu,
$\epsilon/k_\mathrm{B}=124.0\,$K,
and $\sigma=0.3418\,$nm,\cite{Sciver12}
the first quantum correction was found to be effectively zero.
In this case, at  $k_\mathrm{B}T/\epsilon=$ 0.5,
the thermal wavelength is $\Lambda=0.103 \sigma$.
The radial distribution function $g(q)$
is approximately zero for $ q \alt \sigma$,
and the Gaussian factor that arises from symmetrization
is effectively zero for $q \agt \Lambda/\sqrt{2\pi}$.
In the case of argon there is almost no overlap between the two.
As the density is increased,
the non-zero part of the radial distribution function
shifts to smaller separations.
But the classical pressure also rapidly increases,
and it is difficult to see the quantum correction
becoming relatively significant for argon.
Higher densities have not been studied
because the present hypernetted chain algorithm is unsuited
for the solid phase.

The first quantum correction due to non-commutativity
is proportional to the thermal wave length squared,
and it about $200$ times smaller for argon than for helium
at the lowest temperature shown.

%%%%%%%%%%%%%%%%%%%%%%%%%%%%%%%%%%%%%%%%%%%%%%%%%%%%%%%%%%%%%%%%%%
\begin{figure}[t!]
\centerline{
\resizebox{8.5cm}{!}{ \includegraphics*{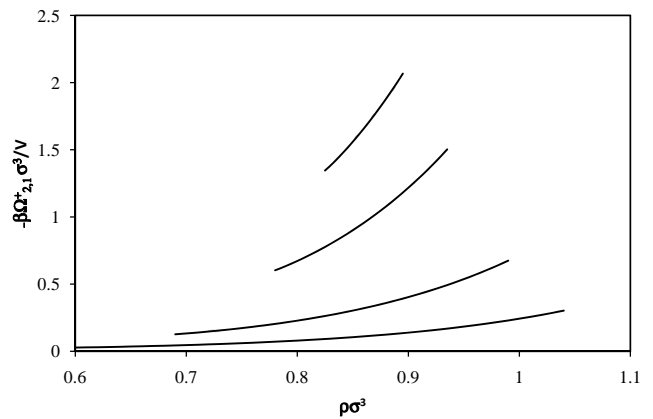} } }
% From My Documents\Projects\QSM-LJ16\LJ\LJxlsx:chart4
\caption{\label{Fig:W21}
Dimer grand potential for bosons
with first correction for non-commutativity,
$-\beta \Omega^+_{2,1} \sigma^3/V$,
with all curves and parameters as in the preceding figure.
} \end{figure}
%%%%%%%%%%%%%%%%%%%%%%%%%%%%%%%%%%%%%%%%%%%%%%%%%%%%%%%%%%%%%%%%%%

Figure~\ref{Fig:W21} shows the leading order quantum correction
due to the combination of non-commutativity and symmetrization,
Eq.~(\ref{Eq:Omega-21}).
This is positive for bosons and negative for fermions.
Interestingly enough,
it is larger in magnitude than is the first correction for symmetrization
alone (cf.\ Fig.~\ref{Fig:W2}).
It is however smaller in magnitude than the first non-zero correction
for non-commutativity alone (cf.\ Fig.~\ref{Fig:W12}).
At low densities toward the liquid spinodal point,
where the total classical pressure is close to zero or negative,
this correction can exceed the total classical pressure.
At the lowest temperature shown
it is comparable to the total classical pressure
along the whole liquid branch isotherm.

%%%%%%%%%%%%%%%%%%%%%%%%%%%%%%%%%%%%%%%%%%%%%%%%%%%%%%%%%%%%%%%%%%
\begin{figure}[t!]
\centerline{
\resizebox{8.5cm}{!}{ \includegraphics*{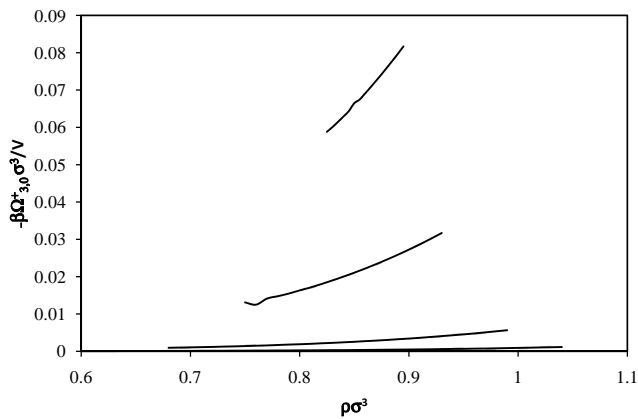} } }
% From My Documents\Projects\QSM-LJ16\LJ\LJxlsx:chart5
\caption{\label{Fig:W3}
Trimer grand potential for bosons, $-\beta \Omega^+_{3,0} \sigma^3/V$,
with all curves and parameters as in the preceding figure.
} \end{figure}
%%%%%%%%%%%%%%%%%%%%%%%%%%%%%%%%%%%%%%%%%%%%%%%%%%%%%%%%%%%%%%%%%%

Figure~\ref{Fig:W3} shows the trimer grand potential,
to be again compared directly to the pressure
$\beta p \sigma^3$.
In this case the quantum correction is identical for bosons and for fermions.
The fact that it is positive means that second order quantum effects
increase the classical pressure.
It should be cautioned that the quantitative results here
depend on the accuracy of the Kirkwood superposition approximation
(cf.\ the discussion of Fig.~\ref{Fig:W12}).

It can be seen that this second quantum correction
due to symmetrization is about a factor of 2
smaller than the first quantum correction shown in Fig.~\ref{Fig:W2}
at the lowest temperature.
At higher temperatures the second quantum correction is
smaller than this in relative terms.
These results allow the tentative conclusion
that the loop expansion is rapidly converging,
at least at these temperatures.
Because the corrections alternate in sign for fermions,
one might expect the series for them to be more slowly
converging than for bosons.

%%%%%%%%%%%%%%%%%%%%%%%%%%%%%%%%%%%%%%%%%%%%%%%%%%%%%%%%%%%%%%%%%%%%%%%%%%
\section{Conclusion}

The quantum states that represent phase space identified here
arise by taking an asymmetric expectation value of
the Maxwell-Boltzmann operator.
This involves two complete basis sets:
plane waves, which localize particle momenta,
and Gaussians, which localize particle positions.
The consequent manipulations of the von Neumann trace
for the partition function are formally exact.

The Maxwell-Boltzmann operator acting on  a plane wave
is here written as the exponential of an effective Hamiltonian,
similar to an expression given by Wigner.\cite{Wigner32}
A subsequent expansion
gives the classical Maxwell-Boltzmann probability function
times a series of terms involving powers of gradients of the potential energy,
essentially the same as that given by Kirkwood.\cite{Kirkwood33}
The classical average of these gives the quantum corrections
due to the non-commutativity of the position and momentum operators.
The present derivation is relatively simple
and it appears straightforward
to obtain the higher order terms in the expansion explicitly.

The  particle interchange symmetrization of the wave function
gives rise to an overlap factor in the grand partition function.
This is expanded and re-summed,
and the resultant permutation loop overlap factors
are simple Gaussians in classical phase space.
Again the terms in the series can be written as classical averages.

One piece of evidence for the validity of the formalism
is the analysis of the ideal gas.
In this system the quantum effects are due entirely
to wave function symmetrization.
The fact that the fugacity series for the pressure found here
agrees with that given in standard texts\cite{Pathria72}
gives one confidence that the approach is fundamentally sound
and free of algebraic errors.

A second piece of evidence is the comparison with
the results of  Wigner\cite{Wigner32} and  Kirkwood\cite{Kirkwood33}
for the leading quantum correction due to non-commutativity
(ie.\ neglecting wave function symmetrization).
Despite the fact that the formulations of the actual quantum states
for classical phase space are quite different,
it turns out that
the average of a function of position,
and the average of the kinetic energy operator,
are the same to leading order in the two approaches.
However whereas Wigner\cite{Wigner32} and Kirkwood\cite{Kirkwood33}
give a phase space probability distribution,
albeit a different one in each case,
this appears to be precluded in the present theory.

The dual expansion obtained here appears computationally tractable.
%and, at least for symmetrization effects, rapidly converging.
All the terms that appear can be expressed as classical equilibrium averages,
and there is no need to compute eigenvalues, eigenfunctions,
or explicitly symmetrized wave functions.
Here  liquid helium was modeled with a Lennard-Jones potential
and the leading order corrections
 were obtained with the hypernetted chain approximation.
No doubt in the future classical Monte Carlo or molecular dynamics simulations,
in which the computational burden is ${\cal O}(N)$,
could be used to obtain higher order corrections
or to treat more challenging cases such as solid state
or electronic systems.
In the present calculations for liquid isotherms,
quantum effects due to non-commutativity
were found to be much larger than the effects
due to wave function symmetrization.
This is unlikely to be a general rule
and it would be interesting to apply the present expansion
to systems where symmetrization effects dominate.

%\newpage
%\section*{References}
%\newpage $\;$ \newpage
%%%%%%%%%%%%%%%%%%%%%%%%%%%%%%%%%%%%%%%%%%%%%%%%%%%%%%%%%%%%%%%%%%%%%%%%%%

%%%%%%%%%%%%%%%%%%%%%%%%%%%%%%%%%%%%%%%%%%%%%%%%%%%%%%%%%%%%%%%%%%%%%%%%%%%

\appendix
\setcounter{equation}{0}
\renewcommand{\theequation}{\Alph{section}.\arabic{equation}}

%\newpage $\;$ \newpage
%%%%%%%%%%%%%%%%%%%%%%%%%%%%%%%%%%%%%%%%%%%%%%%%%%%%%%%%%%%%%%%%%%%%%%%%%
%
\section{Exponential Expansion for the Quantum Weight Factor}
\label{Sec:W=exp(w)}
%
%%%%%%%%%%%%%%%%%%%%%%%%%%%%%%%%%%%%%%%%%%%%%%%%%%%%%%%%%%%%%%%%%%%%%%%%%

In the text, \S \ref{Sec:MBtilde},
the quantum weight $W({\bf \Gamma})$,
which arises from the non-commutativity of the momentum and position operators,
was introduced.
This was expanded in powers of $\hbar$,
with the coefficients determined from a recursion relation
that arises from the temperature derivative of the defining equation.
The structure of the defining equation, Eq.~(\ref{Eq:W-defn}),
suggests that it may be useful to write
\begin{equation}
W({\bf \Gamma}) \equiv e^{w({\bf \Gamma})} .
\end{equation}
The reasons why it is better to deal with $w$ than $W$
will be discussed at the end of this appendix.

With this the temperature derivative of the defining equation~(\ref{Eq:dW/dB})
becomes
\begin{eqnarray}
%\lefteqn{
\frac{\partial w}{\partial \beta }
%} \nonumber \\
& = &
\frac{i\hbar}{m} e^{\beta U-w}  {\bf p} \cdot
\nabla  \left\{e^{w-\beta U }  \right\}
%\nonumber \\  && \mbox{ }
 + \frac{\hbar^2}{2m} e^{\beta U-w}  \nabla^2
\left\{e^{w-\beta U } \right\}
\nonumber \\ & = &
\frac{i\hbar}{m} {\bf p} \cdot \nabla (w-\beta U)
 + \frac{\hbar^2}{2m}
% \nonumber \\ && \mbox{ } \times
\left\{ \rule{0cm}{0.4cm}
\nabla (w-\beta U) \cdot \nabla (w-\beta U)
\right. \nonumber \\ && \left. \mbox{ }
+\nabla^2 (w-\beta U)
\rule{0cm}{0.4cm}\right\} .
\end{eqnarray}

One can expand $w$ in powers of Planck's constant,
\begin{equation}
w \equiv \sum_{n=1}^\infty w_n \hbar^n .
\end{equation}
This begins at $n=1$ because the classical part must vanish,
$W(\hbar=0)=1 \Rightarrow w(\hbar=0)=0$.
Also $W(\beta=0)=1 \Rightarrow w_n(\beta=0)=0$.

The first coefficient satisfies
\begin{equation}
\frac{\partial w_1}{\partial \beta }
=
\frac{-i\beta}{m} {\bf p} \cdot \nabla  U ,
\end{equation}
which gives
\begin{equation}
 w_1
=
\frac{-i\beta^2}{2m} {\bf p} \cdot \nabla  U .
\end{equation}
The second coefficient satisfies
\begin{eqnarray}
%\lefteqn{
\frac{\partial w_2}{\partial \beta }
%} \nonumber \\
& = &
\frac{i}{m} {\bf p} \cdot \nabla w_1
 + \frac{1}{2m}
% \nonumber \\ && \mbox{ } \times
\left\{ \rule{0cm}{0.4cm}
\beta^2 \nabla  U \cdot \nabla  U
%\right. \nonumber \\ && \left. \mbox{ }
-\beta \nabla^2 U
\rule{0cm}{0.4cm}\right\},
\nonumber \\ &&
\end{eqnarray}
which gives
\begin{eqnarray}
%\lefteqn{
w_2
%} \nonumber \\
& = &
\frac{\beta^3}{6m^2}
{\bf p} {\bf p} : \nabla \nabla  U
 + \frac{1}{2m}
% \nonumber \\ && \mbox{ } \times
\left\{ \rule{0cm}{0.4cm}
\frac{ \beta^3}{3}  \nabla  U \cdot \nabla  U
%\right. \nonumber \\ && \left. \mbox{ }
-\frac{ \beta^2}{2}  \nabla^2 U
\rule{0cm}{0.4cm}\right\}.
\nonumber \\ &&
\end{eqnarray}
In general for $n > 2$
\begin{eqnarray}
%\lefteqn{
\frac{\partial w_n}{\partial \beta }
%} \nonumber \\
& = &
\frac{i}{m} {\bf p} \cdot \nabla w_{n-1}
 + \frac{1}{2m}
 \sum_{j=0}^{n-2}  \nabla w_{n-2-j}  \cdot \nabla w_j
\nonumber \\ && \mbox{ }
- \frac{\beta}{m} \nabla w_{n-2}  \cdot \nabla U
%\nonumber \\ && \mbox{ }
 + \frac{1}{2m} \nabla^2 w_{n-2} .
\end{eqnarray}

For $n=3$ one can show that
\begin{eqnarray}
%\lefteqn{
w_3
%} \nonumber \\
& = &
\frac{i\beta^4}{24m^3}
{\bf p} {\bf p} {\bf p} \vdots \nabla\nabla \nabla  U
+ \frac{5i\beta^4}{24m^2}  {\bf p}  (\nabla  U) : \nabla \nabla  U
\nonumber \\ && \mbox{ }
-\frac{i\beta^3}{6m^2} {\bf p} \cdot \nabla \nabla^2 U ,
\end{eqnarray}
and for $n=4$ one has
\begin{eqnarray}
%\lefteqn{
w_4
%} \nonumber \\
& = &
\frac{- i^4 \beta^{5}}{5!m^4} ( {\bf p} \cdot \nabla )^4 U
%\nonumber \\ && \mbox{ }
- \frac{\beta^5}{30m^3}
(\nabla U ){\bf p} {\bf p} \vdots \nabla \nabla \nabla U
\nonumber \\ && \mbox{ }
-\frac{\beta^5}{15m^2}
 (\nabla U)  (\nabla U) :  \nabla\nabla  U
%\nonumber \\ && \mbox{ }
+ \frac{\beta^4}{16m^2}  \nabla U \cdot \nabla \nabla^2 U
\nonumber \\ && \mbox{ }
 + \frac{\beta^4}{48m^3}
{\bf p} {\bf p} : \nabla \nabla \nabla^2 U
%\nonumber \\ && \mbox{ }
+ \frac{\beta^4}{48m^2}
\nabla^2 (\nabla  U \cdot \nabla  U)
 \nonumber \\ && \mbox{ }
- \frac{ \beta^3}{24m^2}  \nabla^2 \nabla^2 U
%\nonumber \\ && \mbox{ }
- \frac{\beta^5}{40m^3}
( {\bf p} \cdot \nabla \nabla U ) \cdot ( {\bf p} \cdot \nabla \nabla U ) .
\nonumber \\
\end{eqnarray}

Dealing with  $w$ has arguably two advantages over $W$.
First, experience shows that the expansion of an exponent converges more quickly
than the expansion of an exponential.
For example keeping only the terms $w_1$ and $w_2$
encompasses the terms $W_1$ and $W_2$,
and an infinite number of higher order terms besides.
Second, $w$ is an extensive variable,
as can be seen by inspection,
whereas $W$ is a sum of terms of every power of the system size.

%%%%%%%%%%%%%%%%%%%%%%%%%%%%%%%%%%%%%%%%%%%%%%%%%%%%%%%%%%%%%%%%%%%%%%%%%
%
\section{Non-Extensive Quantum Weight}
\label{Sec:W=exp(w)2}
\setcounter{equation}{0} \setcounter{subsubsection}{0}
%
%%%%%%%%%%%%%%%%%%%%%%%%%%%%%%%%%%%%%%%%%%%%%%%%%%%%%%%%%%%%%%%%%%%%%%%%%

The point made at the end of the preceding appendix
---that the quantum weight for non-commutativity is not extensive---
has some significant consequences for the analysis in the text.
Although the exponent $w$
may be expanded in powers of $\hbar$,
the fact that it is extensive means that one cannot
realistically linearize the exponential to obtain a useful
expansion of $W$ in powers of $\hbar$ without further work.
%The expansion given in this appendix in terms of averages of $w$
%and its fluctuations
%is arguably more practical
%than that given by Kirkwood.\cite{Kirkwood33}

That $w$ is extensive can be seen from the
original defining equation,
\begin{equation}
e^{-\beta \hat{\tilde{\cal H}}({\bf \Gamma}) } 1
=
e^{-\beta {\cal H}({\bf \Gamma})+w({\bf \Gamma})}.
\end{equation}
Since the operator exponent on the left hand side is extensive,
then so too must be the function exponent on the right hand side.
Alternatively, this can also be seen from the temperature derivative,
\begin{eqnarray}
%\lefteqn{
\frac{\partial w}{\partial \beta }
%} \nonumber \\
& = &
\frac{i\hbar}{m} {\bf p} \cdot \nabla (w-\beta U)
 + \frac{\hbar^2}{2m}
% \nonumber \\ && \mbox{ } \times
\left\{ \rule{0cm}{0.4cm}
\nabla (w-\beta U) \cdot \nabla (w-\beta U)
\right. \nonumber \\ && \left. \mbox{ }
+\nabla^2 (w-\beta U)
\rule{0cm}{0.4cm}\right\} .
\end{eqnarray}
On the right hand side each element of the gradient operator
acts on a single particle. Hence the scalar product with
the gradient operator is extensive.

%%%%%%%%%%%%%%%%%%%%%%%%%%%%%%%
\subsection{Second Order Analysis}

In the first instance one can focus on the monomer grand potential.
(A completely analogous argument can be applied
to the loop grand potentials for $l \ge 2$.)
The monomer grand potential is
$\Omega_1 = -k_\mathrm{B}T \ln \Xi_{1,w} $,
with the monomer grand partition function being
\begin{equation}
\Xi_{1,w}
=
\sum_{N}\frac{z^N }{N! h^{3N}  }
%\nonumber \\ && \mbox{ } \times
\int \mathrm{d}{\bf \Gamma} \;
 e^{ -\beta {\cal H}({\bf \Gamma})} e^{w({\bf \Gamma})} .
\end{equation}

The difference between the monomer grand potential
and the classical grand potential is essentially
just the logarithm of the classical average
of the quantum  weight due to non-commutativity
\begin{equation}
-\beta [\Omega_1-\Omega_{1,0} ]
=
\ln \left< e^{w} \right>_{1,0} .
\end{equation}
The left hand side is extensive,
which is to say that it scales with the volume of the system.
Hence the grand potential density is independent
of the size of the system.
Since $w \propto V$, and since $w(\hbar=0) = 0$,
one can suppose that there exists
a volume small enough that $w \ll 1$.
%(but still large enough that boundary effects remain negligible).
For such a volume one can expand the exponential to obtain
\begin{eqnarray}
\lefteqn{
-\beta [\Omega_1-\Omega_{1,0} ]
} \nonumber \\
& = &
\ln \left< e^{w} \right>_{1,0}
\nonumber \\ & = &
\ln\left< 1 + w + \frac{1}{2} w^2 +\ldots  \right>_{1,0}
\nonumber \\ & = &
\left<  w  \right>_{1,0}
+ \frac{1}{2} \left<  [w-\left<  w  \right>_{1,0}]^2 \right>_{1,0}
+ \ldots
\end{eqnarray}
The first term on the right hand side is extensive.
The second term,
as the fluctuation of an extensive variable, is also extensive.
Presumably, so are the higher order terms  (see \S \ref{Sec:w-Gauss}).
Having obtained the final result,
one can now allow the volume to become as large as desired.

%%%%%%%%%%%%%%%%%%%%%%%%%%%%%%%%%%%%%%%%%%%%%%%%%%%%
\subsubsection{Explicit Second Order Result}

The first order correction vanishes,
$\Omega_{1,1} = \hbar\left<  w_1 \right>_{1,0} =0$,
because $w_1$ is odd in ${\bf p}$.
The second order correction is given by
\begin{subequations}
\begin{eqnarray}
\lefteqn{
-\beta \Omega_{1,2}
} \nonumber \\
& = &
\hbar^2
\left< \frac{1}{2} w_1^2 + w_2 \right>_{1,0}
\nonumber \\ & = &
\hbar^2 \left<
\frac{-1}{2}
\frac{\beta^4}{4m^2} ({\bf p} \cdot \nabla U)^2
+ \frac{\beta^3}{6m^2}
{\bf p} {\bf p} : \nabla \nabla  U
\right. \nonumber \\ && \mbox{ } \left.
 +
\frac{ \beta^3}{6m}  \nabla  U \cdot \nabla  U
%\right. \nonumber \\ && \left. \mbox{ }
-\frac{ \beta^2}{4m}  \nabla^2 U
 \right>_{1,0}
\nonumber \\ & = &
\hbar^2 \left<
\frac{-1}{2}
\frac{\beta^4}{4m^2} m k_\mathrm{B}T \nabla U \cdot \nabla U
+ \frac{\beta^3}{6m^2}
m k_\mathrm{B}T  \nabla^2  U
\right. \nonumber \\ && \mbox{ } \left.
 +
\frac{ \beta^3}{6m}  \nabla  U \cdot \nabla  U
%\right. \nonumber \\ && \left. \mbox{ }
-\frac{ \beta^2}{4m}  \nabla^2 U
 \right>_{1,0}
\nonumber \\ & = &
\hbar^2 \left<
\frac{ \beta^3}{24m}  \nabla  U \cdot \nabla  U
%\right. \nonumber \\ && \left. \mbox{ }
-\frac{ \beta^2}{12m}  \nabla^2 U
 \right>_{1,0}
 \\ & = &
 \left<
\frac{- \hbar^2\beta^2}{24m}  \nabla^2 U
 \right>_{1,0} .
\end{eqnarray}
\end{subequations}
This is the same as the result given in the text,
which in turn agrees with that given by Wigner\cite{Wigner32}
 and by Kirkwood.\cite{Kirkwood33}
The present derivation is more satisfactory than
the others because it explicitly takes into account extensivity.
Also, the higher order terms lend themselves to practical implementation
either as an expansion,
or else as the original exponential combined with umbrella sampling.

%%%%%%%%%%%%%%%%%%%%%%%%%%%%%%%%%%%%%%%%%%%%%%%%%%%%%%%%%%%%%%%%%%
\begin{table}[t!]
\caption{ \label{Tab:1}
Quantum correction for monomers due to non-commutativity,
$ {-\beta \Omega_{1,2}\sigma^3}/{V} $,
for Lennard-Jones models of neon and helium
at $k_\mathrm{B}T/\epsilon=1.5$ and $\rho \sigma^3 = 0.5$.
For comparison, the classical pressure is $0.27575 \pm 0.00080$.
The results are from Metropolis Monte Carlo simulations,
with $50 \times 100$ configurations for averages,
20 trial steps per atom between averages, acceptance rate $\approx 50\%$,
$R_\mathrm{cut} =3.5\sigma$, tail correction added,
and no umbrella sampling,
for a canonical system with $N=$250, 500, or 1000.
The statistical error (68\% confidence level)
in the final digit is shown in parenthesis.
}
\begin{center}
\begin{tabular}{c c c c }
\hline\noalign{\smallskip}
 & 250  &    500  &   1000   \\
\hline \\
&   \multicolumn{3}{c}{Neon}    \\
%  \cline{2-3} \cline{5-6}

a  & -0.03484(8) & -0.03500(5) & -0.03504(3)  \\
b  & -0.03475(5) & -0.03500(3) & -0.03498(2)  \\
%c  & -0.026(2) & -0.0287(4)  & -0.026(2)   \\
d  & -0.0256(6)  & -0.0285(3)  & -0.0300(2)  \\
\\
&   \multicolumn{3}{c}{Helium}    \\
%\hline \\
a    & -0.730(2) & -0.734(1)  & -0.7346(7) \\
b    & -0.728(1) & -0.7338(7) & -0.7333(5)  \\
%c    & 33(2)   & 37(1)      & 46(2) \\
d    & -.38(1) & -0.518(7)  & - \\
\hline
\end{tabular} \\
\end{center}
\end{table}
%%%%%%%%%%%%%%%%%%%%%%%%%%%%%%%%%%%%%%%%%%%%%%%%%%%%%%%%%%%%%%%%%%

The justification for the final equality is an integration by parts,
assuming that surface integrals are negligible.
As can be seen from the Table~\ref{Tab:1},
computer simulations confirm the validity of this assumption.
The results labeled d correspond to
averaging the full exponential without umbrella sampling,
$ {-\beta \Omega_{1,2}^d}
=\ln \left< \exp[ \hbar w_1 + \hbar^2 w_2]  \right>_{1,0}$.
It can be seen that this approach has a significant system size dependence,
and that in the case of helium numerical overflow occurs
for the largest size studied.
For helium at this state point,
the hypernetted chain approximation used in the text gives
$\beta p\sigma^3=0.391$,
${-\beta \Omega_{1,2}^a\sigma^3}/{V} = -0.741$,
and
${-\beta \Omega_{1,2}^b\sigma^3}/{V} = -0.790$.

%%%%%%%%%%%%%%%%%%%%%%%%%%%%%%
\subsubsection{Tail Correction}

The Laplacian of the tail of the Lennard-Jones potential is
\begin{eqnarray}
\nabla^2 U_\mathrm{tail}
& = &
\sum_{j\ne k}
\left\{  u_\mathrm{tail}''(q_{jk})
%\right. \nonumber \\ & & \left. \mbox{ }
+  \frac{ 2 }{ q _{jk} } u_\mathrm{tail}'(q_{jk})
\right\}
\nonumber \\ & = &
-4\epsilon \sum_{j\ne k}
\left\{ \frac{42 \sigma^6}{q_{jk}^8}
-  \frac{12 \sigma^6}{q_{jk}^8} \right\} .
\end{eqnarray}
Hence the tail of the quantum correction is
\begin{eqnarray}
-\beta \Omega_{1,2}^\mathrm{tail}
& = &
\frac{-  \hbar^2 \beta^2}{24m} \left<  \nabla^2 U_\mathrm{tail} \right>_{1,0}
\nonumber \\ & = &
\frac{-  \hbar^2 \beta^2}{24m}
4\pi \rho^2 V (-4\epsilon)
\int_{q_\mathrm{cut}}^\infty \mathrm{d}q \, q^2
 \frac{30 \sigma^6}{q^8}
 \nonumber \\ & = &
\frac{- 4\pi \rho^2 V \hbar^2 \beta^2}{24m}
(-4\epsilon)
  \frac{6 \sigma^6}{q^5_\mathrm{cut}} .
\end{eqnarray}

%%%%%%%%%%%%%%%%%%%%%%%%%%%%%%%
\subsection{Higher  Order Analysis}

%%%%%%%%%%%%%%%%%%%%%%%%%%%%%%%%%%%%%%%
\subsubsection{Fourth Order Correction}

The classical fluctuation of the weight exponent
$w({\bf \Gamma})$  is
\begin{equation}
\Delta_w \equiv w - \left<  w  \right>_{1,0} .
\end{equation}
Define a phase space constant that is
a series of averages of powers of the fluctuation
\begin{eqnarray}
D(\Delta_w)  & \equiv &
\left<  w  \right>_{1,0}
+ \frac{1}{2}\left<  \Delta_w^2  \right>_{1,0}
+ \frac{1}{3!} \left<  \Delta_w^3  \right>_{1,0}
\nonumber \\ &&   \mbox{ }
+ \frac{1}{4!} \left<  \Delta_w^4  \right>_{1,0}
- \frac{1}{8} \left<  \Delta_w^2  \right>_{1,0}^2 .
\end{eqnarray}
This is constant in phase space,
and so it can be taken in and out of averages.
The average of the exponential of $w$ less this particular constant is
\begin{eqnarray}
\lefteqn{
\left<  e^{ w-D(\Delta_w) }  \right>_{1,0}
} \nonumber \\
& = &
\left<    \exp
\left\{ \Delta_w
- \frac{1}{2}\left<  \Delta_w^2  \right>_{1,0}
- \frac{1}{3!} \left<  \Delta_w^3  \right>_{1,0}
\right. \right. \nonumber \\ && \left. \left.  \mbox{ }
- \frac{1}{4!} \left<  \Delta_w^4  \right>_{1,0}
+ \frac{1}{8} \left<  \Delta_w^2  \right>_{1,0}^2
 \right\}   \right>_{1,0}
\nonumber \\ & = &
1 + \left<  \Delta_w  \right>_{1,0}
- \frac{1}{2}\left<  \Delta_w^2  \right>_{1,0}
- \frac{1}{3!} \left<  \Delta_w^3  \right>_{1,0}
- \frac{1}{4!} \left<  \Delta_w^4  \right>_{1,0}
\nonumber \\ && \mbox{ }
+ \frac{1}{8} \left<  \Delta_w^2  \right>_{1,0}^2
%\nonumber \\ && \mbox{ }
+ \frac{1}{2}
\left< \left[  \Delta_w - \frac{1}{2}\left<  \Delta_w^2  \right>_{1,0}
 \right]^2  \right>_{1,0}
\nonumber \\ && \mbox{ }
+ \frac{1}{3!} \left<
\left[  \Delta_w - \frac{1}{2}\left<  \Delta_w^2 \right>_{1,0} \right]^3
\right>_{1,0}
%\nonumber \\ && \mbox{ }
+ \frac{1}{4!} \left<  \Delta_w^4  \right>_{1,0} + \ldots
\nonumber \\ & = &
1 +  \left<  {\cal O}( \Delta_w^5 )  \right>_{1,0} .
\end{eqnarray}
This enables expansions to $ {\cal O}(\hbar^4)$
to be obtained relatively painlessly.
For example,
the monomer potential is
\begin{eqnarray}
e^{-\beta [ \Omega_1-\Omega_{1,0} ]}
& = &
\frac{\Xi_1}{\Xi_{1,0}}
\nonumber \\ & = &
\left<    e^w  \right>_{1,0}
\nonumber \\ & = &
e^{ D(\Delta_w) } \left<  e^{ w-D(\Delta_w) }  \right>_{1,0}
\nonumber \\ & = &
e^{ D(\Delta_w) }
\left[ 1 + \left<  {\cal O}( \Delta_w^5 )  \right>_{1,0} \right].
\end{eqnarray}
Hence
\begin{equation}
-\beta  \Omega_1
=
-\beta  \Omega_{1,0}
+ D(\Delta_w) +  {\cal O}( h^5 ) ,
\end{equation}
assuming that $w$ itself has been obtained to at least fourth order.

More generally, for obtaining averages and loop potentials,
for a phase function $A$ define the fluctuation
\begin{equation}
\Delta_{Aw}
\equiv
w - \left<  w  \right>_{1,A}
\equiv
w  - \frac{ \left<  A w  \right>_{1,0} }{ \left<  A  \right>_{1,0} }  .
\end{equation}
Define the phase space constant
\begin{eqnarray}
D(\Delta_{Aw})  & \equiv &
\left<  w  \right>_{1,A}
+ \frac{1}{2}\left<  \Delta_{Aw}^2  \right>_{1,A}
+ \frac{1}{3!} \left<  \Delta_{Aw}^3  \right>_{1,A}
\nonumber \\ &&   \mbox{ }
+ \frac{1}{4!} \left<  \Delta_{Aw}^4  \right>_{1,A}
- \frac{1}{8} \left<  \Delta_{Aw}^2  \right>_{1,A}^2 .
\end{eqnarray}
One has
\begin{eqnarray}
%\lefteqn{
\left< A e^{ w }  \right>_{1,0}
%} \nonumber \\
& = &
\left< A \right>_{1,0} \left< e^{ w }  \right>_{1,A}
\nonumber \\ & = &
\left< A \right>_{1,0}
e^{D(\Delta_{Aw})}
\left< e^{ w - D(\Delta_{Aw}) }  \right>_{1,A}
\nonumber \\ & = &
\left< A \right>_{1,0}
e^{D(\Delta_{Aw})}
[1 +  {\cal O}(\hbar^5 )  ] .
\end{eqnarray}
This result gives directly the monomer average of
of a function (operator) of the position configuration $A({\bf q})$.

For the loop potential one need only take
$A \Rightarrow X^{(l)}$.
With this the loop potential $l \ge 2 $ can be expanded as
\begin{eqnarray}
-\beta \Omega_l
& = &
\frac{  \left<  X^{(l)}   e^w  \right>_{1,0}
}{ \left<   e^w  \right>_{1,0} }
\nonumber \\ & = &
\frac{ \left<   X^{(l)}  \right>_{1,0} \left<  e^{w}  \right>_{1,X^{(l)}}
}{ \left<   e^w  \right>_{1,0} }
\nonumber \\ & = &
\left<   X^{(l)}  \right>_{1,0}
e^{ D(\Delta_{X^{(l)}w}) - D(\Delta_w) }
+  {\cal O}( h^5 ) .
\end{eqnarray}

The full average of a function of position is
given by Eq.~(\ref{Eq:<O(r)>-W}) (with $W \Rightarrow e^w$),
\begin{eqnarray}
\lefteqn{
\left<  O({\bf r}) \right>_{\mu,V,T}
}  \\
& = &
\frac{\langle O e^w \rangle_{1,0}}{\langle e^w \rangle_{1,0}}
\prod_{l=2}^\infty \exp
% \nonumber \\ &&   \mbox{ } \times
\left<X^{(l)}
\left[
\frac{Oe^w}{\langle Oe^w \rangle_{1,0}}
 -  \frac{ e^w }{\langle e^w \rangle_{1,0}}
 \right] \right>_{1,0}
 .\nonumber
\end{eqnarray}
One need only replace $A$ by an appropriate function
in each average on the right hand side
to obtain a relatively simple expression that is valid to $ {\cal O}( h^4 )$.

%%%%%%%%%%%%%%%%%%%%%%%%%%%%%%
\subsubsection{Gaussian Form} \label{Sec:w-Gauss}

One can assume,
with a confidence approaching certainty in the thermodynamic limit,
that the value of an extensive physical variable is Gaussian distributed.

In the present case this means that
\begin{equation}
\wp(w) =
\frac{1}{\sqrt{2\pi\sigma^2}}
e^{ - \Delta_w^2/2\sigma^2 } ,
\end{equation}
where $\Delta_w \equiv w - \langle w \rangle_{1,0}$
and $\sigma = \sqrt{\langle \Delta_w^2 \rangle_{1,0}}$.
Hence
\begin{eqnarray}
\left\langle e^w \right\rangle_{1,0}
& = &
\int \mathrm{d} w \; \wp(w) e^w
\nonumber \\ & = &
\frac{1}{\sqrt{2\pi\sigma^2}}
\int \mathrm{d} w \;
e^{ \langle w \rangle_{1,0} }
e^{ -( \Delta_w - \sigma^2  )^2 /2\sigma^2 }
e^{ \sigma^2/2 }
\nonumber \\ & = &
e^{ \langle w \rangle_{1,0} }
e^{ \langle \Delta_w^2 \rangle_{1,0}/2 } .
\end{eqnarray}
This is consistent with the leading terms derived explicitly above.
Obviously one can replace the subscript $1,0$ by $1,A$,
and $\Delta_w$ by $\Delta_{Aw}$,  as appropriate.

%%%%%%%%%%%%%%%%%%%%%%%%%%%%%%%%%%%%%%%%%%%%%%%%%%%%%%%%%%%%%%%%%%%%%%%%%%
\end{document}